\documentclass[12pt,letterpaper,notitlepage]{article}

\usepackage{floatrow}
\usepackage[strict,backend=bibtex]{biblatex-chicago}
\usepackage{endnotes}
\usepackage{graphicx}
\usepackage{subfigure}
\usepackage{bm}
\usepackage{multirow}
\usepackage{amssymb}
\usepackage{amsmath}

\usepackage[paperwidth=215.9mm,paperheight=279.4mm,centering,hmargin=2.54cm,vmargin=2.54cm]{geometry}

\usepackage{setspace}
\doublespace

\newcommand{\capdef}{}
\newcommand{\mycaption}[2][\capdef]{\renewcommand{\capdef}{\sf #2}%
        \caption[{\sf #1}]{{\sf  #2}}} 

\renewcommand{\endnote}[1]{\footnote{#1}}
\bibliography{references}
\begin{document}

% page numbers bottom-center
\pagestyle{plain}
\thispagestyle{empty}
%%%%%%%%%%%%%%%%%%%%%%%%%%%%%%%%%%%%%%%%%%%%%%%%%%%%%%%%%%%%%%%%%%%%%%%%%%%%
\vspace*{1.5cm}

\begin{center}
\textbf{\Large Antineutrino reactor safeguards -- a case study}
\end{center}

\vspace*{.8cm}

{\begin{center} {{\bf
        Eric Christensen,
                Patrick Huber$^\dagger$,
                Patrick Jaffke
                }
                }
\end{center}}
{\it
\begin{center}
       Center for Neutrino Physics, Virginia Tech, Blacksburg, VA 24061, USA
\end{center}}
\vspace*{0.5cm}

%%%%%%%%%%%%%%%%%%%%%%%%%%%%%%%%%%%%%%%%%%%%%%%%%%%%%%%%%%%%%%%%%%%%%%%%%%%%
\begin{abstract}
Antineutrinos have been proposed as a means to safeguard nuclear
reactors for more than 30 years and there has been impressive
experimental progress in antineutrino detection that makes this method
increasingly practical for use by the International Atomic Energy
Agency. In this paper we conduct, for the first time, a case study of
the application of antineutrino safeguards to a real-world scenario --
the North Korean nuclear crisis in 1994. We derive detection limits to
a partial or full core discharge in 1989 based on actual IAEA
safeguards access and find that two independent methods would have
yielded positive evidence for a second core with very high confidence.
To generalize our results, we provide detailed estimates for the
sensitivity to the plutonium content of various types of reactors,
including most types of plutonium production reactors, based on
detailed reactor simulations. A key finding of this study is that a
wide class of reactors with a thermal power of 0.1-1\,GW$_\mathrm{th}$
can be safeguarded achieving IAEA goals for quantitative sensitivity
and timeliness with antineutrino detectors right outside the reactor
building. This type of safeguards does not rely on the continuity of
knowledge and provides information about core inventory and power
status in real-time.
\end{abstract}

\vspace{0.5cm}
\vfill
{\footnotesize{$^\dagger$\makebox[1.cm]{Email:}\sf pahuber@vt.edu}}

\newpage

%%%%%%%%%%%%%%%%%%%%%%%%%%%%%%%%%%%%%%%%%%%%%%%%%%%%%%%%%%%%%%%%%%%%%%%%%%%

\tableofcontents

\newpage

\section{Introduction}

The first use of nuclear weapons in 1945, at the end of World War II,
had a profound and permanent impact on foreign relations and
international security. While initially there was some hope that the
secrets of the manufacture of nuclear weapons would remain exclusively
in the hands of the United States, the Soviet Union tested its first
nuclear device in 1949. Several times during the Cold War, the world
stood at the brink of nuclear armageddon. It was only due to the
strong commitment of political leaders on both sides and the high
degree of professionalism in the armed forces that the disaster of a
nuclear war could be averted~\endnote{\fullcite{abyss}. For a different
  perspective, why we the Cold War remained cold, see for
  instance~\fullcite{Sagan}.}. During the Cold War, nuclear security
was essentially a bipolar issue between the United States and the
Soviet Union; other players like Great Britain (1952), France (1960),
and China (1964) appeared at the fringes but did not play a major role
for most parts.  To some degree independently of Cold War politics,
Israel and South Africa launched a nuclear weapons program in the
early 60s and 70s, respectively, which in both cases was triggered by
unique national security needs -- a small minority population
surrounded by hostile neighbors, which in turn resulted in a rather
unusual alliance~\autocite{alliance}. South Africa, quite remarkably,
relinquished its nuclear weapons and the associated infra-structure
towards the end of the apartheid regime in 1991~\autocite{doyle}.
India's nuclear program was launched quite early and presumably was a
direct response to its deteriorating relations with Pakistan. Also
confrontations with China over territories in the Himalayas in
combination with China obtaining a permanent seat on the UN Security
Council contributed significantly to the decision to go nuclear.
Obviously, India's possession of nuclear weapons since 1974, then
created a perceived need for nuclear armament in Pakistan, which first
tested a nuclear device in 1998. The last country to join the circle
of nuclear armed nations was the Democratic People's Republic of Korea
(DPRK) with its first nuclear test in 2006.

The Treaty on the Non-proliferation of Nuclear Weapons (NPT) was
opened for signature in 1968, entered into force in 1970, and on May
11, 1995 the NPT was extended indefinitely~\autocite{NPT}. The NPT is,
with exception of the UN charter, the most widely accepted
international treaty to date. Currently, 190 states are party to the
NPT. The legal mechanisms for IAEA safeguards, as set out in
article~III of the NPT, are bi-lateral agreements between individual
member states and the IAEA. These so-called comprehensive safeguards
agreements have been put into force by all but 12 of the
non-nuclear-weapons states~\autocite{NPTstatus}.  The Additional
Protocol was introduced in response to the failure of the regular
safeguards scheme to provide timely indication of Saddam Hussein's
nuclear weapons program before the first Gulf War in 1990. The
Additional Protocol in particular provides IAEA inspectors with the
right to collect environmental samples at locations outside of
declared facilities and to obtain access to sites which have not been
declared as nuclear facilities but are suspected to be. These
provisions close an important gap in the regular safeguards scheme,
which relies on a state's declaration of nuclear facilities and
materials.  139 states have signed the Additional Protocol, and 117
states have put it into force~\autocite{iaea}.  The regular safeguards
scheme could only confirm the correctness of a state's declaration of
nuclear activities; with the Additional Protocol, the completeness of
the declaration can also be addressed\endnote{In principle, special
  inspections could also provide the means to verify the completness,
  independent of whether the Additional Protocal is in force.}. It can
be argued that the correctness aspect of safeguards is working quite
well. No case of diversion of fissile material has been documented at
safeguarded facilities, which presumably is due to the fact that a
potential proliferator deems the risk of discovery to be unacceptably
high~\autocite{Zykov2012}. However, the completeness aspect remains
troubling, especially for those states which have not put the
additional protocol into force, such as Iran.

 Neutrinos were postulated by Wolfgang Pauli in 1930 and have been
 experimentally discovered by Clyde Cowan and Fred
 Reines~\autocite{Cowan:1992xc} in 1956 using neutrinos\endnote{To be
   precise, a reactor is a source of electron antineutrinos. In the
   interest of brevity and given the fact that here we are dealing
   only with electron antineutrinos, we will use the term neutrino,
   instead.} from the Savannah River reactor. Neutrinos are nearly
 massless, electrically neutral, spin 1/2 particles and play a central
 role in the electroweak Standard Model of particle physics. Neutrinos
 participate only in weak interactions and therefore possess unusual
 penetrating power -- no practical means to attenuate or to shield
 neutrinos are known. Neutrinos are copiously produced in the
 beta-decays of fission fragments and this makes nuclear reactors the
 most powerful artificial neutrino source. The basic concept to
 monitor nuclear reactors using neutrinos was proposed by Borovoi and
 Mikaelyan in 1978~\autocite{Borovoi:1978}. There have been a number of
 quantitative studies of the level of accuracy at which the plutonium
 content in a reactor can be determined using
 neutrinos~\autocites{Bernstein:2001cz}{Nieto:2003wd}{Huber:2004xh}{Misner:2008}{Bernstein:2009ab}{Bulaevskaya:2010wy}{Hayes:2011ci}{huberINMM},
 and different authors seem to come to different conclusions. Closer
 inspection of those results reveal that very different assumptions
 about detector capabilities are made and also the level of
 statistical analysis, particularly in terms of rates versus spectral
 information, is very different. These differences likely account for
 the variety of opinions on the feasibility and quantitative accuracy
 of neutrino safeguards. In particular, the assumptions about detector
 capabilities seem to be strongly influenced by earlier safeguards
 detector deployments and do not reflect modern state-of-the-art
 neutrino detectors. We will discuss these issues in detail in
 section~\ref{sec:neutrinos}. We will not discuss the application of
 neutrinos for long-range detection of nuclear activities\endnote{for a
 recent review on this topic, see \fullcite{Jocher:2013gta}}.

Here, we present the first case study of a real safeguards scenario --
the first nuclear crisis in the DPRK in
1994~\autocite{GoingCritical}. Earlier studies of neutrinos for
safeguards are, to a large degree, based on pre-conceived notions of
how safeguards of a particular reactor type work, and thus do not
allow critical examination of the strength and weaknesses of neutrino
safeguards as compared to more conventional means. In many cases, this
comparison seems to disfavor neutrinos, not because neutrinos do not
offer any new capabilities, but because the conventional techniques
are specifically designed to work well in those standard
scenarios~\autocite{Zykov2012}. On the other hand, inventing scenarios
in which the standard methods fail brings about the criticism that
these scenarios are artificial, unrealistic, and contrived for the
sole purpose of demonstrating the usefulness of neutrinos. In the rare
case that any of these scenarios would reflect an actual concern of
the professional safeguards community, it is far from obvious that
anyone in that community would want to admit it.  Therefore, what is
needed is a real-world case in which conventional methods did not
yield the desired outcome for the IAEA and for which sufficient
information is publicly available to perform a detailed technical
analysis. The first North Korean nuclear crisis fits this bill on all
accounts and we, therefore, have chosen it as our sandbox to explore
the abilities and limitations of neutrino safeguards. Despite the
brief discussion of the impact neutrino safeguards might have had on
the unfolding of history in section~\ref{sec:matters}, the main thrust of
this study is not an attempt at counter-factual history but to
demonstrate that under real-world constraints and boundary conditions,
neutrino safeguards can provide a decisive advantage over conventional
techniques, in particular, with a view of the next nuclear crisis in
some other part of the world.

This paper is organized as follows: the technical aspects of neutrino
safeguards and the general principles applicable to a wide range of
reactor types and situations are presented in section~\ref{sec:neutrinos}
. In particular, figure~\ref{fig:scaling} is one of our main results. In
section~\ref{sec:crisis}, we present a summary of the first North Korean
nuclear crisis based mostly on historic records. The specific features
of the North Korean nuclear program relevant to this study are
summarized in section~\ref{sec:dprkpu}. In section~\ref{sec:dprknu}, we
apply the techniques developed in section~\ref{sec:neutrinos} to the
particular problems posed by the North Korean nuclear program and
provide detailed quantitative analysis for four detector deployment
options. The resulting improvements in the quantitative understanding
of the DPRK's nuclear program and in particular the assessment of the
veracity of the initial declaration of the DPRK to the IAEA are
discussed in Sec~\ref{sec:appl1994}. Also, a critical comparison to
conventional techniques is offered. We summarize in
section~\ref{sec:summary}. Appendices~\ref{sec:5MWSCALE} to~\ref{sec:IRT}
provide the details of our reactor simulations.

%%%%%%%%%%%%%%%%%%%%%%%%%%%%%%%%%%%%%%%%%%%%%%%%%%%%%%%%%%%%%%%%%%%%%%%
\section{Antineutrino reactor safeguards}
%%%%%%%%%%%%%%%%%%%%%%%%%%%%%%%%%%%%%%%%%%%%%%%%%%%%%%%%%%%%%%%%%%%%%%
\label{sec:neutrinos}

The fact that nuclear reactors are powerful neutrino sources was
realized soon after nuclear reactors became practical. Neutrinos are
not directly produced in nuclear fission but result from the
subsequent beta-decays of the neutron-rich fission fragments. On
average there are about 6 neutrinos per fission emitted and thus, for
one gigawatt of thermal power a flux of about
$10^{20}\,\mathrm{s}^{-1}$ neutrinos is produced. The total number of
emitted neutrinos is proportional to the total number of fissions in
the reactor. Moreover, the distribution of fission fragments, and
hence their beta-decays, are different for different fissile
isotopes. Thus, careful neutrino spectroscopy should provide
information not only on the total number of fissions but also about
the fission fractions of the various fissile isotopes contained in the
core. The basic concepts~\autocite{Borovoi:1978} of both power
monitoring and observing the plutonium content of a reactor were
experimentally demonstrated in pioneering work performed by a group
from the Kurchatov Institute lead by Mikaelyan. They deployed a
neutrino detector of about 1\,$\mathrm{m}^3$ volume at the Rovno
nuclear power plant. For the power measurement, an agreement with the
thermal measurements was found to within
2.5\%~\autocite{Korovkin:1988} and the effect due to a changing
plutonium content was demonstrated~\autocite{Klimov:1990}; more
recently the quantitative accuracy has been studied as
well~\autocite{Huber:2004xh}.  This allows one to determine the
plutonium content and power level of the reactor core {\it in situ} at
a standoff distance of 10's of
meters~\autocite{Klimov:1994,huberINMM}. The practical feasibility of
reactor monitoring using neutrinos has also been demonstrated using a
small, tonne-size detector at the San Onofre power station, called
SONGS~\autocite{Bernstein:2008tj}.

%%%%%%%%%%%%%%%%%%%%%%%%%%%%%%%%%%%%%%%%%%%%%%%%%%%%%%%%%%%%%%%%%%%%
\subsection{Neutrino detection}

Beginning with the discovery of the neutrino, inverse beta-decay (IBD)
has been the workhorse of reactor neutrino experiments
\begin{equation}
\label{eq:ibd}
\bar\nu_e + p \rightarrow n + e^+
\end{equation}
In IBD, an electron antineutrino interacts with a proton to produce a
neutron and a positron. Due to the mass difference of a neutron and a
proton as well as the mass of the positron, this process has an
approximate energy threshold of $(m_n-m_p+m_e)c^2=1.8$\,MeV. The
positron will go on to annihilate with a nearby electron producing a
pair of 511\,keV gamma rays. This energy deposition is typically
detected together with the kinetic energy of the positron $E_e$ and
thus the visible energy in detector, $E_\mathrm{vis}=E_e+2\times
511\,\mathrm{keV}$. There is a one-to-one correspondence between
neutrino energy and the positron energy $E_\nu=E_e+1.8\,$MeV. Here we
neglect the recoil energy of the neutron which is much smaller than
the energy resolution of even the best neutrino detectors. Therefore,
a measurement of $E_\mathrm{vis}$ directly translates into a
measurement of the neutrino energy $E_\nu$.

The reaction in equation~\ref{eq:ibd} also results in a neutron, which in
itself is invisible to the detector, but will slow down in collisions
with the detector material and eventually undergo neutron capture. A
careful choice of the nucleus on which the neutron captures allows 
tailoring this signature. Common neutron capture agents are gadolinium,
e.g. Daya Bay~\autocite{An:2012eh} or lithium, e.g.
Bugey~\autocite{Declais:1994su}. In the case of gadolinium, the signature
of neutron capture is the emission of several gamma rays with a total
energy of 8\,MeV, whereas in the case of lithium the signature is the
production of an alpha particle and a $^3$H nucleus. The slowdown and
capture of the neutron requires a characteristic time, allowing for
what is called a delayed coincidence: there is a primary energy
deposition from the positron followed somewhat later by a neutron
capture signal. This delayed coincidence is key to separate neutrino
events from backgrounds. The neutron capture cross sections of both
gadolinium and lithium are much larger than of any of the other
detector materials. Therefore, even small concentrations at a level of
a percent or less will result in the majority of neutron captures
occurring on those nuclei.

Eventually, all signatures will result in ionization and this
ionization is detected by using organic scintillator which can be
either liquid or solid. The organic nature of the scintillator
provides the free protons for the interaction in
equation~\ref{eq:ibd}. Recently, there have been three
experiments~\autocite{Abe:2011fz,An:2012eh,Ahn:2010vy} aimed at
fundamental physics employing gadolinium-doped liquid scintillator at
a large scale of several 10 tonnes without any safety incidents and
excellent long-term stability. Specifically, throughout this paper we
consider a 5\,t detector based on organic scintillator corresponding
to $4.3\times10^{29}$ target protons. A real detector will not have
100\% efficiency and to obtain the same number of events a larger
detector will be needed. Many neutrino detectors with efficiencies
above 50\% have been built and thus even a realistic detector yielding
the same event numbers would be less than 10\,t. A standard 20 feet
intermodal shipping container has an interior volume of
33.1\,$\mathrm{m}^3$ and a net load capacity of 28.2\,t, thus even a
10\,t neutrino detector fits easily within such a container together
with its support systems. The neutrino spectrum is divided in energy
from 1.8\,MeV to 8\,MeV in bins of 0.2\,MeV width, which at 4\,MeV
approximately corresponds to $10\%/\sqrt{E}$ resolution, which is
similar to the resolution of recent
experiments~\autocite{Abe:2011fz,An:2012eh,Ahn:2010vy}. We checked
that a resolution half as good would yield virtually identical
results. For the IBD cross section we use the result of Vogel and
Beacom~\autocite{Vogel:1999zy} corrected for a neutron lifetime of
878.5\,s\endnote{This values is taken from~\autocite{Serebrov:2004zf}
  and is very close to the value of 880\,s currently recommended by
  the Particle Data Group,~\autocite{PDG}. It should be mentioned that
  there still are measurements deviating significantly from that value
  by several standard deviations, see e.g.~\autocite{Yue:2013qrc}. }.  For all
measurements at reactors, the standoff is 20\,m, which for both of the
considered reactors would allow for deployment outside the reactor
building. Such a detector at this standoff would typically register
about 5,000 events per year for a reactor operating at
1\,MW$_\mathrm{th}$ throughout that year.

%%%%%%%%%%%%%%%%%%%%%%%%%%%%%%%%%%%%%%%%%%%%%%%%%%%%%%%%%%%
\subsection{Reactor flux models}
\label{sec:fluxes}

More than 99\% of the power in reactors, in a uranium fuel cycle, is
produced in the fission of four isotopes: uranium-235, plutonium-239,
uranium-238, and plutonium-241. A reactor with fresh fuel starts with
only fissions in the uranium isotopes and plutonium is produced via
neutron capture on uranium-238 as the burn-up increases. The total
neutrino flux from a reactor $\phi$ can be written as
\begin{equation}
\phi(E)=\sum_I f_I S_I(E)\,,
\end{equation}
where $f_I$ is the fission rate in isotope $I$ and $S_I(E)$ is the
neutrino yield for the isotope $I$. The thermal power of the reactor
is also given in terms of the fission rates
\begin{equation}
\label{eq:power}
P_\mathrm{th}=\sum_I f_I p_I\,,
\end{equation}
where $p_I$ is the thermal energy release in one fission of the
isotope $I$; we use the values for $p_I$ given by
Kopeikin~\autocite{Kopeikin:2004cn}.  In order to be able to
disentangle the contributions of the four isotopes, we need to know
the neutrino yields $S_I$. These neutrino yields, in principle, are
given by the neutrino spectra $\nu_k(E)$ of each fission fragment $k$
and the cumulative fission yield for each fragment, $Y_k^I$,
\begin{equation}
S_I(E)=\sum_k Y_k^I \nu_k(E)\,,
\end{equation}
where $k$ typically runs over about 800 isotopes. In practice, we do
not know the neutrino spectrum of a given fission fragment, but have
only information regarding the beta spectrum and in many cases this
knowledge is inaccurate, incomplete, or entirely missing. Even for a
well known beta spectrum, significant complications arise from the
conversion of a beta spectrum into a neutrino spectrum since each
individual beta-decay branch has to be treated separately. As a result,
a direct computation of the neutrino yields $S_I$ via the summation of
all individual neutrino spectra will be of limited
accuracy~\autocite{Mueller:2011nm,Fallot:2012jv}, but in many cases is the
only available method.

A more accurate method is based on the measurement of the integral
beta spectrum of all fission
fragments~\autocite{VonFeilitzsch:1982jw,Schreckenbach:1985ep,Hahn:1989zr,Haag:2013raa}
and subsequently the neutrino spectrum can be reconstructed from those
measurements~\autocite{Huber:2011wv}. This method is less dependent on
nuclear data about individual fission fragments but is not entirely
free from uncertainties related to effects of nuclear
structure~\autocite{Huber:2011wv,Hayes:2013wra}.

We need to point out that the problem of neutrino yields has recently
received significant scrutiny. Until the 2011 work by a group from
Saclay~\autocite{Mueller:2011nm}, the results by Schreckenbach et al.
~\autocite{VonFeilitzsch:1982jw,Schreckenbach:1985ep,Hahn:1989zr},
obtained in the 1980s at the Institut Laue-Langevin in Grenoble were
considered the gold standard. The Saclay group, in preparation of the
Double Chooz neutrino experiment~\autocite{Abe:2011fz}, revisited the
previous results in an attempt to reduce the uncertainties. Instead,
they found a upward shift of the central value of the average yield by
about 3\% while the error budget remained largely unchanged. This
result, in turn, requires a reinterpretation of a large number of
previous reactor neutrino experiments, since this changes the expected
number of events. Together with the changes of the value of the
neutron lifetime~\autocite{Greene} and corrections from so-called
non-equilibrium effects, the previous experiments appear to observe a
deficit in neutrino count rate of about 6\%; this is called the
\emph{reactor antineutrino anomaly} and was first discussed by Mention
et al.~\autocite{Mention:2011rk}. The initial result on the flux
evaluation and the 3\% upward shift was independently confirmed by one
of the authors~\autocite{Huber:2011wv}. A plausible explanation could
come in the form of a new particle, a sterile neutrino, which is not
predicted by the Standard Model of particle physics. Given the
far-flung consequences of the existence of this sterile neutrino a
considerable level of research activity ensued\endnote{for a recent
  review, see~\fullcite{Abazajian:2012ys}}.

In table~\ref{tab:fluxes} the event rate predictions for various flux
models are compared for the four fissile isotopes. The ENSDF flux
model is based on thermal neutron fission yields of uranium-235,
plutonium-239, and plutonium-241 from the JEFF database,
version~3.1.1~\autocite{jeff}; the fast neutron fission yield of
uranium-238 from the ENDF-349 compilation conducted at Los Alamos
National Laboratory~\autocite{lanl}; and on the beta-decay information
contained in the Evaluated Nuclear Structure Data File (ENSDF)
database, version~VI~\autocite{ensdf}. The neutrino spectrum is
computed following the prescription of
Huber~\autocite{Huber:2011wv}. Our ENSDF model represents a very crude
summation calculation and we reproduce the measured total beta
spectra~\autocite{VonFeilitzsch:1982jw,Schreckenbach:1985ep,Hahn:1989zr,Haag:2013raa}
to within about 25\%. A state-of-the-art summation calculation is
given by Fallot {\it et al.}~\autocite{Fallot:2012jv}, where great
care is taken to replace the ENSDF entries with high quality
experimental data where available and to use a carefully selected mix
of databases. This model reproduces the measured total beta
spectra~\autocite{VonFeilitzsch:1982jw,Schreckenbach:1985ep,Hahn:1989zr,Haag:2013raa}
to within 10\%. Finally, a direct deconvolution of the neutrino
spectra from the total beta data was performed by
Huber~\autocite{Huber:2011wv} for the isotopes uranium-235,
plutonium-239, and plutonium-241, which to this date represents the
most accurate neutrino yields for those isotopes. We note that the
absolute values differ significantly between models, but once we
normalize the predictions for total rate and mean energy to that of
uranium-235, the predictions become very similar. In other words, the
difference in neutrino yield and mean energy between the fissile
isotopes is consistently predicted by the various flux models -- which
is not surprising given that these differences have their origin in
the fission yields.
\begin{table}[t]
\sf
\begin{center}
\begin{tabular}{l ccc|ccc|ccc}
&\multicolumn{3}{c|}{ENSDF} & \multicolumn{3}{c|}{Fallot} & \multicolumn{3}{c}{Huber} \\ \cline{2-10}
& rate & $\langle E\rangle$ & $\langle E\rangle$  & ratio & $\langle E\rangle$&  $\langle E\rangle$ & ratio &$\langle E\rangle$ &    $\langle E\rangle$\\
&ratio&$[\mathrm{MeV}]$&ratio&ratio&$[\mathrm{MeV}]$&ratio&ratio&$[\mathrm{MeV}]$&ratio\\

 \hline
 uranium-235  & 1 &4.48&1 & 1 &4.28& 1 & 1 & 4.25 & 1 \\
 uranium-238  & 1.53 &4.59 & 1.024 & 1.56 &4.45 & 1.040 &    & &  \\
 plutonium-239 & 0.64 &4.26&0.950 & 0.65 &4.13 &0.965  & 0.66 & 4.04 & 0.951\\
 plutonium-241 & 0.93&4.47&0.998 & 0.90 &4.23&0.988 & 0.91 &4.13 & 0.971
\end{tabular}
\mycaption{\label{tab:fluxes} Rates and mean energies $\langle E \rangle
  $ for a 1\,MW$_\mathrm{th}$ reactor in a 1\,t detector at a standoff
  of 10\,m measuring for 1 year for each individual isotope, assuming
  that only this isotope is fissioning. The three different flux
  models are explained in the text. Ratios are given relative to
  uranium-235.}
\end{center}
\end{table}

In practice, the current errors of any flux model are significant and
a set of calibration measurements at reactors of known fissile content
is likely required to mitigate the effect of these uncertainties,
particularly in view of the reactor antineutrino anomaly. A proof of
concept at a theoretical level for these calibrations has been
performed~\autocite{huberINMM}.  On the experimental side, the Daya
Bay collaboration has demonstrated the ability to cross-calibrate a
set of 8 neutrino detectors to within better than
0.5\%~\autocite{DayaBay:2012aa}.

\subsection{Reactor physics}
\label{sec:reactors}

\begin{figure}
\includegraphics[height=0.4\textwidth]{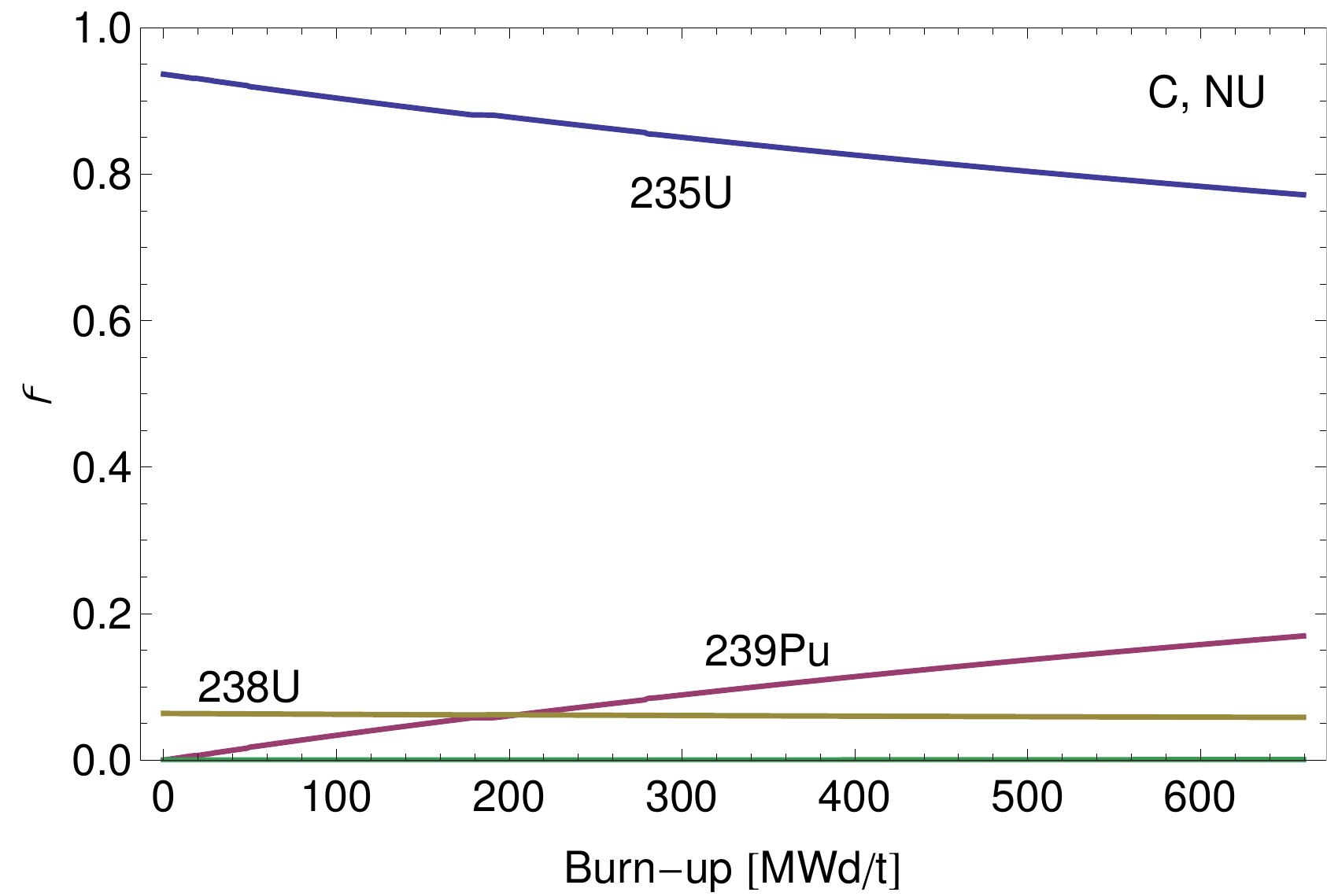}~~%
\includegraphics[height=0.39\textwidth]{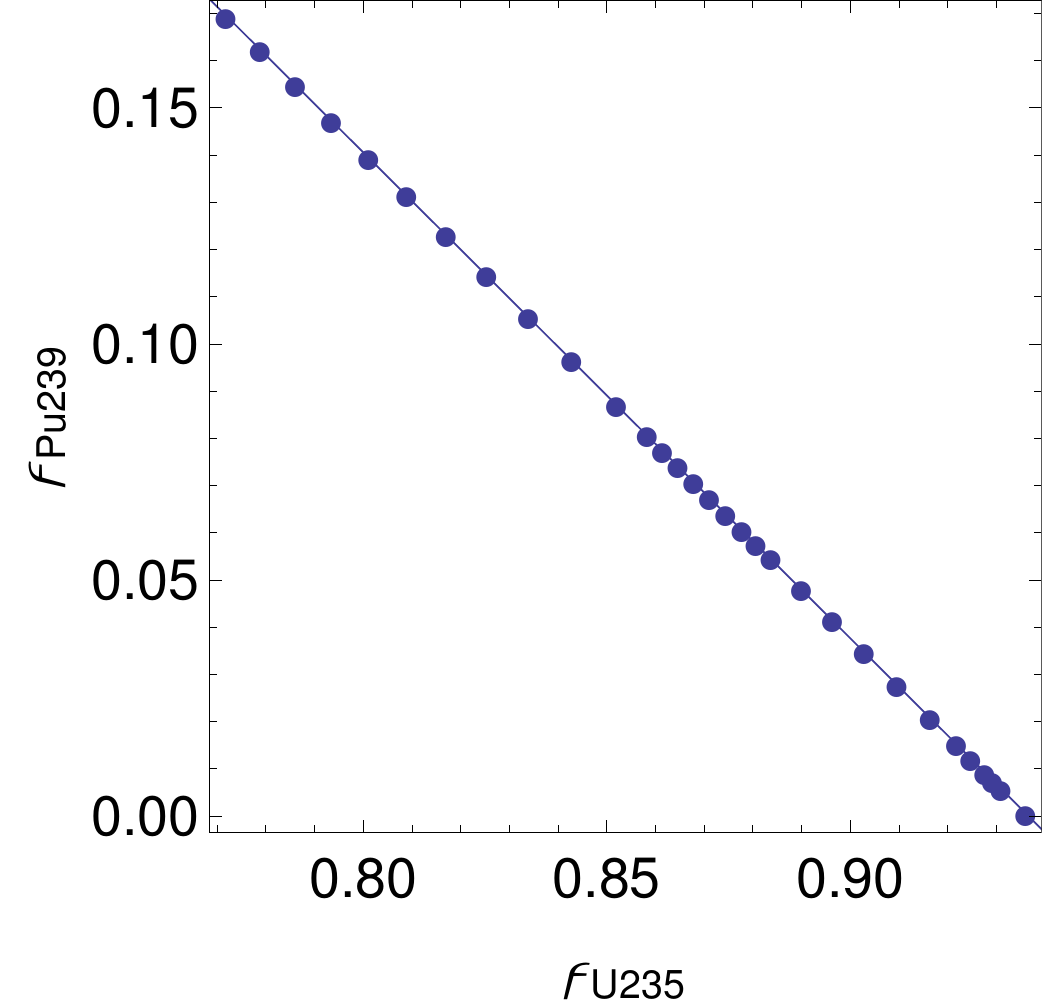}
\mycaption{\label{fig:reactorphysics}The left hand panel shows the
  evolution of the fission fractions in a graphite moderated natural
  uranium fueled reactor as a function of burn-up. The right hand
  panel shows the anti-correlation of the fission fractions in
  uranium-235 and plutonium-239.}
\end{figure}

The connection between fission rates and mass inventory requires a
more detailed look at the reactor physics inside the core; our
ultimate goal is to infer mass inventories.  For a neutron flux which
is constant in time and space, the fission rate and mass of a given
fissile isotope have a simple linear relationship
\begin{equation}
\label{eq:fissions}
f_I=\phi_n\,\sigma_I\,m_I\,,
\end{equation}
where $m_I$ is the mass of isotope $I$, $\sigma_I$ is the energy
averaged fission cross section and $\phi_n$ is the neutron
flux. Throughout the evolution of the core, all factors on the right
hand side of equation~\ref{eq:fissions} will change. Due to burn-up
effects, the mass $m_I$ will change and the neutron flux typically
will be adjusted to compensate for changes in reactivity while
maintaining constant power. The accumulation of fission fragments will
change the neutron absorption, which, in turn, alters the neutron
energy spectrum; the cross section $\sigma_I$ will evolve as well. The
evolution of the isotopic content can be described by a set of Bateman
equations and neutron transport methods can be used to recompute the
relevant cross sections. We have performed evolution or burn-up
calculations for several reactor types using the SCALE software
suite~\autocite{scale}.  For the further discussion it is useful to
introduce fission fractions $\digamma_I$, which are defined by
\begin{equation}
\label{eq:fractions}
\digamma_I=\frac{f_I}{\sum_I f_I}\,\quad\text{with}\quad \sum_I\digamma_I=1\,.
\end{equation}
This definition has the advantage that the problem can be phrased
independently of reactor power. For illustration, the time evolution
of the $\digamma_I$ for a graphite moderated, natural uranium fueled
reactor is given in the left hand panel of
figure~\ref{fig:reactorphysics}, where the fission fractions are shown
as a function of the burn-up. $\digamma_\mathrm{Pu241}$ is very close
to zero in this type of reactor and therefore is not visible in this
figure.
The fission rate in uranium-238 stays constant since the amount of
uranium-238 in the reactor changes very little with time.  There is a
clear anti-correlation between the fission fractions in uranium-235 and
plutonium-239. The anti-correlation is nearly exact as shown in the right
hand panel of figure~\ref{fig:reactorphysics} and we will make use of
this later. In this context, it turns out that the burn-up is a useful
variable which allows a summary of the reactor inventory with a single
number. Burn-up measures the number of fissions which have occurred
per unit of fuel mass or, in other terms, the amount of energy
extracted; the unit for burn-up is MWd/t. For example, 1\,tonne of
fuel producing 5\,MW for 1 day yields a burn-up of 5\,MWd/t; the same
burn-up would be obtained by 1\,tonne of fuel running at 1\,MW for 5
days. Neglecting radioactive decays, the isotopic composition of both
samples would be identical since the total number of fissions which
took place is the same. As a result, the reactor core evolution is, to
a very high degree of accuracy, a function of only the burn-up. That
is, details of the power history, like innage factors and shut downs,
have only a minor impact on the reactivity and fission fractions. The
amount of plutonium produced depends on the details of the reactor
operations and so does the resulting neutrino signal. Therefore, we
need to have a reasonably accurate model of the reactor power history,
which in turn serves as input for a detailed reactor physics
calculation. Neutrino emission is a result of radioactive decay of
fission fragments and therefore, fuel of the same burn-up will have,
to very good approximation, the same isotopic composition and will
produce the same distribution and amount of neutrinos at a given power
level. Therefore, our ability to predict the neutrino emission over
time relies on an accurate model of the burn-up as a function of time.

This burn-up calculation also allows for the study of the time evolving
relation between fission fractions and mass inventory as given
in equation~\ref{eq:fissions}. We find, to very good accuracy, this is a
linear relationship and the time evolution of the proportionality
constant $\phi_n \sigma_I$, throughout the fuel cycle, is very
small. Using a fixed value for $\phi_n \sigma_\mathrm{Pu239}$,
throughout the reactor cycle, induces a root mean square error of 2\%
for plutonium mass determinations for a graphite moderated reactor
and errors of similar size for the other reactor types considered
later. This type of sensitivity study needs to be performed
for each reactor type and design. Also, the actual values of
$\phi_n\sigma_I$ have to be determined for each specific case.

%%%%%%%%%%%%%%%%%%%%%%%%%%%%%%%%%%%%%%%%%%%%%%%%%%%%%%%%%%%%%%%%%%%%%%%%
\subsection{Plutonium content determination}
\label{sec:pucontents}

The difference in the spectral neutrino yield, of the
four fissile isotopes, can be used to disentangle the contribution of each
of those isotopes to the total neutrino flux. In order to do so, we
set up a binned $\chi^2$-analysis, where the event rate in each bin
$n_i$ is given as
\begin{equation}
n_i=N\,\sum_I f_I \int_{E_i-\Delta E/2}^{E_i+\Delta E/2 } dE\,\sigma(E)\,S_I(E)\,,
\end{equation}
where $E_i$ is central energy of bin $i$, $\Delta E$ is the bin width
and $\sigma(E)$ is the IBD cross section. $N$ is an overall
normalization constant set by the detector mass, or number of free
protons, detection efficiency, and time interval of data taking. In order to
compute the event rates $n_i$, we have to specify the four fission
rates
$\mathbf{f}=(f_\mathrm{U235},f_\mathrm{U238},f_\mathrm{Pu239},f_\mathrm{Pu241})$. We
denote the true or input values for our calculation by a
superscript 0, i.e. the true fission rates are $\mathbf{f}^0$ and,
in the same way, we will denote the $n_i$ computed for the true values
$\mathbf{f}^0$ as $n_i^0$. We define the $\chi^2$-function as
\begin{equation}
\chi^2(\mathbf{f}):=\sum_i \frac{\left(n_i(\mathbf{f})-n_i^0\right)^2}{n_i^0}\,,
\end{equation}
This $\chi^2$-function will be zero for $\mathbf{f}=\mathbf{f}^0$. The allowed region for $\mathbf{f}$ is obtained by requiring that
\begin{equation}
\chi^2(\mathbf{f})\leq \chi^2_c\,,
\end{equation}
where the critical value $\chi^2_c$ is determined from a $\chi^2$
probability distribution with, in this case, 4 degrees of freedom. If
we are only interested in the total number of fissions in plutonium
given by $f_\mathrm{Pu}=f_\mathrm{Pu239}+f_\mathrm{Pu241}$, the
following marginalized function has to be used
\begin{equation}
\bar\chi^2(f_\mathrm{Pu})=\min_{f_\mathrm{U235},f_\mathrm{U238},\kappa} \chi^2(f_\mathrm{U235},f_\mathrm{U238},(1-\kappa)f_\mathrm{Pu},\kappa f_\mathrm{Pu})\,,
\end{equation}
and in this case, since we are interested only in the single parameter
$f_\mathrm{Pu}$ the number of degrees of freedom is 1. Similarly,
we can define a corresponding single parameter function for the
measurement of reactor power.

\begin{figure}[t]
\centering \includegraphics[width=0.7\textwidth]{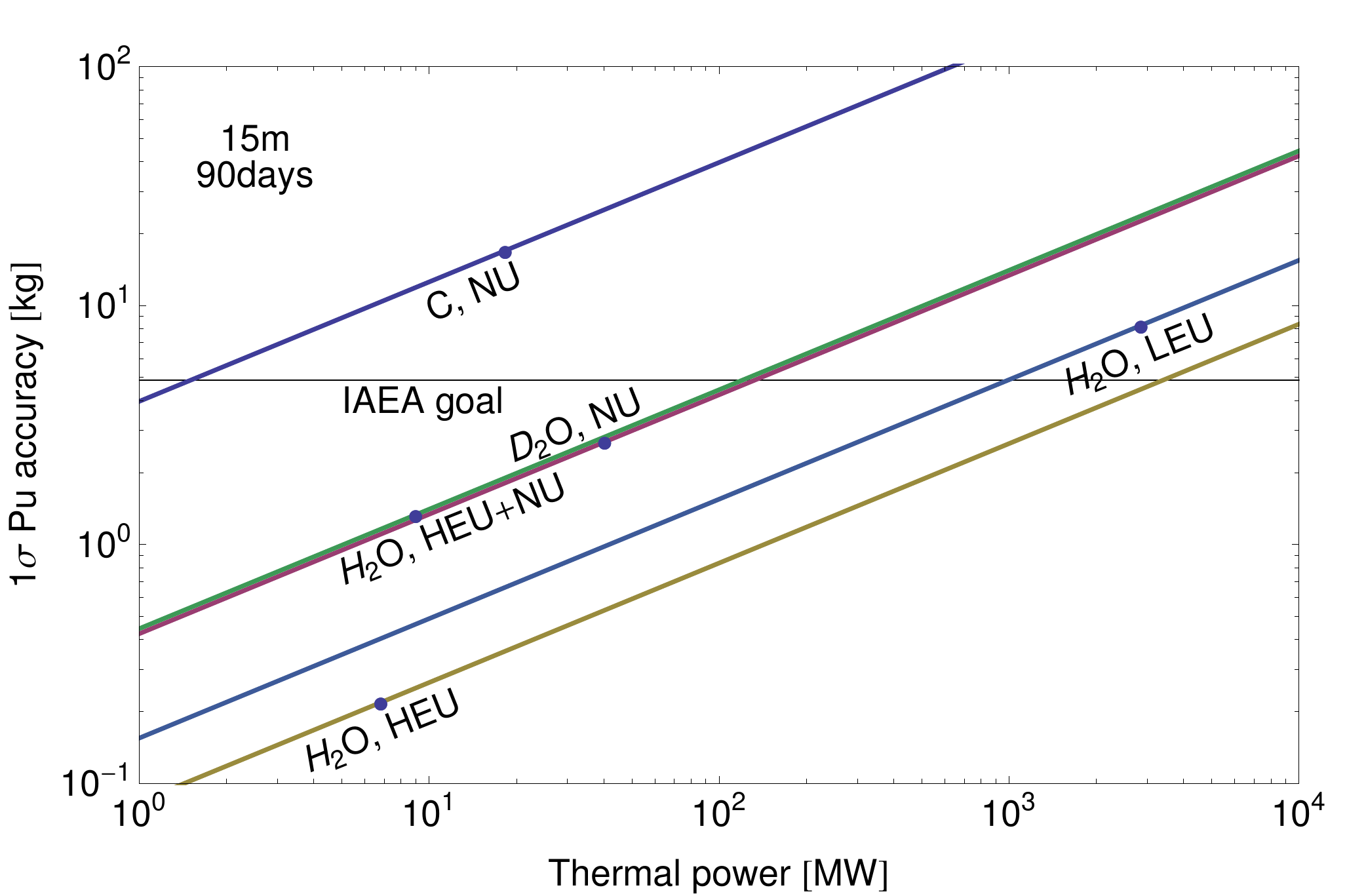}
\mycaption{\label{fig:scaling} Absolute accuracy in the determination
  of the plutonium content based on the measurement of the neutrino
  spectrum as a function of the thermal power of the reactor. The
  different lines stand for different types of reactors as indicated
  by the labels: the first term indicates the type of moderator,
  whereas the second part denotes the fuel type, natural uranium (NU),
  low enriched uranium (LEU) and highly enriched uranium (HEU). This
  figure assumes a 5\,t detector, a standoff of 15\,m from the reactor
  core, and 90\,days of data taking. The horizontal line labeled
  ``IAEA goal'' indicates the accuracy which corresponds to the
  detection of 8\,kg of plutonium at 90\% confidence level.}
\end{figure}

To relate a measured value of $f_\mathrm{Pu}$ to the mass inventory, a
reactor physics simulation is required. $f_\mathrm{Pu}$ will be
proportional to the plutonium mass, $m_\mathrm{Pu}$, in the reactor
\begin{equation}
\gamma=\frac{m_\mathrm{Pu}}{f_\mathrm{Pu}}\,,
\end{equation}
where $\gamma$ is the proportionality constant.  Therefore, a
measurement of $f_\mathrm{Pu}$ translates into a determination of
$m_\mathrm{Pu}$. $\gamma$, in turn, depends on the details of the
reactor physics as well as the instantaneous reactor thermal power;
note that according to equation~\ref{eq:fissions}, $\gamma=1/(\phi_n
\sigma_\mathrm{Pu}$) and thus is inverse to the neutron flux density
$\phi_n$. The determination of $m_\mathrm{Pu}$ and its connection to
$\gamma$ is clearly illustrated in figure~\ref{fig:scaling}, where we
show the accuracy in the determination of $m_\mathrm{Pu}$ for a
variety of reactor types as a function of the thermal power.
This figure is based on a full calculation of the reactor burn-up,
where ``C, NU'' corresponds to a graphite moderated reactor running on
natural uranium and the dot on this line is the 5\,MW$_\mathrm{e}$
reactor, whose simulation details are explained in
appendix~\ref{sec:5MWSCALE}. ``$\mathrm{H}_2$O, HEU'' and
``$\mathrm{H}_2$O, HEU + NU'' correspond to the IRT with drivers only
and to the IRT with drivers and targets, respectively. The details of
the simulation are explained in appendix.~\ref{sec:IRT}. The case
``$\mathrm{H}_2$O, LEU'' is computed for a typical pressurized light
water reactor. We have taken a power history from one such reactor,
with a total fuel load of 72.4\,MTU\endnote{MTU stands for metric
  tonne of uranium and is often synonymous with metric tonnes of
  heavy metal (HM), where heavy refers to all actinides including
  plutonium.} enriched to 3.7\%.  The case ``$\mathrm{D}_2$O, NU''
describes a heavy water moderated reactor running on natural uranium
modeled on a CANDU design with a 8.6\,MTU natural uranium fuel
load and running at 40\,MW$_\mathrm{th}$. The 40\,MW$_\mathrm{th}$
point on this line resembles, in many aspects, the Iranian reactor at
Arak~\autocite{Arak} and the accuracy would be at the level of 2.7\,kg
within 90 days. The details of the calculations for the
``$\mathrm{H}_2$O, LEU'' and ``$\mathrm{D}_2$O, NU'' can be found in
appendix~\ref{sec:CANDULEU}. The horizontal line corresponds to a sensitivity to 8\,kg plutonium within 90 days, which is the stated IAEA goal\endnote{The
    IAEA uses this interpretation of the term “significant quantity”
    as the design basis for planning routine inspections in declared
    facilities.  Any amount of fissile material diversion, or
    undeclared production, would be sufficient to warrant suspicion
    and follow-up activities to determine whether or not
    non-compliance might be considered by the IAEA Board of
    Governors.}.

For all of those quite different reactor types the accuracy of a
$m_\mathrm{Pu}$ measurement can be described by the following simple
relation
\begin{equation}
\label{eq:scaling}
\delta m_\mathrm{Pu}=1.942 \,\mathrm{kg} \left(\frac{\gamma}{10^{-16}\mathrm{kg}\,\mathrm{s}}\right)\left(\frac{L}{\mathrm{m}}\right)\left(\frac{P_{th}}{\mathrm{MW}}\right)^{1/2}\left(\frac{\mathrm{tonnes}}{M}\right)^{1/2}\left(\frac{\mathrm{days}}{t}\right)^{1/2}\,,
\end{equation}
where $L$ is the standoff of the neutrino detector, $P_\mathrm{th}$ is
the average thermal reactor power, $M$ is the detector mass in tonnes
(assuming $8.65\times 10^{28}$ protons per tonne), and $t$ is the
length of the data taking period. Table~\ref{tab:scaling} lists the
corresponding values of $\gamma$, and
\begin{table}[h!]
\sf
\begin{tabular}{cccccc}

reactor type&C, NU & $\mathsf{H}_2$O, HEU & $\mathsf{H}_2$O, HEU+NU & $\mathsf{D}_2$O, NU & $\mathsf{H}_2$O, LEU\\
\hline
$\gamma\,[10^{16}\,\mathsf{kg}\,\mathsf{s}]$&2.889&0.064&0.337&0.299&0.108
\end{tabular}
\mycaption{\label{tab:scaling} The values of $\gamma$ for a number of
  reactor types.}
\end{table}
using those values, equation~\ref{eq:scaling} reproduces the results of the
full calculation within a few percent. For graphite moderated
reactors, we find that the resulting $\delta m_\mathrm{Pu}$ is
significantly larger, by a factor of at least 8.5, than for any other
reactor type we have investigated. As we will show in the following,
the fact that neutrino safeguards still yield meaningful results and
are applicable for this reactor type is a testimony to the great
versatility and power of this technique.

For most reactor running conditions, the variation in $\gamma$ is very
small and depends only very weakly, at the level of a few percent, on
burn-up and reactor history. This implies that our result most likely
will hold up even for detailed 3-dimensional reactor physics
calculations, taking into account spatial burn-up variations.

We further observe, that for reactors with a thermal power in excess
of 1\,GW$_\mathrm{th}$, which is the bulk of all reactors globally
used for electricity production, this approach to safeguards will have
difficulties in meeting the IAEA goal of detection of 1 significant
quantity, which for plutonium is 8\,kg, within 90
days\endnote{Plutonium in irradiated fuel is a so-called
  \emph{indirect use nuclear material} and the precise IAEA goal is a
  90\% or higher confidence level detection of the diversion of 1
  significant quantity within 90 days, according
  to~\fullcite{IAEAglossary}.}. On the other hand, neutrino safeguards
is quite straightforward for research, small modular reactors, and
plutonium production reactors.

As discussed in the previous section, the fission fractions and thus
the fission rates are \emph{not} independent from each other but are
coupled by the physics inside the reactor; for an illustration, see
the right hand panel of figure~\ref{fig:reactorphysics}. In trying to
determine the plutonium mass inventory, we can make use of these
correlations. Basically, reactor physics determines how the fission
rates evolve together with burn-up. Therefore, a reactor model will
provide the fission rates as a function of burn-up. This allows a
rephrasing of the fitting problem in terms of one independent quantity
-- the burn-up. The result of the analysis will be a value for burn-up
and some error bounds and since the reactor model also provides all
the mass inventories as a function of burn-up, a measurement of the
burn-up translates into a measurement of the core inventory and the
errors can be determined by standard error propagation. In the case of
the graphite moderated reactor this reduces the error in plutonium
mass determination by roughly 50\%; for details, see
section~\ref{subsection:5MW}. One potential drawback is the reliance on a
reasonably accurate reactor model. In cases where there is reliable
design information and the key operating parameters are known the
burn-up model will reproduce the core inventory to within the 5-10\%
range, which for most purposes will be a small extra contribution to
the overall error budget. In those cases, where the reactor design and
operating parameters have to be considered as unknown or the knowledge
is deemed unreliable, a fit to fission fractions and power should be
performed. The loss in sensitivity is moderate compared to the
increase in reliability of the result.

%%%%%%%%%%%%%%%%%%%%%%%%%%%%%%%%%%%%%%%%%%%%%%%%%%%%%%%%%%%%%%%%%%%%%
\section{Summary of the 1994 crisis}
%%%%%%%%%%%%%%%%%%%%%%%%%%%%%%%%%%%%%%%%%%%%%%%%%%%%%%%%%%%%%%%%%%%%%
\label{sec:crisis}

The DPRK is rather unique in many regards, including its use of a
nuclear weapons program as a bargaining tool.  The direct tactical use
of its small and crude nuclear arsenal against the U.S., or its
regional allies like South Korea or Japan, is presumably deterred by
the threat of U.S. retaliation. It is a serious concern that North
Korea may share its nuclear know-how, materials, or even a fully
functional weapon with third parties, but the fear of the likely
attribution in case of a nuclear incident and the accompanying U.S.
reaction may counteract this risk~\autocite{hecker}. So far, North
Korea obtained the largest benefit from its nuclear adventures by
offering to abstain in the future. For the use as a bargaining tool,
it is desirable to create a large degree of ambiguity about the type
and scope of nuclear activities. At the same time, there are
indications that North Korea was surprised by the level of information
IAEA could glean from environmental sampling and allowing the IAEA to
employ this method in the intial inspections may have been a serious
miscalculation on the side of North Korea~\autocite{Baradei}. The
three nuclear tests in 2006, 2009, and 2013 have removed a great deal
of ambiguity about the kind and goal of North Korea's nuclear
activities, but they shed no light on the scope of activities and the
size of the resulting arsenal. New concerns have surfaced relating to
the uranium enrichment program~\autocite{HeckerUranium}.

\begin{figure}
\includegraphics[width=\textwidth]{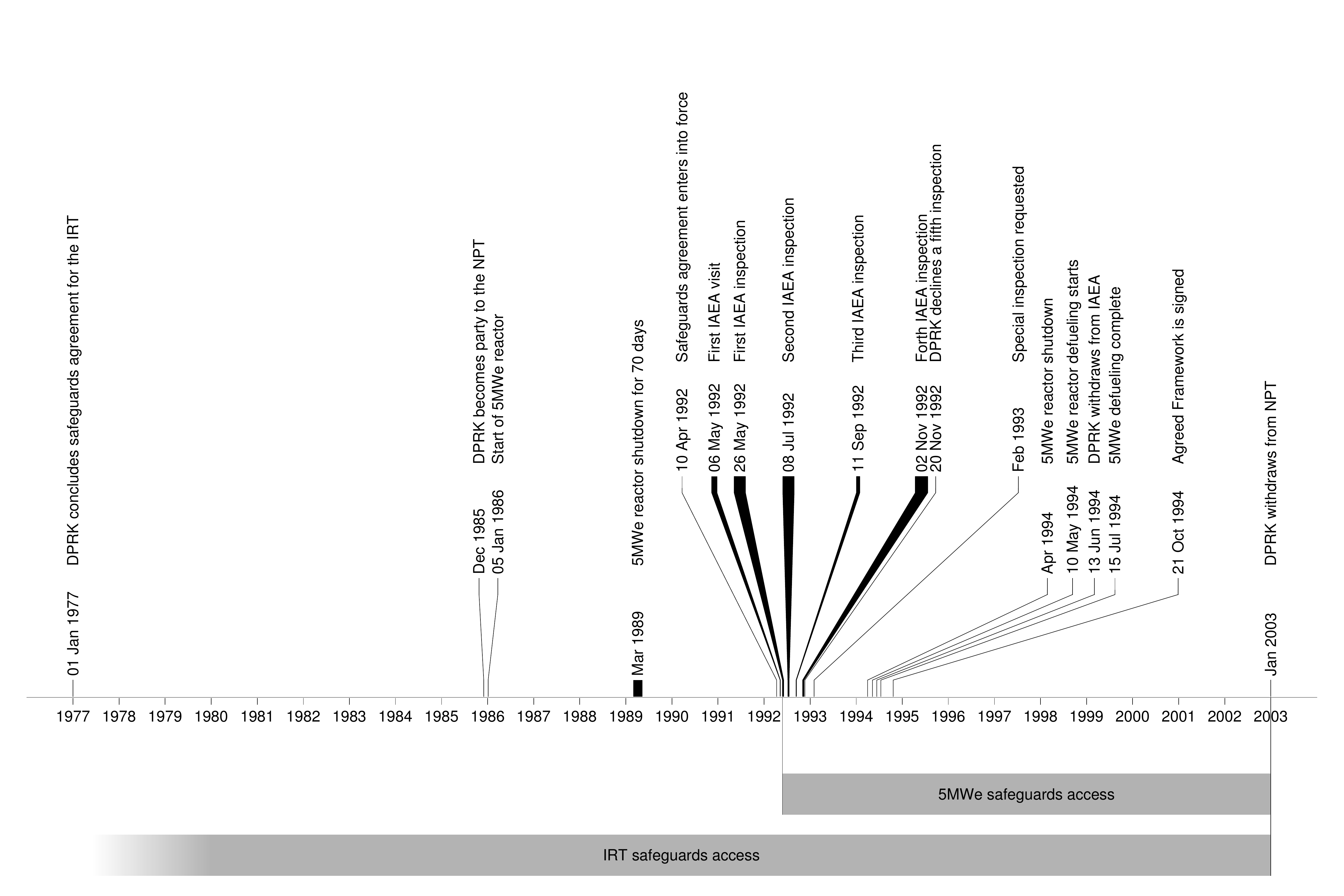}
\mycaption{\label{fig:timeline} Time line of events}
\end{figure}

The DPRK signed the NPT on December 12, 1985; a safeguards agreement
entered into force on April 10, 1994; and notice of withdrawal from
the treaty was given on January 10, 2003~\autocite{iaeahistory}. On
February 26, 1993, the IAEA called for special inspections, which in
retrospect may have been counterproductive~\autocite{Baradei}, to
resolve the discrepancies found during the first safeguards
inspections in 1992. The issue of contention was the amount of
plutonium the DPRK had separated from spent nuclear fuel -- North
Korea declared it produced about 90\,g~\autocite[p. 269]{Oberdorfer},
but IAEA data allowed for the possibility of a much larger amount,
maybe as much as 14\,kg~\autocite{AlbrightBulletin}, which would be
sufficient to build two or more nuclear bombs.  On March 12, 1993, the
DPRK declared its intention to leave the NPT by June 12, 1993 after
being threatened with special inspections but was persuaded by the
U.S. on June 11 not to do so. A detailed representation of the time
line is given in figure~\ref{fig:timeline}.

IAEA safeguards ended in 2003 and therefore we would like to focus on
the time before 2003. Would antineutrino reactor safeguards have been
able to reduce the pre-2003 ambiguities about the DPRK's plutonium
production program and what would the potential impact of this
information on the development of the crisis have been?

North Korea joined the NPT in 1985, but it took until spring of 1992
for a safeguards agreement to enter into force. North Korea started
its $5\,\mathrm{MW}_e$ reactor at Yongbyon in 1986. In 1989 there was
a 70 day shutdown, providing an opportunity to unload between 50-100\%
of the spent fuel in the core. In its initial declaration to IAEA in
1992, North Korea indicated that they ran a one-time reprocessing
campaign in 1990 that resulted in 90\,g of plutonium from a
limited number of damaged fuel rods removed during the 1989 shutdown.
The results of IAEA environmental sampling conducted during the first
safeguards inspection in 1992, however, indicated at least three
campaigns of reprocessing in 1989, 1990, and
1991~\autocite{AlbrightBulletin} which in turn admits the hypothesis
that a significant fraction of the spent fuel had been removed in 1989
and subsequently reprocessed. As a result, a larger amount of
separated plutonium may have been obtained by the DPRK, possibly
sufficiently large to build two or more nuclear bombs. Given the
ramifications of these findings, IAEA Director General Hans Blix
insisted on a definitive resolution of this question as a precondition
to declare the DPRK to be in compliance with its commitments under the
NPT. In particular, finding and sampling the reprocessing waste
streams was a priority for IAEA, eventually triggering the request for
special inspections~\autocite{HeinonenInterview}. The diplomatic
exchange between IAEA and the DPRK dragged on in parallel with
negotiations between the DPRK and the U.S.; the latter eventually
leading to the Agreed Framework. In April 1994 North Korea forced the
issue by beginning to unload spent fuel from the reactor core. An
analysis of the gamma-radiation of spent fuel taken at known positions
in the reactor core would have resolved the question of how much spent
fuel was discharged in 1989 and whether the North Korean declaration
was correct. Knowing the original position of a sample in the reactor
core is crucial for this analysis, since the fission rate is higher in
the center than at the edge of the core; for technical details of this
method see section~\ref{sec:conventional}. However, the unloading
proceeded very fast and it appears as if the operators took deliberate
steps to obliterate any information about the original position of
each fuel element in the reactor; effectively, IAEA inspectors could
only observe the unloading but were unable to take any meaningful
measurements of any individual fuel elements as they were being
removed, or to make a connection between the fuel elements and the
core locations they had occupied. As a result, crucial evidence was
denied to the IAEA and on June 2, 1992 Blix declared that the ability
to resolve the issue had been ``seriously
eroded''~\autocite{AlbrightBulletin}. The fuel discharged in 1994 was
canned using U.S. equipment and subsequently was put into storage and
was under IAEA surveillance until 2003, when the DPRK declared its
withdrawal from the NPT. The 1994 crisis was resolved by the so called
\emph{Agreed Framework} under which the DPRK halted any plutonium
production and fuel reprocessing in exchange for the promise to obtain
two pressurized light-water reactors at not
cost~\autocite{GoingCritical}.  The \emph{Agreed Framework} unraveled
in 2003 and eventually, in 2006, North Korea conducted its first
nuclear test explosion.

%%%%%%%%%%%%%%%%%%%%%%%%%%%%%%%%%%%%%%%%%%%%%%%%%%%%%%%%%%%%%%%%%%%%%%%%
\section{Plutonium production in the DPRK}
%%%%%%%%%%%%%%%%%%%%%%%%%%%%%%%%%%%%%%%%%%%%%%%%%%%%%%%%%%%%%%%%%%%%%%%%
\label{sec:dprkpu}

\begin{figure}
\begin{center}
\includegraphics[width=\textwidth]{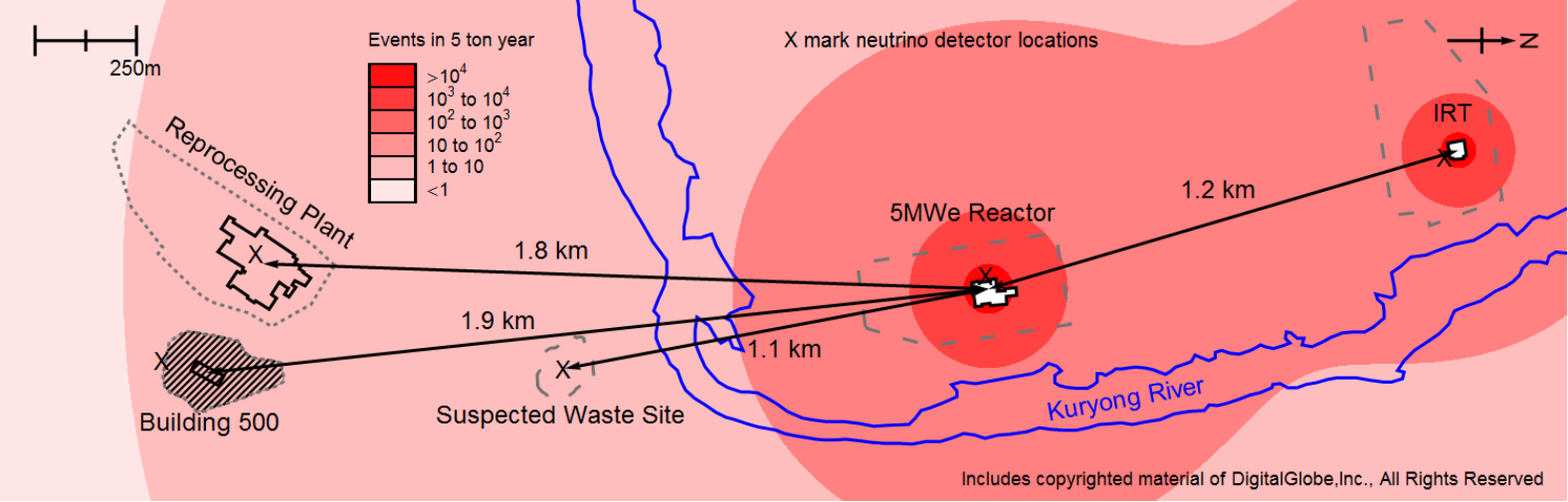}
\end{center}
\mycaption{A map of relevant boundaries and geographies of the Yongbyon
  nuclear facility. Contours show expected inverse beta-decay event
  rates for a 5\,t detector over the course of a year. X's mark
  the location of various neutrino detectors used in the paper. The
  satellite image on which this map is based was taken on May 16, 2013
  by GeoEye-1.}
\label{fig:NeutrinoMap}
\end{figure}

A comprehensive account of the history of the North Korean nuclear
program is provided by Hecker~\autocite{hecker,Hecker2010}. For our
purposes, three facilities are relevant: a Soviet supplied research
reactor with a power of around 8\,MW$_\mathrm{th}$, the IRT; a
graphite moderated reactor with a thermal power of approximately
20\,MW$_\mathrm{th}$, which generally is referred to by its electrical
power, hence the name 5\,MW$_\mathrm{e}$; and the Radiochemical
Laboratory, which is a reprocessing facility which allows for the
extraction of plutonium from the spent fuel from the
5\,MW$_\mathrm{e}$ reactor. These facilities and their relative
locations are shown in figure~\ref{fig:NeutrinoMap}.  Features such as
the river and relevant buildings are outlined, neutrino detector
locations are marked, and IBD event rate iso-contours are shown.

In the 1960s, the IRT was supplied by the Soviet
Union~\autocite{Hecker2010}. This reactor is a light-water moderated
reactor running on highly enriched uranium, with enrichment from 10\%
to 80\%~\footfullcite[p.\,148]{puzzle}. The Soviet Union also provided the HEU
fuel until its own demise in the 1990s. With this reactor, the Isotope
Production Laboratory, a facility for handling irradiated materials,
was provided. The nominal power of this reactor is
8\,MW$_\mathrm{th}$. Using the laboratory, early, small scale
plutonium separation experiments may have been conducted with fuel or
targets irradiated in this reactor~\autocite[p.\,92]{puzzle}.

North Korea started serious fuel cycle activities in the 1980s and the
plan was to build and operate three gas-cooled, graphite moderated,
natural uranium fueled reactors. A $5\,\mathrm{MW}_\mathrm{e}$ and
$50\,\mathrm{MW}_\mathrm{e}$ reactor were foreseen for the Yongbyon
site and a $200\,\mathrm{MW}_\mathrm{e}$ power reactor was planed at
Taechon.  The design followed the British Magnox design, where Magnox
is the name of the alloy used for the fuel cladding: magnesium
non-oxidizing. The thermal power of Magnox reactors is typically 4-6
times higher than the above quoted electrical power, so they are much
less efficient than, for instance, pressurized light-water reactors,
where this factor is closer to 3.  Apart from efficiency, the choice
of Magnox has another severe drawback: Magnox fuel cladding corrodes
in contact with water such that long term storage of spent fuel under
water is not possible. This makes encapsulation or some level of
reprocessing essentially mandatory~\autocite{AlbrightBulletin}.  The
attractive features of this design are its simplicity and that it does
not require uranium enrichment or the use of exotic moderators like
heavy water. So, this reactor type was well adapted to North Korean
indigenous industrial capabilities. At the same time, Magnox reactors
were originally designed as dual-use facilities to produce both
electricity \emph{and} weapons-grade plutonium.

The amount of plutonium produced in a reactor can be estimated if the
integrated neutron flux, which is proportional to the total energy
produced, is known, or equivalently if a \emph{complete} history of
the reactor power is available.  It turns out that all uncertainty
about the produced amounts of plutonium center on the issue of the
completeness, and to a lesser degree, the uncertainty of the record of
the power history. To obtain the produced plutonium in usable form,
the reactor has to be shut down\endnote{In principle, Magnox reactors
  can be refueled under load, but the DPRK seems not to have mastered
  this technology at that time.}, the irradiated fuel rods removed, and
the plutonium then needs to be chemically separated from the spent
fuel at the Radiochemical Laboratory. The location of the various
facilities can be seen in figure~\ref{fig:NeutrinoMap}.

\begin{figure}
\includegraphics[width=0.7\textwidth]{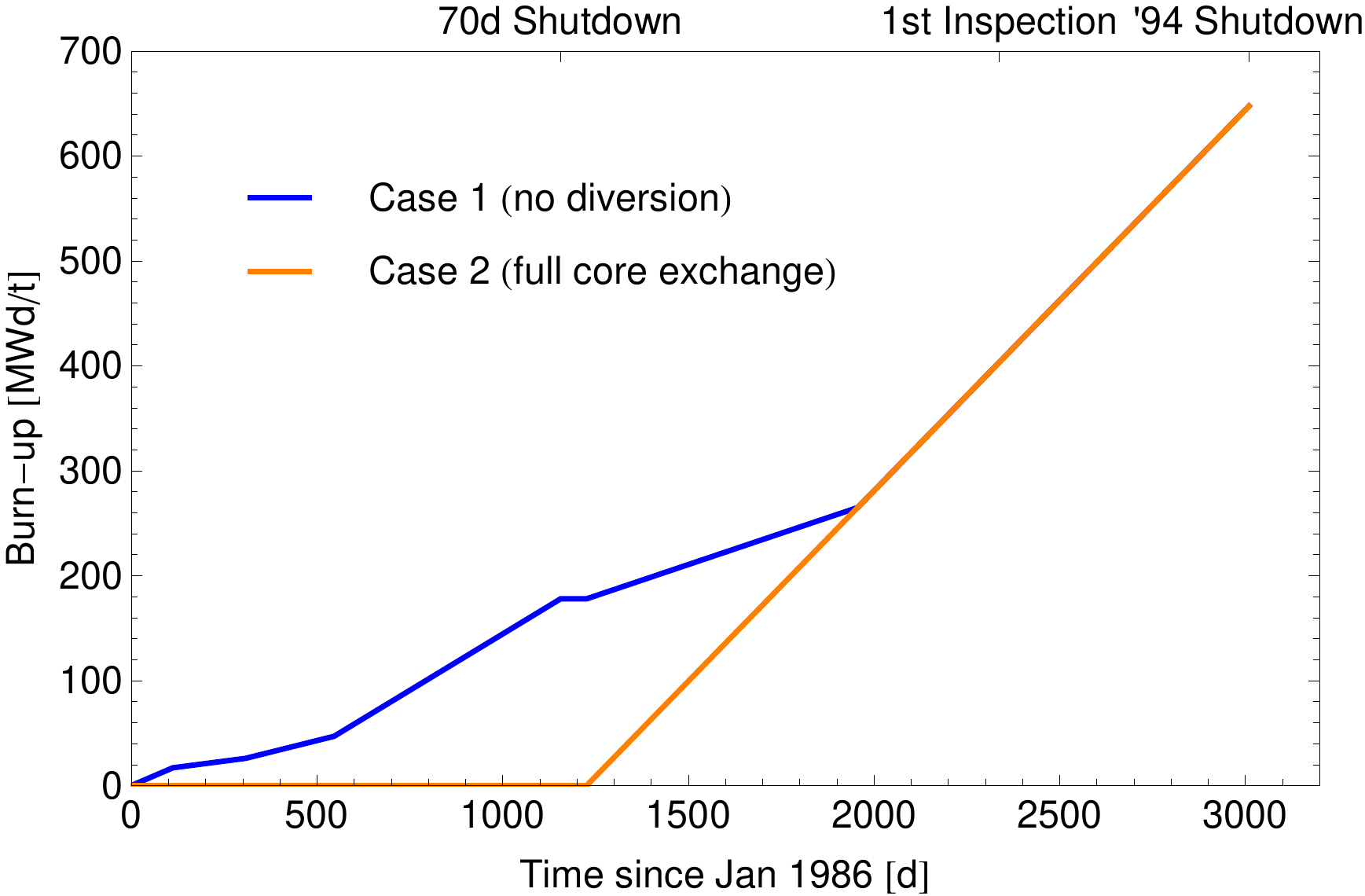}
\mycaption{\label{fig:AlbrightBU} Burn-up of the fuel in the
  5\,MW$_\mathrm{e}$ reactor as function of time measured in days
  since January 1, 1986.  The blue curve is based on the values
  declared by the DPRK, i.e. no major refueling has taken place
  in 1989. The orange curve is derived assuming that the full core has
  been replaced with fresh fuel in 1989. Figure adapted from
  \shortcite{puzzle}.}
\end{figure}

The time evolution of the burn-up for the 5\,MW$_\mathrm{e}$ is shown
in figure~\ref{fig:AlbrightBU} which has been adapted
from~\citetitle{puzzle} and is deemed
accurate~\autocite{HeinonenBurnup}. The information in this figure is
the backbone of the analysis presented here and our quantitative
results are based on this information.  The blue curve is based on the
declarations made by the DPRK and, thus, the assumption is that no
major refueling has taken place in 1989. The orange curve is derived
assuming that the full core has been replaced with fresh fuel in 1989
under the constraint of arriving at the same final burn-up.
These numbers can be readily converted into reactor thermal power
levels using the fact that there are approximately 50 tonnes of
uranium in this reactor~\autocite{puzzle}. The power levels then form
the input for a detailed calculation of the reactor isotopic
composition and fission rates for the various fissile isotopes. The
software  we used to compute the relevant reactor parameters is
called SCALE~\autocite{scale} and is considered a standard method for
this type of problem. The details of the calculation can be found in
appendix~\ref{sec:5MWSCALE}.

%%%%%%%%%%%%%%%%%%%%%%%%%%%%%%%%%%%%%%%%%%%%%%%%%%%%%%%%%%
\section{Neutrinos in the DPRK case}
\label{sec:dprknu}

The basic analysis techniques developed in the previous section can
now be applied to the specific situation in the DPRK in the time frame
of 1986-1994. The central question for the international community,
after the initial discrepancies appeared in 1992, was how much
plutonium the DPRK had separated. The lower bound on this quantity is
represented by assuming that the DPRK's initial declaration to IAEA
was quantitatively correct, i.e. only 90\,g of plutonium were
separated from a few hundred damaged fuel elements discharged and
replaced during the 1989 shutdown. The upper bound on the amount of
separated plutonium is obtained by assuming that the full core with a
burn-up of approximately 200\,MWd/t was discharged in 1989, containing
8.8\,kg of plutonium and that this full core was subsequently
reprocessed. The North Korean scientists could have produced
additional plutonium over a long period of time by irradiating natural
uranium targets in the IRT, resulting in roughly 0.5\,kg of plutonium
per 250 day run of the IRT. The limited amount of Soviet supplied fuel
and the fact that the IRT was under IAEA safeguards from 1977 on
limits the amount the DPRK could have produced via that route to less
than 1\,kg~\autocite{HeinonenIRT}. Other authors~\autocite[p.\,120]{puzzle}
estimate the theoretical upper limit of the amount of plutonium to be as
large as 4\,kg.

As far as the 5\,MW$_\mathrm{e}$ reactor is concerned, at the time of
the first IAEA inspection in 1992, the burn-up and reactor power were
the same for both the extreme cases (see
figure~\ref{fig:AlbrightBU}). Therefore, our analysis will include the
hypothetical scenario where neutrino safeguards were applied before
and after the 1989 shutdown\endnote{In principle, the amounts of the
  long-lived isotopes strontium-90, ruthenium-106, and cerium-144 will
  be different between the two irradiation histories which leads to
  differences in the low energy neutrino spectrum below
  3.6\,MeV. However, extensive calculations show that the resulting
  event rate differences in 1992 are too small to be reliably
  detected.}. The specific unique capability represented by neutrino
safeguards in this case derives from the ability to measure the power
history and burn-up \emph{independently} -- any mismatch indicates a
fuel diversion.

In the PUREX process for reprocessing, the fission fragments remain in
the aqueous phase and therefore will end up in the waste. Some of
these fission fragments produce neutrinos above IBD threshold even
after a considerable time interval has elapsed, which we will refer to
as long-lived isotopes (LLI), in particular: strontium-90 with a half-life
of 28.9\,y, ruthenium-106 with a half-life of 372\,d, and cerium-144 with
a half-life of 285\,d. These three isotopes have large direct fission
yields and are produced in amounts which are proportional to
the number of total fissions and thus are accurate tracers of
burn-up. Detecting neutrinos from LLI is a direct method to find
reprocessing wastes and, in principle, also yields an estimate of the
amount of plutonium separated. Given the high penetrating power of
neutrinos, this method is equally applicable to buried wastes.

Finally, neutrinos can travel arbitrary distances, and thus a neutrino
detector deployed for safeguarding the IRT would also be sensitive to
neutrinos from the 5\,MW$_\mathrm{e}$, especially during times when
the IRT is shut down. This signal will allow a remote power measurement
which can distinguish the two cases shown in figure~\ref{fig:AlbrightBU}.

%%%%%%%%%%%%%%%%%%%%%%%%%%%%%%%%%%%%%%%%%%%%%%%%%
\subsection{5\,MW$_\mathrm{e}$ reactor}
\label{subsection:5MW}

\begin{figure}
\begin{center}
\includegraphics[width=0.5\textwidth]{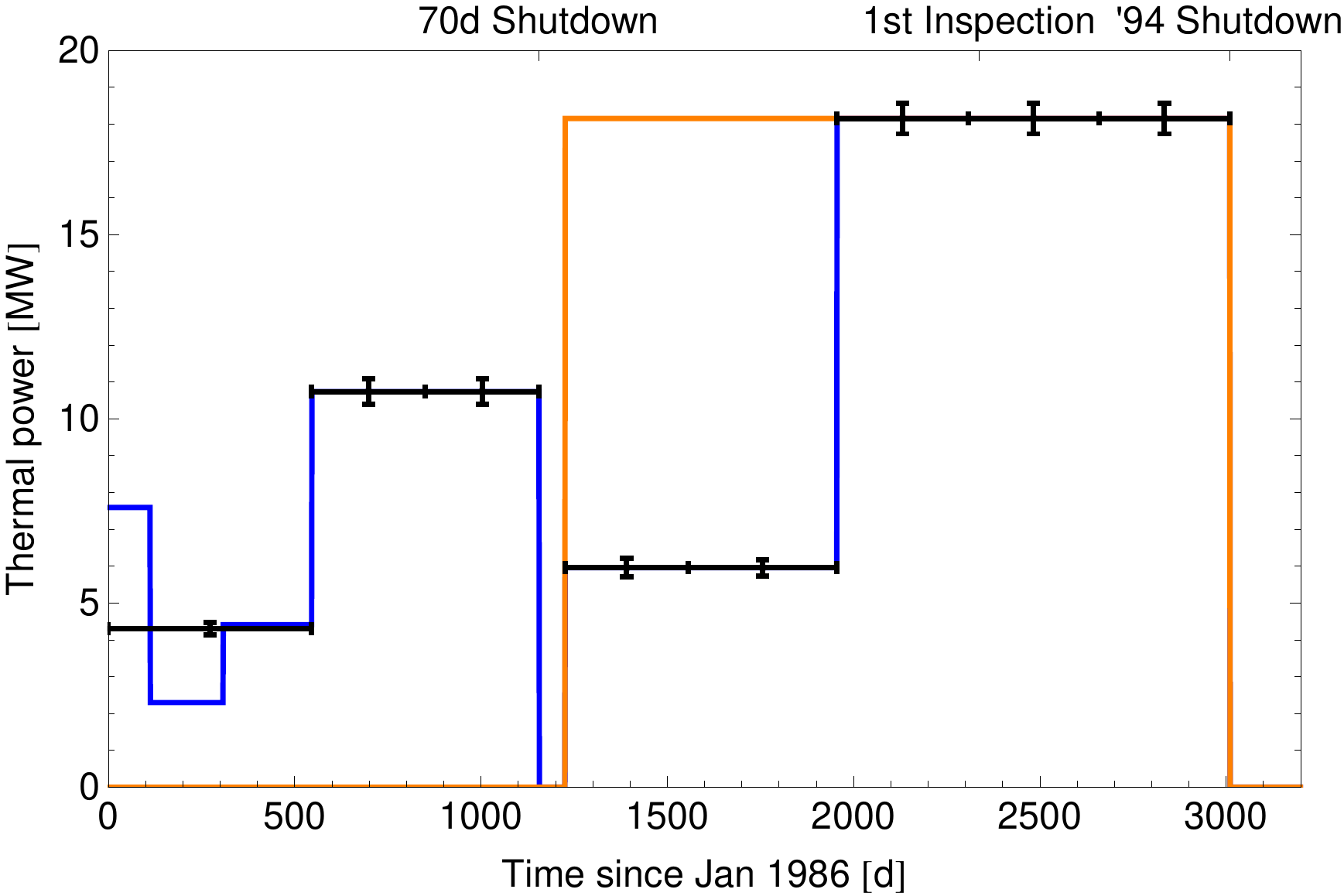}%
\includegraphics[width=0.5\textwidth]{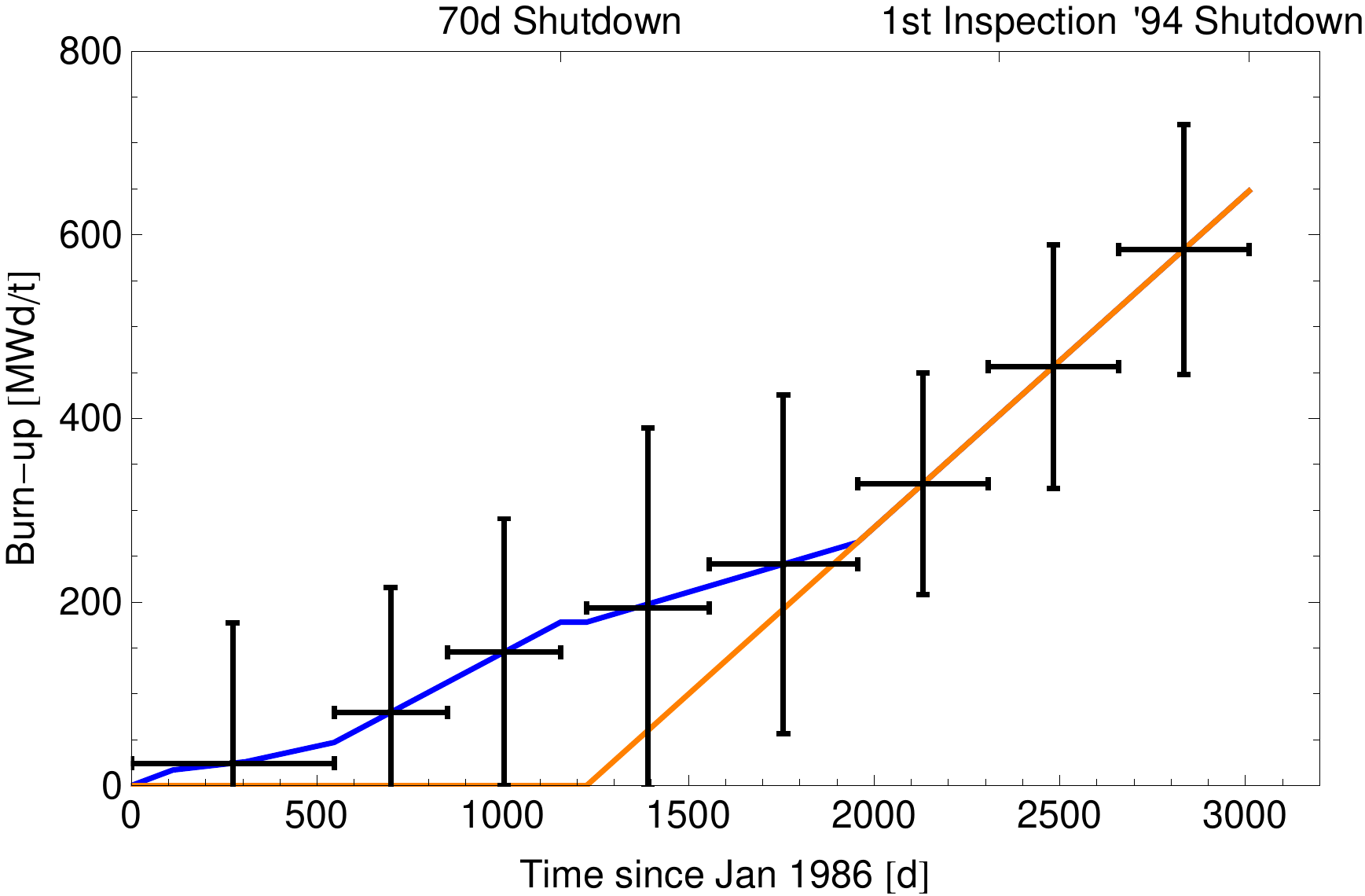}
\end{center}
\mycaption{\label{5MWPowerHistory} In the left hand panel, 1$\,\sigma$
  sensitivities to reactor power are shown for varying data collection
  periods using a 5\,t detector at 20\,m standoff from the
  5\,MW$_\mathrm{e}$ reactor. Fission fractions are free parameters in
  the fit. In the right hand panel, 1$\,\sigma$ sensitivities to
  burn-up are shown, where power is a free parameter in the fit. The
  blue curve shows the history under the assumption of no
  diversion. The orange curve shows history for the case of a full
  core discharge in 1989.}
\end{figure}

In the following analysis, sensitivities to power, burn-up, and
plutonium content are determined based on the declared power
history. This history is displayed as blue curves in the various
figures in this section. Comparisons are made to a hypothetical
undeclared core swap to a fresh reactor core during the 70 day
shutdown period, displayed as orange curves. The difficulty in
determining the difference between the two curves lies in the fact
that after 1992, power and burn-up are the same. As seen in
figure~\ref{fig:5mwfissionrates}, after the 1st inspection, all the
fission rates from the four primary fissioning isotopes are identical
with or without diversion. For the following analyses, a standard 5\,t
detector at 20\,m standoff from the reactor is used, which for a data
taking period of one year corresponds to about 95,000 events.

A power sensitivity computation is first considered. The analysis is
done using the following $\chi^2$-function
\begin{equation} \label{eq:powerchi2}
\chi^2 =\sum_{i}\frac{1}{n_i^0}.\left[\left(N\,P_\mathrm{th}\,\sum_{I} \digamma_I S_{I,i}\right)-n_i^0\right]^2\,,
\end{equation}
where $\digamma_I$ is the fission fraction for isotope $I$, $n_i^0$ is
the measured number of neutrino events in energy bin $i$, and
$S_{I,i}$ is the neutrino yield in energy bin $i$ for isotope
$I$. $P_{th}$ is the thermal power and $N$ is a normalization
constant. Moreover, the fission fractions $\digamma_I$ are subject to a
normalization constraint as given in equation~\ref{eq:fractions}.

The resulting 1$\,\sigma$ sensitivities are shown in the left hand
panel of figure~\ref{5MWPowerHistory}. This analysis assumes precise
knowledge of the distance from the reactor to the detector and
treats them both as points. Any uncertainty in the geometric
acceptance will directly relate into an uncertainty of the
normalization constant, $N$, and thus into an uncertainty in the power
$P_\mathrm{th}$. Neglecting this potential source of systematic
uncertainty, a power accuracy of around 2\% can be achieved.

\begin{figure}[t]
\includegraphics[width=0.5\textwidth]{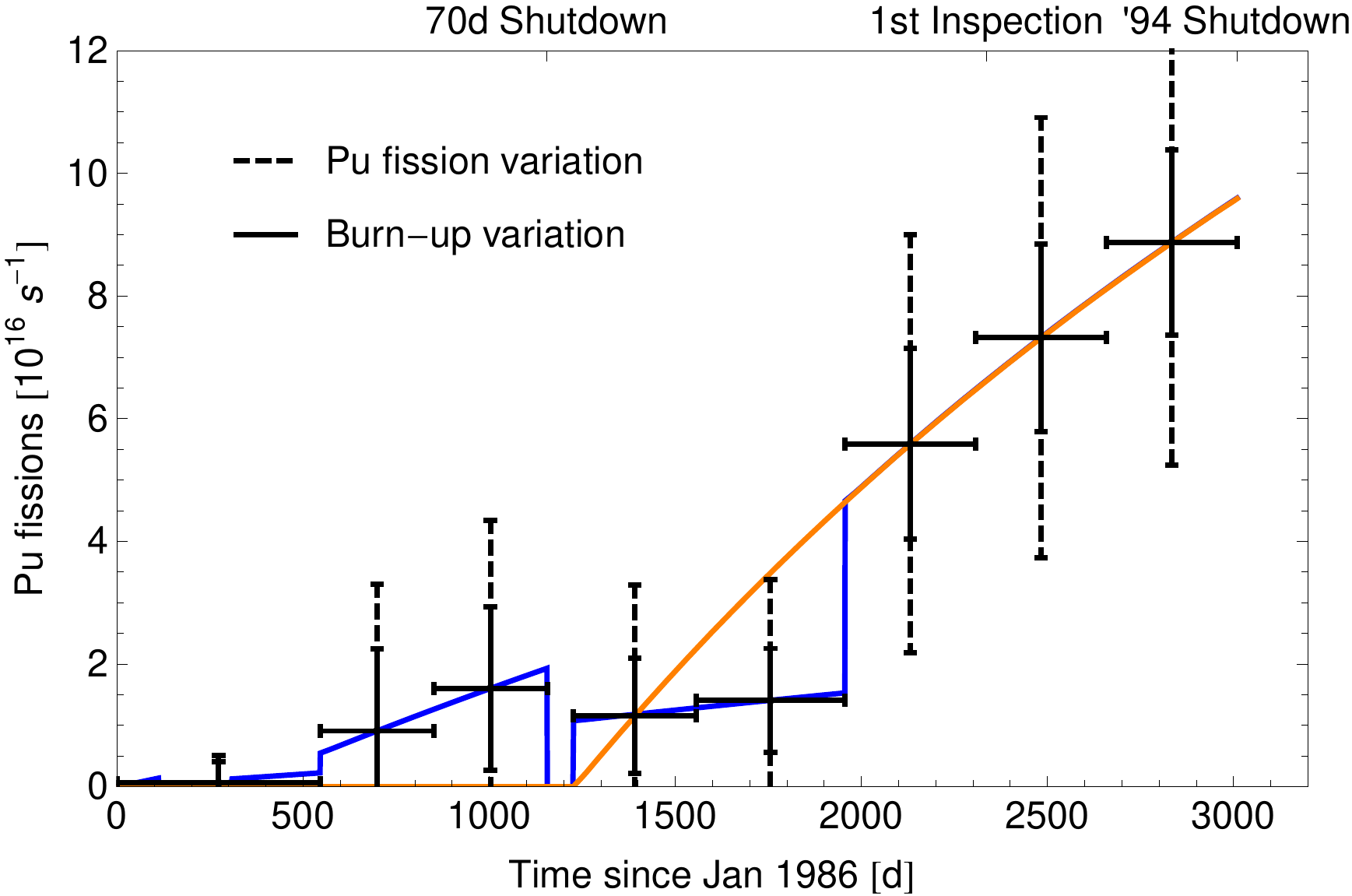}%
\includegraphics[width=0.5\textwidth]{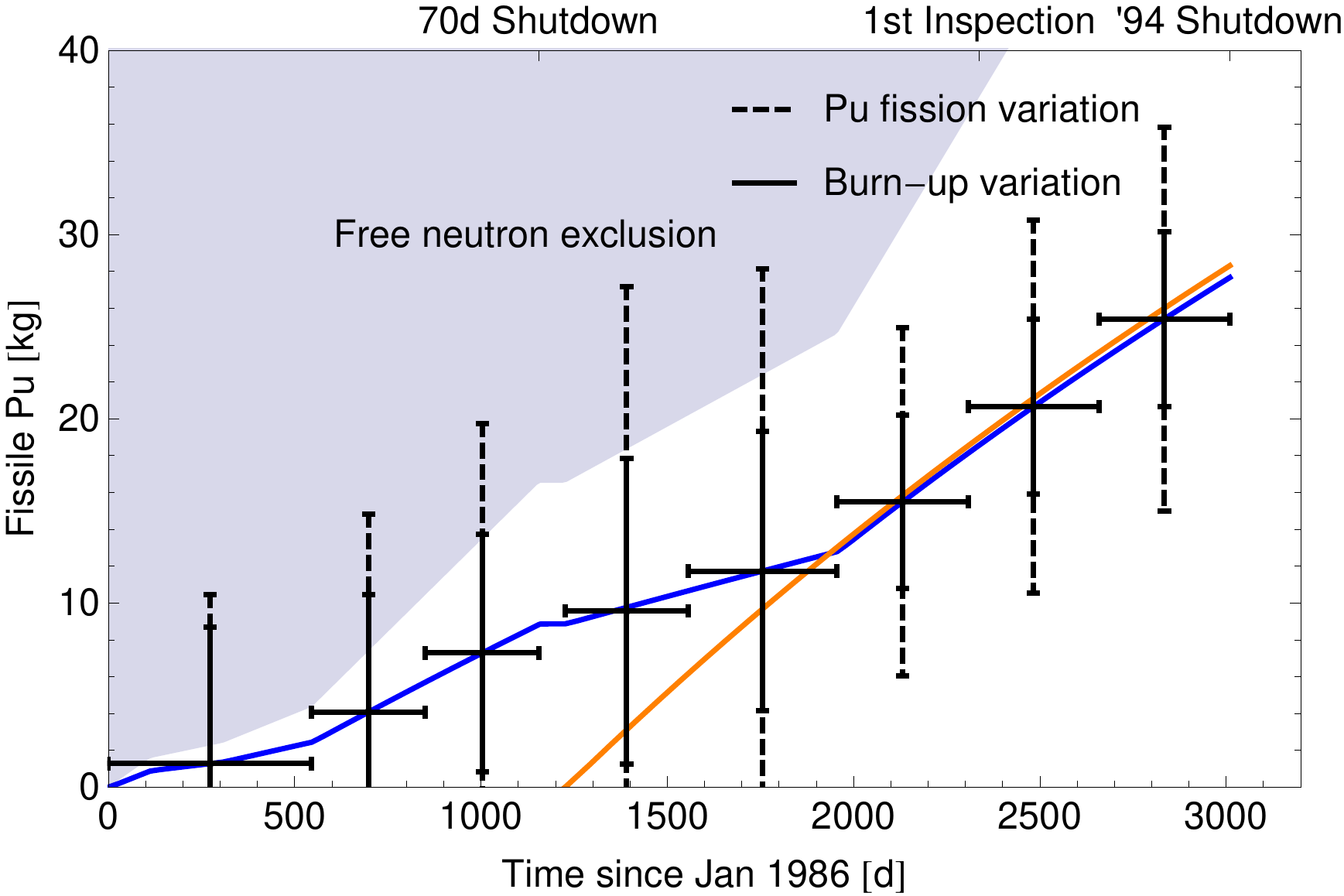}
\mycaption{\label{5MWPu239History} 1\,$\sigma$ sensitivities to
  plutonium are shown for varying data collection periods using a 5\,t
  detector at 20\,m standoff from the 5\,MW$_\mathrm{e}$ reactor.  The
  blue curve shows the plutonium-239 history under the assumption of no
  diversion. The orange curve shows the plutonium-239 history if there
  had been diversion. Black dashed error bars show the 1$\sigma$
  sensitivity by measuring the plutonium fission rates with uranium
  fission rates and reactor power free in the fit. Solid black error
  bars show the 1$\sigma$ sensitivity determined by constraining the
  burn-up using a reactor model.  The left plot shows the errors on
  absolute plutonium fission rates and the right plot show the
  corresponding errors for plutonium mass with a shaded exclusion
  region from the assumption that all neutrons not needed for fission
  are available for the production of plutonium.}
\end{figure}

A similar analysis can be done to determine the sensitivities for
burn-up, $BU$, using equation~\ref{eq:powerchi2}. In this circumstance,
$P_\mathrm{th}$ is free in the fit and the fission fractions
$\digamma_I$ are now functions of burn-up, determined by a reactor
core simulation as described in appendix~\ref{sec:5MWSCALE}. The results
of this analysis are shown in the left hand panel of
figure~\ref{5MWPowerHistory}. Burn-up across the history of the reactor
has an error of $\sim 100$\,MWd/t.  Closely related to the burn-up is
the amount of plutonium in the nuclear reactor. This analysis is done
again using equation~\ref{eq:powerchi2}. This time, $P_\mathrm{th}$ as well
as $\digamma_\mathrm{U235}$ and $\digamma_\mathrm{U238}$ are free
parameters as well as the relative contribution of the two plutonium
fission rates, $\kappa$, and the resulting sensitivities are shown as
dashed black lines in figure~\ref{5MWPu239History}. Alternatively, one
can use the burn-up sensitivity to constrain the plutonium content as
well. After computing burn-up errors, a reactor model is used to
compute the change in plutonium fissions. This is shown as the solid
black error bars. These errors are given both in terms of raw
plutonium fissions in the left hand panel as well as the corresponding
plutonium masses in the right hand panel. In the right hand panel, a
very naive exclusion region is shown for comparison. It assumes that
each of the 1.7 neutrons per fission not being used to sustain the
chain reaction is instead available to produce more plutonium. This
limit is shown as the shaded region in the right hand plot.

%%%%%%%%%%%%%%%%%%%%%%%%%%%%%%%%%%%%%%%%%%%%%%%%%
\subsection{IRT reactor}

\begin{figure}[t]
\includegraphics[width=0.5\textwidth]{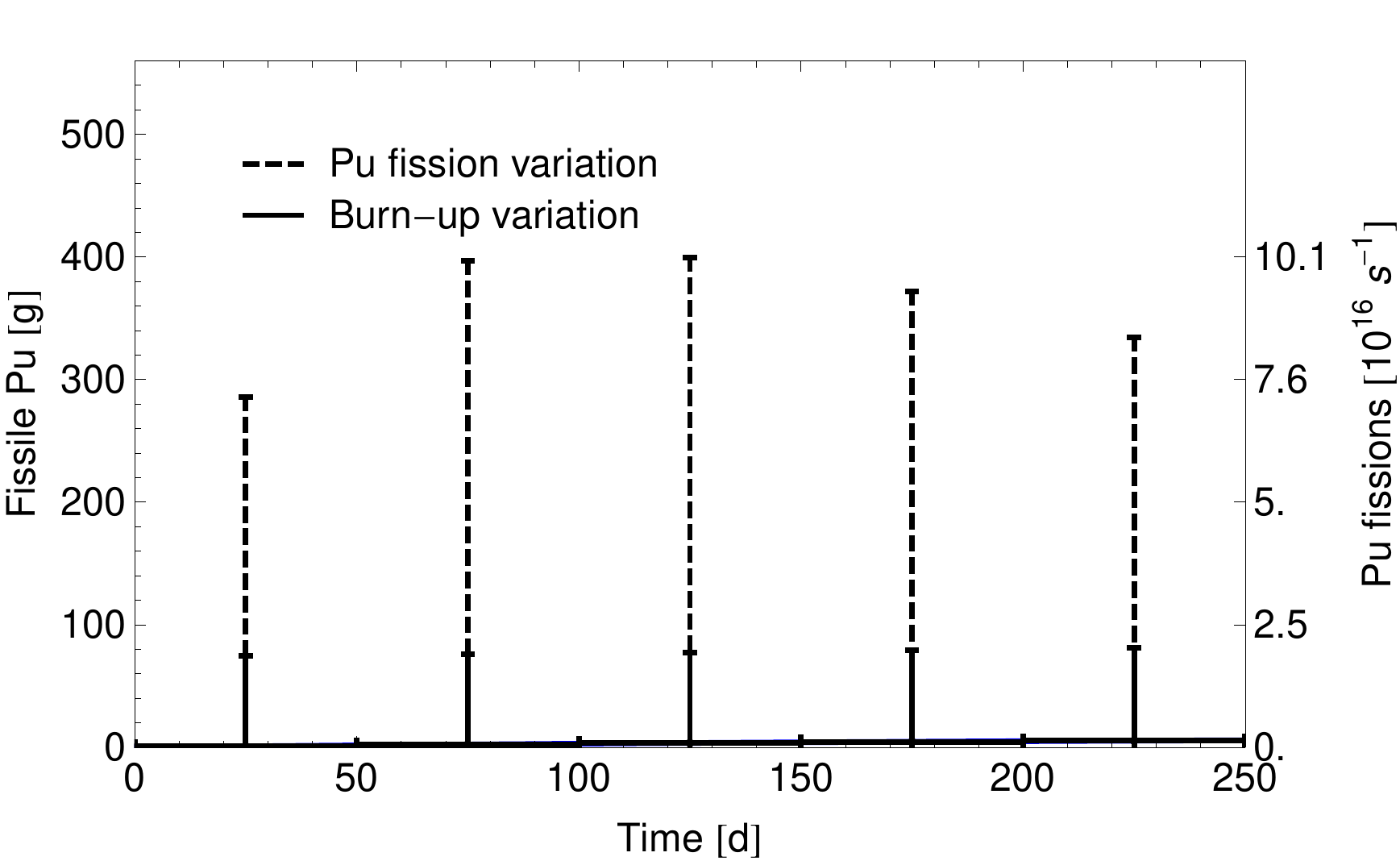}%
\includegraphics[width=0.5\textwidth]{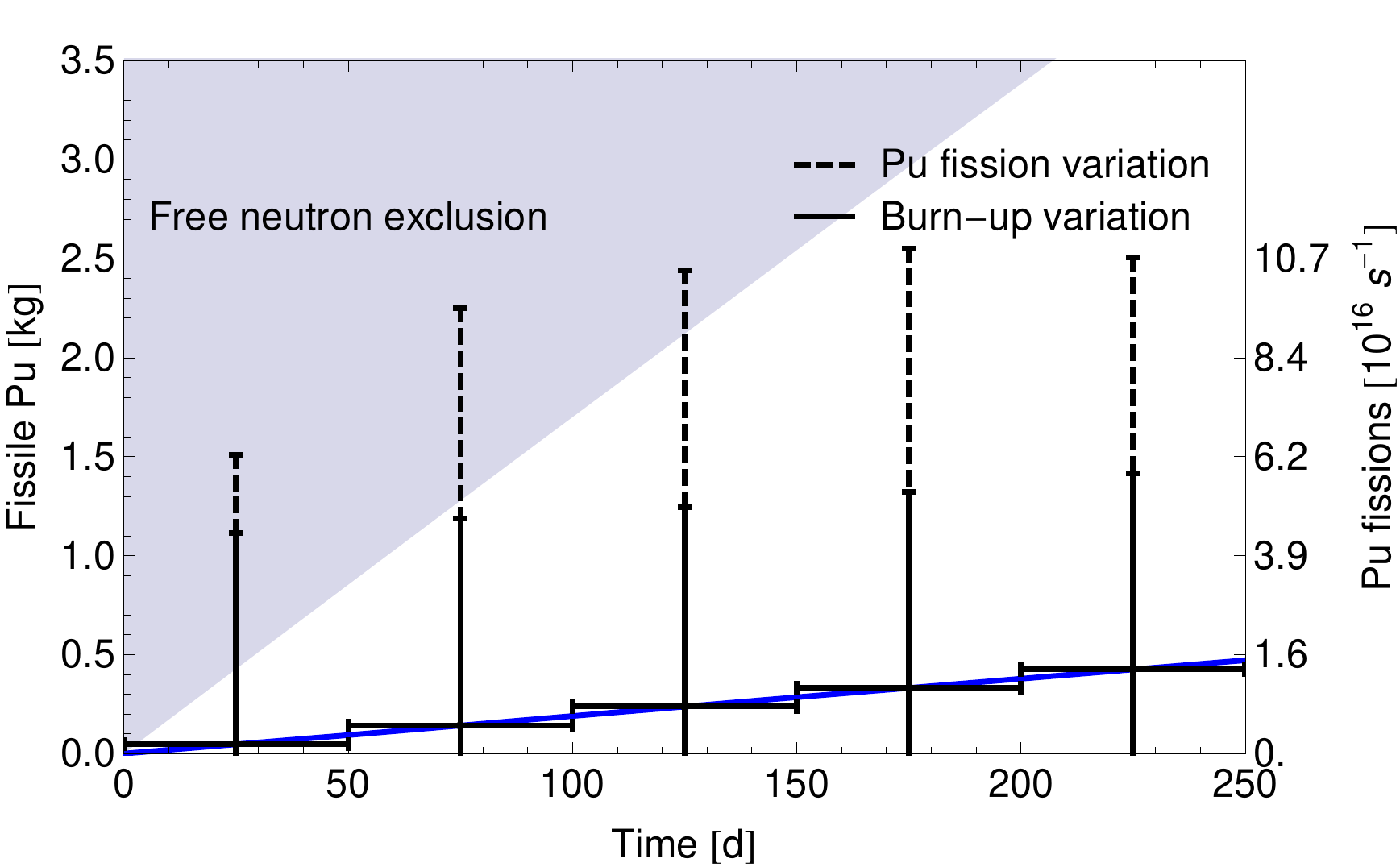}
\mycaption{\label{IRTPuHistory} 1\,$\sigma$ sensitivities to reactor
  plutonium fissions are shown for 50 day collection periods using a
  5\,t detector, 20\,m away from the IRT reactor. Black dashed error
  bars show the 1$\sigma$ sensitivity resulting from measuring the
  plutonium fission rate with with uranium contributions and power
  free in the fit. The solid black error bars show the 1\,$\sigma$
  sensitivity determined using a burn-up model. The left plot shows
  driver only results and the right plot shows results for driver and
  targets combined.}
\end{figure}

The IRT is assumed to run for a 250 day period followed by a 100 day
shutdown~\autocite[pp.\,148-165]{puzzle}, and the fission rates are
computed in appendix~\ref{sec:IRT} and shown in
figure~\ref{fig:IRTfissrates}. The natural uranium targets provide much
more uranium-238, changing the fission fractions substantially and
allowing an order of magnitude increase in plutonium-239 production and
fissions. As with the $5\,\mathrm{MW}_\mathrm{e}$ reactor, it is
assumed that a 5\,t neutrino detector is placed 20\,m away from this
reactor. The $\chi^2$ from equation~\ref{eq:powerchi2} can be used to
determine the thermal power to within 0.6\,MW  in each 50 day
period. All other things the same, the addition of targets will
increase the power output of the reactor. As long as the detector
distance and mass were sufficiently well known, the errors would
be small enough to clearly notice the power difference caused by the
addition of breeding targets. At the same time, it would be trivial
for the operator to adjust the power with targets to remain the same
as without targets, which would reduce plutonium production by about
25\%.

The 1\,$\sigma$ errors on plutonium content can be determined by
measuring the fission rates using the $\chi^2$ prescription in
equation~\ref{eq:powerchi2} and then converting these to a plutonium mass
using equation~\ref{eq:fissions}. Alternatively, one can determine the
burn-up in conjunction with a reactor model and then infer the errors
on plutonium mass inventory. The results of the analysis are shown in
figure~\ref{IRTPuHistory}. Similar error bars are found on raw plutonium
fission rates, with and without the targets. Note, that these are also
similar to the 5\,MW$_\mathrm{e}$ reactor results. Very different,
however, are the sensitivities to the mass of plutonium. In the case
with only drivers, a neutrino detector would be sensitive to tens of
grams of plutonium. With both the drivers and targets, there is an
order of magnitude increase in the errors into the hundreds of grams
of plutonium. The difference is even more pronounced in comparison to
the $5\,\mathrm{MW}_\mathrm{e}$ reactor, where plutonium mass
sensitivities in the multi-kg range are obtained. Despite similar
sensitivities for plutonium fission rates, the sensitivity to core
inventory is strikingly different for the reasons explained in detail
in section~\ref{sec:pucontents}. The neutron flux density in the fuel
containing the plutonium is very different for the two configurations
of the IRT with and without the breeding targets. Since the change in
plutonium fission rates is relatively small between these two
configurations, we have to conclude that neutrino safeguards is not
effective in determining which configuration is used. Therefore, the
spread in plutonium mass predictions between the two configurations
has to be taken as error, which is 0.36\,kg, over one 250 day run.
Taking the upper end of the range of plutonium produced in the
IRT~\autocite[p.\,120]{puzzle} of 4\,kg, we see that this requires about
8-10 reactor cycles. Since the errors from a neutrino measurement
between each cycle are statistically independent we find the total
error from a neutrino measurement taking 8 cycles to be
$0.36\,\mathrm{kg}\sqrt{8}=1.0\,\mathrm{kg}$. In the more realistic
case of no plutonium production in the IRT this measurement translates
into an upper bound of the same size from this source.

%%%%%%%%%%%%%%%%%%%%%%%%%%%%%%%%%%%%%%%%%%%%%%%%%
\subsection{5\,MW$_\mathrm{e}$ reactor power measurement at IRT}

\begin{figure}[t]
\begin{center}
\includegraphics[width=0.5\textwidth]{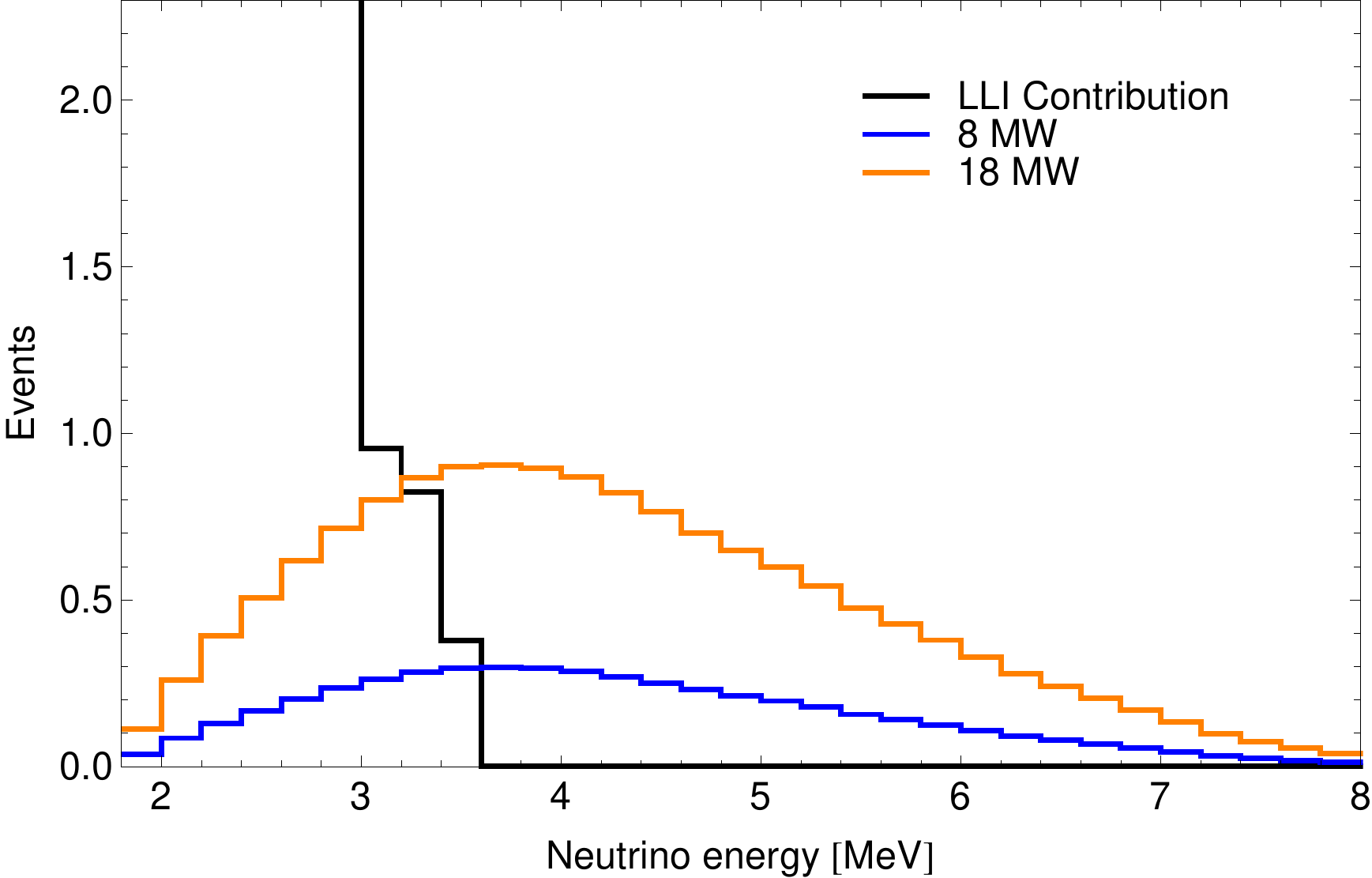}%
\includegraphics[width=0.5\textwidth]{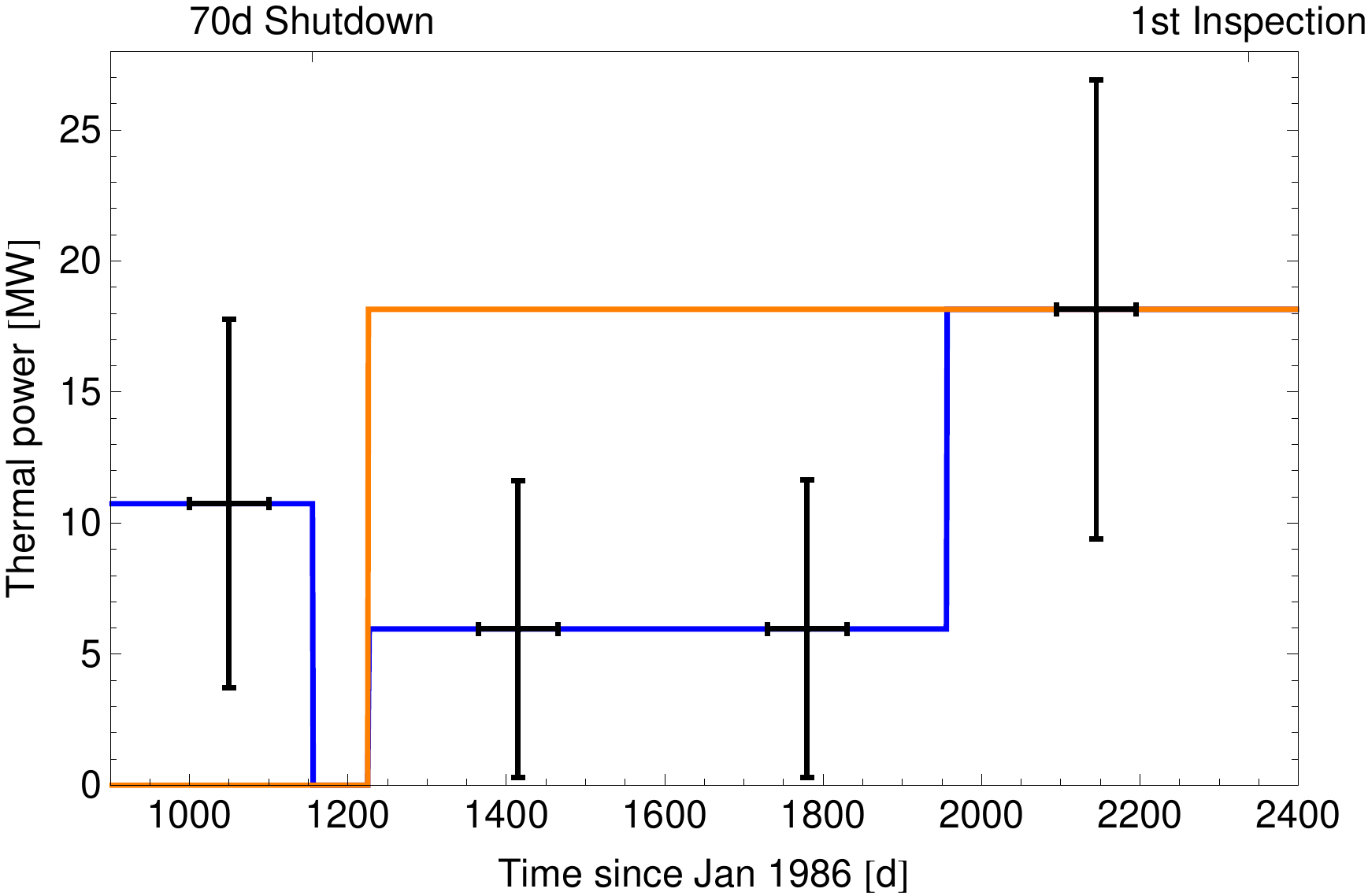}
\end{center}
\mycaption{ In the left hand panel events are shown for 200 days of data
  collection 20\,m from the shut down IRT reactor and
  1.2\,km from the running 5\,MW$_\mathrm{e}$ reactor. The IRT is
  assumed to only contribute to the detected neutrino spectrum through
  its long lived isotopes shown in black. The
  5\,MW$_\mathrm{e}$ reactor is assumed to be running either at
  the declared 8\,MW$_\mathrm{th}$, as shown in blue, or at 
  18\,MW$_\mathrm{th}$, as shown in orange. The right hand panel shows 
  the 1$\sigma$ sensitivities to reactor power
  resulting from this measurement. The blue curve shows the power
  history under the assumption of no diversion. The orange curve shows
  the power history if there had been diversion.
\label{5MWPowerAtIRTEvents}}
\end{figure}

An additional benefit of having a neutrino detector at the IRT reactor
is that it would also be sensitive to neutrinos from the
5\,MW$_\mathrm{e}$ reactor. This is particularly useful during times
when the IRT is shut down, which happens for approximately 100 days
every year~\autocite[pp.\,148-149]{puzzle}. This will yield two
measurement periods of 100 days each for the reactor power of the
5\,MW$_\mathrm{e}$ reactor during the crucial time, after the 70\,d
shutdown and before the first inspection, where the declared power was
low, around 8\,MW$_\mathrm{th}$, but would have been as high as
18\,MW$_\mathrm{th}$, in order to bring the second core to the same
final burn-up, see figure~\ref{fig:AlbrightBU}.

Data collection is assumed to start shortly after an IRT shutdown at a
point where all but the long-lived neutrino producing isotopes have
decayed away, leaving only the LLI: strontium-90, ruthenium-106, and
cerium-144. This occurs on the order of days. The number of atoms for
each of the LLI was computed using SCALE and is shown in
table~\ref{table:IRTLLI}.
\begin{table}[h]
\sf
\begin{center}
\begin{tabular}{cccc}
Isotope & strontium-90 & ruthenium-106 & cerium-144 \\ \hline
Amount (atoms) & $3.4\times10^{23}$ & $2.8\times10^{22}$ & $2.5\times10^{23}$
\end{tabular}
\mycaption{\label{table:IRTLLI}Number of long-lived isotope atoms
  assumed shortly after IRT shutdown.}
\end{center}
\end{table}
As in the previous sections, we use a 5\,t detector at 20\,m standoff
from the IRT and 1.2\,km from the 5\,MW$_\mathrm{e}$ reactor, see
figure~\ref{fig:NeutrinoMap}. Data is collected over two 100 day periods
and the detected spectrum is shown in the left hand panel of
figure~\ref{5MWPowerAtIRTEvents}. The signal event numbers are small and
therefore we use the appropriate Poisson log-likelihood to define the
$\chi^2$-function
\begin{equation} \label{eq:powerdeviance}
\chi^2
=2\sum_{i}[n_i\,\mathrm{log}\frac{n_i}{n_i^0}-(n_i-n_i^0)]\quad\text{with}\quad
n_i = N\,P_\mathrm{th}\,\sum_I\digamma_I\,S_{I,i} + LLI_i\,,
\end{equation}
where $LLI_i$ is the long lived isotope contribution in the bin
$i$. Resulting sensitivities are shown in the right hand panel of
figure~\ref{5MWPowerAtIRTEvents}. This corresponds to an uncertainty of
about 3.8\,MW$_\mathrm{th}$ during the periods of interest. The
difference in reactor power for a second core would be detected at
3.2\,$\sigma$.

This result implies that a larger detector could be used to safeguard
several reactors in a larger area. In particular, a detector that is
sensitive to direction could identify the reactor that contributed the
neutrino and get several power measurements simultaneously. Also,
without the need to be close to a reactor, it could be placed
underground allowing for greater background reduction~\autocite{Jocher:2013gta}.

%%%%%%%%%%%%%%%%%%%%%%%%%%%%%%%%%%%%%%%%%%%%%%%%%
\subsection{Waste detection}

In addition to directly monitoring reactors, neutrino detectors can be
used for detection of nuclear waste. With sufficient insight of where
waste might be disposed, a nearby neutrino detector can see the
signature of LLI, even after years of
storage. Table~\ref{table:WasteAtoms} lists the number of atoms of
each of the three primary LLI that would be expected in the waste at
the point in time of the first inspection, roughly 3 years after the
70 day shutdown. In the following analysis, it is assumed that the
complete core was removed during the 70 day shutdown and the resulting
reprocessing wastes are stored together in one of three locations: the
``suspected waste site'', building 500, or the Radiochemical
Laboratory~\autocite{puzzle,HeinonenWaste}. All three locations are
shown in figure~\ref{fig:NeutrinoMap}. For building 500, we assume
that we can not deploy inside the hatched area, since this facility
was declared to be a military installation exempt from safeguards
access~\autocite[pp.\,149-154]{puzzle}. The resulting standoff distances
are shown in table~\ref{table:WasteLocations}.
\begin{table}[h]
\sf
\begin{center}
\begin{tabular}{cccc}
Isotope & strontium-90 & ruthenium-106 & cerium-144 \\ \hline
Amount (atoms) & $1.2\times10^{24}$ & $1.4\times10^{22}$ &$ 3.7\times10^{22}$
\end{tabular}
\mycaption{\label{table:WasteAtoms}Number of long-lived isotopes at day
  2251 for a complete reactor core removed at day 1156 and stored for
  3 years.}
\end{center}
\end{table}

\begin{figure}[t]
\includegraphics[width=0.5\textwidth]{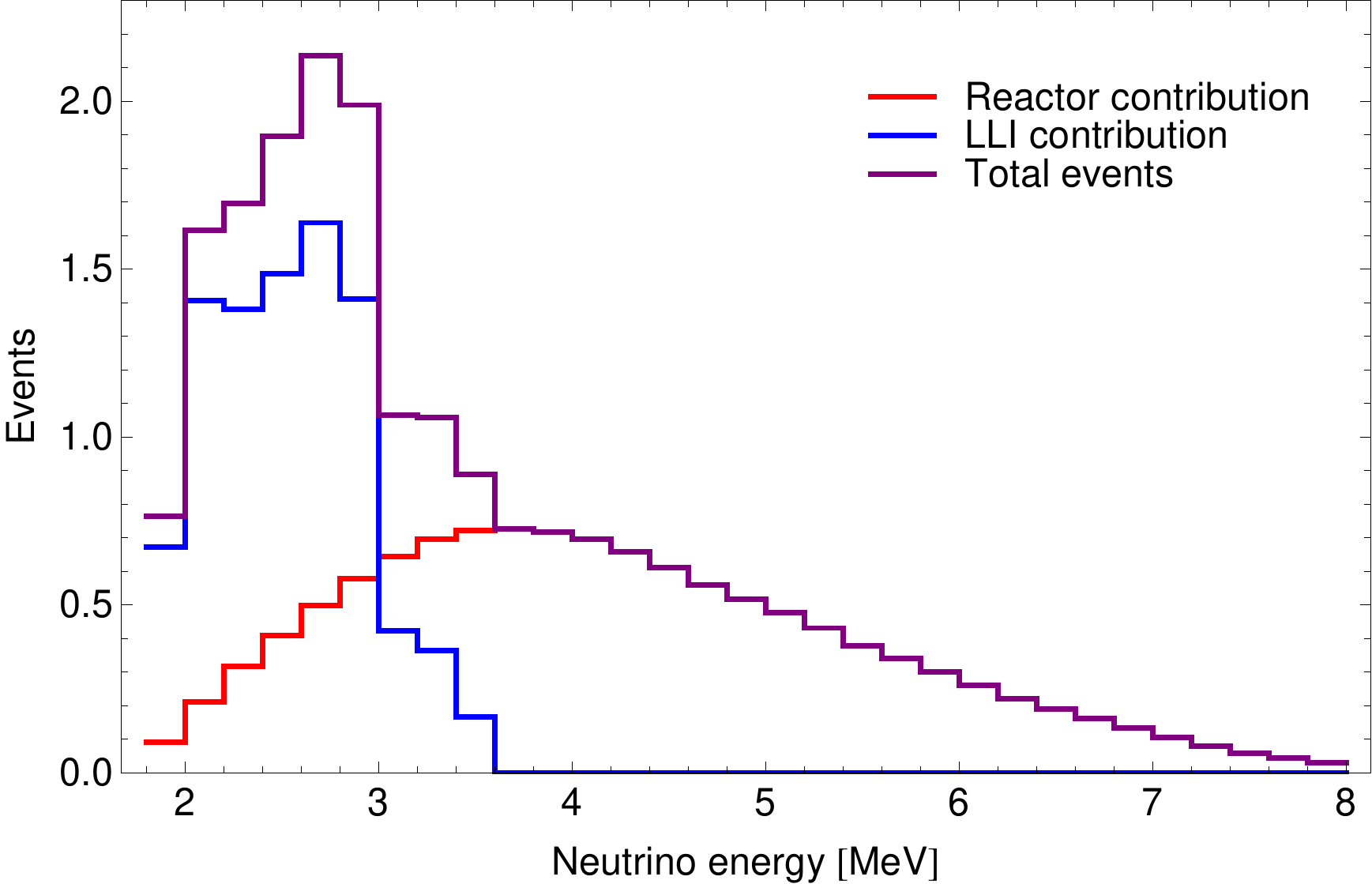}
\mycaption{Total event rates are shown in purple for 1 year of
  integrated data collection starting in 1992 with a 5\,t detector
  25\,m from spent fuel and 1.83\,km from the 5\,MW$_\mathrm{e}$
  reactor. The reactor contribution to total event rates are shown in
  red and long lived isotope contributions shown in blue.}
\label{LLI3}
\end{figure}

Due to the low event statistics, a Poisson log-likelihood is used, as
in equation~\ref{eq:powerdeviance}, with the difference that the reactor
events from the 5\,MW$_\mathrm{e}$ are now background and the signal
are the $LLI_i$.  Table~\ref{table:WasteLocations} summarizes the
results for each location. Figure~\ref{LLI3} shows the event rate
spectrum in the most promising of the setups considered, the case of
the reprocessing plant. It is found that a detector around 25\,m from
the waste and 1.8\,km from the 5\,MW$_\mathrm{e}$ reactor would have a
2\,$\sigma$ signal after 55 days of data collection. The strongest
contributor to detection capability is the distance from the source.

\begin{table}[h]
\sf
\begin{center}
\begin{tabular}{ccccccc}
Location & Reactor & Fuel  & Reactor  & Fuel  & $\chi^2$ & 2$\sigma$ Time [y]\\
 & Distance [m] &  Distance [m] &  Events &  Events &  & \\
 \hline
Building 500 & 1980 & 80 & 10.1 & 0.9 & 0.34 & $\ge$10\\
Suspected Waste Site & 1060 & 25 & 35.3 & 8.9 & 8.22 & 0.33\\
Reprocessing Plant & 1830 & 25 & 11.8 & 8.9 & 16.95 & 0.15\\
Reprocessing Plant & 1800 & 100 & 12.2 & 0.6 & 0.12 & $\ge$10
\end{tabular}
\mycaption{\label{table:WasteLocations} Events are integrated over 1
  year with a 5\,t detector. The waste corresponds to a complete
  reactor core discharged in 1989 during the 70 day shutdown. Long
  lived isotopes are decayed 3 years before the measurement
  starts. The expected time to achieve  a 2\,$\sigma$ detection is given in the last column.}
\end{center}
\end{table}

%%%%%%%%%%%%%%%%%%%%%%%%%%%%%%%%%%%%%%%%%%%%%%%%%%%%%%%%%%%%%%%%%%%%%%%%%%%%

\subsection{Continuous neutrino observations}

In applying neutrino safeguards, like in conventional safeguards, we
can use individual measurements taken at different times and apply
them in combination to infer what actually happened. The initial
declaration to IAEA by the DPRK admits two extreme cases: both  a
very minor discharge of a few hundred fuel elements and a complete
core discharge in 1989 would yield the same overall core configuration
at the time inspectors arrived in 1992, see
figure~\ref{fig:AlbrightBU}.

The real strength of a neutrino detector is evident when it can
measure over the history of a reactor. As seen in
figure~\ref{5MWPowerHistory}, such a detector is capable of being very
sensitive to reactor power. Thus, if a neutrino detector was present
for the lifetime of the reactor, the declared power would have to
match the measured power at all times and, since the burn-up is just
the time integrated thermal reactor power, the burn-up could be
inferred from a complete power history. At the same time, a burn-up
measurement, in contrast to an inferred burn-up value, can also be
derived from a neutrino measurement, provided a reliable, but not
necessarily very detailed or accurate, reactor model is available.  As
will be shown in section~\ref{sec:conventional}, given that the bulk
quantities in terms of burn-up are the same between the two scenarios,
all conventional methods which can address the issue of the second
core also rely on a reactor model.  The diversion scenario that has
been considered relies heavily upon the ability to adjust the power
relative to the declared power so that both the power and burn-up
match at a later time. In the presence of a neutrino detector, the
difference in burn-up will be frozen between the declared burn-up and
the actual burn-up of the new core. The fact that a neutrino detector
can simultaneously measure power, as well as fission fractions, is
what allows it to detect this difference in burn-up. To determine
sensitivity to such a situation, a modified version of
equation~\ref{eq:powerchi2} is used
\begin{equation} \label{eq:powerburnupchi2}
\chi^2 = \sum_{t} \sum_i \frac{1}{n_{i,t}^0}
\left[\left(1+\alpha_{\rm{detector}}\right)\,P_\mathrm{th}^t \sum_I\digamma_I(BU^t)S_{I,i}-n_{i,t}^0\right]^2 + \left(\frac{\alpha_{\rm{detector}}}{\sigma_{\rm{detector}}}\right)^2.
\end{equation}
where $t$ is indexing the time interval for which a measurement is
available. $\alpha_{\rm{detector}}$ is a detector normalization
parameter with uncertainty $\sigma_{\rm{detector}}$. $P_{\rm{th}}^t$
is the average reactor power in each time bin $t$. $\digamma_I$ are
the fission fractions which are a function of the burn-up in each time
bin $t$, $BU^t$. The burn-up as a function of time is given by
\begin{equation}
BU^t=\left(\sum_{\tau=1}^{t-1} \frac{P_\mathrm{th}^\tau \Delta\tau}{M_\mathrm{core}}\right) + BU^0
\end{equation}
where $\Delta\tau$ is the width of the time bin, $BU^0$ the initial
burn-up at the start of data taking and $M_\mathrm{core}$ the mass of
the reactor core in terms of fuel loading. If this initial burn-up
$BU^0$ is well known, as it would be if data collection began at
start-up, such an analysis greatly reduces the uncertainty in the
total plutonium budget. In table~\ref{table:TotalPuBudget}, the total
error budget is given through the use of this method, labeled ``method
2'', and is shown compared to the results if only the burn-up but
\emph{not} the power history is measured based on the results of the
previous sections, labeled ``method 1''. For method 2 we assumed that
reactors start with a well known composition, that is $BU^0=0$ and a
detector related uncertainty $\sigma_\mathrm{detector}=1\%$ is
achievable and all the $P_\mathrm{th}^t$ are free parameters in the
fit. In the case of the 5\,MW$_\mathrm{e}$ reactor, for both analyses,
the question is what is the maximum change in $BU^x$ during the 70 day
shutdown. The value of $BU^x$ is translated into the resulting
plutonium mass sensitivity by using the reactor model. It is clear
that method~1 is less accurate but does not rely on continuity of
knowledge whereas method~2 is much more accurate but requires
continuity of knowledge. Method~2 still offers a significant advantage
compared to conventional methods by providing its results in real-time
and not only at some later, unspecified time in the future.

\begin{table}[t]
\sf
\begin{center}
\begin{tabular}{cccc|cc|cc}
\multicolumn{2}{c}{\multirow{3}{*}{Reactor}} & \multicolumn{2}{c|}{Final} & \multicolumn{2}{c}{Method 1, 1$\sigma$} & \multicolumn{2}{|c}{Method 2, 1$\sigma$}\\
\hline
\multicolumn{2}{c}{} & Burn-up & Pu & Burn-up & Pu & Burn-up & Pu\\
\multicolumn{2}{c}{} & [MWd/t] & [kg] & [MWd/t] & [kg] & [MWd/t] & [kg]\\
\hline
\multicolumn{2}{c}{IRT/run with targets} & 3550 & 0.47 & 3520 & 0.47 & 39 & 0.01\\
\hline
5\,MW$_\mathrm{e}$ from& Core 1 &178&8.83& 178 & 9.5$^\ast$ &\multicolumn{2}{c}{\multirow{2}{*}{N/A}}\\
 1st inspection& Core 2 & 648 & 27.7 & 95 & 3.29 &\multicolumn{2}{c}{}\\
\hline
5\,MW$_\mathrm{e}$ from& Core 1 &178 & 8.83 & 138 (83) & 6.68 (3.76$^\dagger$ ) & 43 (1.9) & 2.12 (0.11)\\
start-up& Core 2 & 648 & 27.7 & 52 (66) & 1.81 (2.30$^\dagger$ ) & 6.7 (6.9) & 0.23 (0.24)\\
\hline
\multicolumn{2}{c}{5\,MW$_\mathrm{e}$ Core 3} & 307 & 14.6 & 51 & 2.17 & 3.2 & 0.14 \\
\hline
\multicolumn{2}{c}{5\,MW$_\mathrm{e}$ Core 4} & 255 & 12.3 & 53 & 2.36 & 2.7 & 0.12 \\

\end{tabular}
\mycaption{ \label{table:TotalPuBudget} Plutonium content and
  1$\sigma$ uncertainties are given for two analysis techniques for
  both the IRT and 5\,MW$_\mathrm{e}$ reactors. Due to the inability
  to reliably detect the presence of targets in the IRT reactor, they
  are assumed to be in the reactor. The detection capability is given
  for each 250 day run of the IRT. The 5\,MW$_\mathrm{e}$ reactor
  plutonium error is a combination of removed plutonium that may have
  occurred during the 70 day shutdown and the final plutonium content
  in the reactor at the 1994 shutdown. The quantities are independent
  if data is only taken after the 1st inspection and correlated if
  taken from start-up. The flat burn-up analysis adds a fixed burn-up
  to each time bin and the final plutonium error is the final
  plutonium difference between the burn-up increased data and the
  expected data. The power constrained analysis assumes the starting
  fuel composition is known and the burn-up is given by the
  integration of the power with an assumed 1$\%$ detector
  normalization uncertainty. The plutonium error is the maximum
  plutonium difference attainable through power increases and fuel
  removal (in the case of the 5\,MW$_\mathrm{e}$ reactor). Values are
  given for 1$\sigma$ sensitivities for maximizing the plutonium available
  for core 1 or core 2 respectively. Parenthesis are for uncertainties
  in cores using only data from the respective section. Core 3 and
  core 4 are additional fuel loads that are irradiated in the
  5\,MW$_\mathrm{e}$ reactor post-1994 according
  to~\protect\cite{Stock2012} and are added for
  completeness.\protect\newline {\footnotesize $^\ast$\,Using
    uncertainty from~\protect\cite{puzzle}\protect\newline
    $^\dagger$\,These two numbers are anti-correlated with a
    correlation coefficient of -0.962.}}
\end{center}
\end{table}

For completeness we also list the plutonium mass sensitivities from
the indirect method and the detection of reprocessing wastes in
table~\ref{table:TotalPuIndirect}.
\begin{table}
\sf
\begin{center}
\begin{tabular}{cccc}
\multicolumn{2}{c}{\multirow{2}{*}{}} & Core 1 burn-up [MWd/t] & Core 1 Pu [kg]\\
\hline
\multicolumn{2}{c}{Parasitic measurement} & 51 & 2.55\\

\multicolumn{1}{c}{\multirow{2}{*}{Waste measurement}}& Suspected waste site& 56 & 2.76\\

&Reprocessing plant& 34 & 1.67\\
\end{tabular}
\end{center}
\mycaption{\label{table:TotalPuIndirect} 1\,$\sigma$ uncertainties on
  the discharged plutonium for core 1 for the IRT parasitic
  measurement and for the detection of high-level reprocessing waste.}
\end{table}

%%%%%%%%%%%%%%%%%%%%%%%%%%%%%%%%%%%%%%%%%%%%%%%%%%%%%%%%%%%%%%%%%%%%%%%%%%%%
\subsection{Impact of backgrounds}
\label{sec:backgrounds}

So far in this analysis, we have neglected backgrounds not related to
neutrino emissions. The main backgrounds in inverse beta-decay
detectors are: accidentals, where two uncorrelated events caused by
ambient radiation in the detector accidentally fulfill the delayed
coincidence requirements in both time and energy; fast neutron induced
backgrounds, where a fast neutron enters the detector without leaving
trace and scatters off a proton, which then is confused with the
primary energy deposition of a positron, and subsequently the neutron
thermalizes and captures like a genuine neutron from inverse
beta-decay; $\beta$-n backgrounds, where interaction with cosmic ray
muons produces a short-lived radioactive isotope which decays by
beta-delayed neutron emission, which mimics a neutrino event. The rate
of accidentals is determined by the rate of ambient radioactive
decays. Fast neutrons are a result of cosmic ray interactions in
materials surrounding the detector and thus depend on the rate of
cosmic ray muons; the same is true for $\beta$-n
backgrounds. Therefore, the measured background rates due to those two
sources have to be scaled from the underground location, where most
current neutrino detectors are located, to the surface. Neutrino
detectors are commonly put underground precisely to reduce these two
sources of backgrounds, since deep underground the cosmic muon flux is
strongly attenuated. The scaling of the number of background events is
not purely given by the muon flux, but also, to some degree, depends
on the average muon energy~\autocite{Abe:2012ar}, the scaling is given
by
\begin{equation}
\label{eq:bgscaling}
R\propto \phi_\mu \langle E_\mu \rangle^\alpha\,,
\end{equation}
where $\phi_\mu$ is the muon flux and $\langle E_\mu \rangle$ is the
average muon energy which, at the surface, are
$127\,\mathrm{m}^{-2}\,\mathrm{s}^{-1}$ and $4\,\mathrm{GeV}$,
respectively~\autocite{PDG}. $\alpha$ ranges from $0.7-0.9$ depending
on the type of background. Using the numbers measured in the Double
Chooz experiment at a depth corresponding to 300\,meter water
equivalent (mwe)\endnote{Overburden is commonly quoted in units of
  meter water equivalent (mwe), typically 1\,m of rock/soil
  corresponds to about 2-3\,mwe.}, we can scale to a surface deployed
detector and find $1\,\mathrm{d}^{-1}\,\mathrm{t}^{-1}$ fast neutron
events and $43\,\mathrm{d}^{-1}\,\mathrm{t}^{-1}$ $\beta$-n events,
where 1 tonne is assumed to have the composition of CH$_2$. These
rates exceed the accidental rates by a large factor and therefore we
can neglect the accidental backgrounds. This scaling is tested against
several data sets from different experiments spanning a depth range
from $850-120$\,mwe and the scaling is found to be accurate within a
factor of two~\autocite{Abe:2012ar}. At very shallow depths of less
than 10\,mwe, the hadronic component of cosmic radiation is
non-negligible and the scaling relation in equation~\ref{eq:bgscaling}
is probably no longer valid.  With this caveat in mind, we show in
table~\ref{tab:BackgroundSuppression} the noise to signal ratios for
the various detector deployment scenarios. Values smaller than 1
indicate that the current detector technology is likely to be
sufficient and values larger than 1 indicate that improvements in
background rejection are needed.
\begin{table}[h]
\sf
\begin{center}
\begin{tabular}{ccc}
Source & Fast neutron suppression & $\beta$-n suppression\\
\hline
5MW$_{e}$ & 0.07 & 0.21\\
IRT & 0.14 & 0.43\\
IRT parasitic & 260 & 1050\\
Waste & 740 & 3080
\end{tabular}
\mycaption{\label{tab:BackgroundSuppression} Noise to signal ratios for
  a surface deployed detector.}
\end{center}
\end{table}
The required rejection factor can be reduced significantly by
providing a moderate overburden of 10-20\,mwe, which, in principle,
can be engineered into the detector support structure.

Fortunately, there is a significant on-going experimental effort in
several countries to address the R\&D for neutrino
detectors with greatly improved background rejection. These
initiatives are motivated by the search for a new particle called a
sterile neutrino~\autocite{Abazajian:2012ys} through the use of neutrinos from
reactors with detectors placed within meters of the reactor core. The
close proximity to a reactor core results in a high-background
environment which can include a significant flux of fast neutrons and
high-energy gamma-rays from the reactor itself. Almost all reactor
sites under consideration offer only very minimal overburden of
10\,mwe or less. Therefore, these experiments face essentially the
same level of problems in terms of signal to noise conditions as
safeguards detectors would under the conditions outlined in this
paper. Specifically, there are, to name but a few, the PROSPECT
collaboration in the U.S.~\autocite{Djurcic:2013oaa}, the
DANSS~\autocite{Alekseev:2013dmu} project and
NEUTRINO-4~\autocite{Serebrov:2013yaa} in Russia, the STEREO project in
France, and the SOLID project in Belgium. Each experiment has chosen a
unique approach to address the challenges of a high noise to signal
ratio and some experiments have already reached the prototype
stage. 

%%%%%%%%%%%%%%%%%%%%%%%%%%%%%%%%%%%%%%%%%%%%%%%%%%%%%%%%%%%%%%%%%%%%
\section{Application to the 1994 crisis}
%%%%%%%%%%%%%%%%%%%%%%%%%%%%%%%%%%%%%%%%%%%%%%%%%%%%%%%%%%%%%%%%%%%%%
\label{sec:appl1994}

%%%%%%%%%%%%%%%%%%%%%%%%%%%%%%%%%%%%%%%%%%%%%%%%%%%%%%%%%%%%%%%%%%%%%
\subsection{Conventional methods}
\label{sec:conventional}

The events in 1994 put a premium on understanding the actual history
of the North Korean plutonium program; a vivid interest
in this problem remains in the aftermath. The actual text
of the Agreed Framework states

\begin{quote}
{\it [\ldots] before delivery of key nuclear components, the DPRK will come
into full compliance with its safeguard agreement with IAEA
(INFCIRC/403), [\ldots] with regard to verifying accuracy and
completeness of the DPRK's initial report on all nuclear material
[\ldots]} \hfill Section IV, Paragraph 3
\end{quote}

In 1994, and even today, there has been a need to resolve the question
of whether there was significant reprocessing prior to 1992. It comes
as no surprise that actual methods, relying on more conventional
means, were devised. We will briefly review those conventional methods.

The unloading of the 5\,MW$_\mathrm{e}$ core in June 1994 provided a
crucial opportunity to acquire data that would allow a determination
of whether there was a partial or complete core unloading in 1989. The
DPRK, aware of this possibility, tried to prevent IAEA from gaining
this information by unloading the core very quickly and, by doing so,
made it impossible to infer the exact position of a fuel element
inside the core.  Unfortunately, there is little published on the
details of how a measurement would have proceeded and we have to rely
on interviews with experts. The basic concept of the method is to map
out the three dimensional burn-up distribution inside the reactor
core. If the declaration by the DPRK, that only a few hundred damaged
fuel elements were replaced in 1989 were true, then there should be a
discontinuity in the burn-up distribution at the position of the
replaced fuel elements~\autocite{HeinonenInterview,GeshReid}. On the
other hand, if more than those few hundred were replaced then
discontinuities would show up at many more locations. If the whole
core was replaced a continuous distribution would emerge. Overall,
there are about 8,000 fuel elements and the goal is to find a
discrepancy concerning as few as several hundred fuel
elements. Therefore, a sizable sample of about 300 fuel elements is
required~\autocite[p.\,170]{GoingCritical}. There are two principal
methods to determine the burn-up of spent fuel: one is destructive
sampling with subsequent isotopic analysis and the other is to measure
the characteristic radioactivity emanating from a spent fuel
element. Destructive sampling was (and is) exceedingly difficult in
this context~\autocite{GeshReid}. The other possibility to measure
burn-up relies on measuring gamma-emission from mostly cesium-137,
which is a good proxy for burn-up. According to an expert from Los
Alamos National Laboratory who was closely involved in the 1994 DPRK
issue, this technique would provide burn-up errors below 5\% if good
quality calibration data existed~\autocite{Menlove}. This method, in
principle, has been calibrated on British Magnox fuels. Applying this
type of measurement to several hundred of the spent fuel elements and
knowing their location in the core presents a viable method to
reconstruct the three dimensional burn-up distribution. None of the
interviewed experts was willing to make a statement as to what level
of precision, in terms of partial core reloads and extracted plutonium
amounts, would have resulted. However, we can put a lower limit on the
achievable error by assuming that the overall systematic errors are
less than 5\% and the errors for individual fuel elements are in the
1-5\% range~\autocite{Menlove}. Therefore, the overall accuracy very
roughly should be in 1-5\% range.

This estimate coincides with the accuracy of a method which is based
not on the sampling of spent fuel but instead on sampling the graphite
moderator in the reactor. The idea is elegant and simple; the graphite
moderator is in the reactor for the entire lifetime of the reactor and
thus it will be exposed to \emph{all} the neutrons produced throughout
the history of reactor operation. Since plutonium results from neutron
capture on uranium-238, the total amount of plutonium produced is
strictly proportional to the number of all neutrons. Even reactor
grade graphite contains traces of other elements like boron or
titanium and both these elements have stable isotopes, specifically
boron-11 and titanium-49, which result from neutron
capture. Therefore, the ratios boron-10/boron-11 and
titanium-48/titanium-49 will decrease with the total neutron
fluence. This graphite isotope ratio method (GIRM) was first proposed
by Fetter in 1993~\autocite{Fetter} and subsequently developed in
considerable detail at Pacific Northwest National Laboratory. In an
actual application, samples from the graphite would be taken at a few
hundred strategically chosen points throughout the core and the
isotope ratios would be determined by mass spectroscopy. This data
then can be used to reconstruct a three dimensional neutron fluence
distribution which then, in turn, can be converted to the total amount
of plutonium produced in the reactor through its entire lifetime. This
method was experimentally verified at a British
reactor~\autocite{Trawsfynydd} with an accuracy in the 1-5\%
range~\autocite{Heasler}. This method is quite invasive and requires
extensive cooperation between the national authorities and operator of
a reactor in question, and the state or organization carrying out the
testing. It has the advantage that it is tamper resistant and the
historical record, in the form of the graphite moderator of the
5\,MW$_\mathrm{e}$ reactor, is still available.

We have reviewed two methods: one based on gamma-ray emission of
spent fuel and the other on isotope ratios in the graphite
moderator. Both methods rely on several hundred samples collected
across the reactor and a subsequent reconstruction of three
dimensional distributions of either burn-up or neutron fluence. For
the first method we can only provide a rough estimate of accuracy,
whereas for GIRM the errors have been experimentally determined; the
rough estimate and the experimentally determined error are in the same
range of 1-5\%. The method relying on spent fuel measurements
ultimately requires additional information, e.g. that a few
hundred rods were exchanged in 1989, to make inferences about
alternative core histories which lead to the same total burn-up.  The
GIRM method, on the other hand, can directly address the cumulative
plutonium production in the core and thus does not rely on additional
information. Comparing this to the results that can be obtained from
neutrino measurements, see table~\ref{table:TotalPuBudget}, we see that
neutrino measurements, which do not rely on knowing the core history,
reach about 15\% accuracy based on the burn-up of a given core.
Again, for alternative fuel histories which lead to the same burn-up,
this method requires additional information. If the whole fuel cycle
can be monitored the neutrino accuracy improves to about 1\%. The
overall accuracy of neutrino measurements falls in the same general
range as those achievable by conventional means. Each of the
techniques considered here has different requirements for additional
information to resolve equal-burn-up alternative fuel histories, only
GIRM can resolve those without extra information. 

Another marked difference is the level of intrusiveness, which in
descending order goes from GIRM, which requires drilling sizable holes
into the moderator; to the sampling of spent fuel, which requires
considerable access during refueling; and then to neutrino monitoring,
which only requires access to the exterior of the reactor
building. Also, neutrinos are the only method providing essentially
real-time information, a significant advantage in the context of
break-out scenarios. Up to the point of the break-out, all information
is available, whereas for the conventional methods, the information
can be obtained only at very specific points in time, well after the
actual plutonium production took place. If the break-out happens
between the time of plutonium production and the time when
conventional means can be applied, no information is obtained at all
-- like in the case of the DPRK.

%%%%%%%%%%%%%%%%%%%%%%%%%%%%%%%%%%%%%%%%%%%%%%%%%%%%%%%%%%%%%%%%
\subsection{Neutrinos}
\label{sec:matters}

Based on the quantitative results and the time-line of events in 1994,
see figure~\ref{fig:timeline}, the following scenario may have been
put into effect:
\begin{itemize}
\item The IRT is under full neutrino safeguards with a dedicated
  5 tonne detector from 1978 on, which is located outside the IRT
  reactor building at the southern wall.
\item The 5\,MW$_\mathrm{e}$ is under full neutrino safeguards with a
  dedicated 5 tonne detector from May 1992 on, which is located outside
  the 5\,MW$_\mathrm{e}$ reactor building at the western wall.
\item A search for neutrino emissions from the reprocessing waste is
  initiated in November 1992. Three 5 tonne detectors are deployed: one
  at the reprocessing plant;
  one at the suspected waste site, located above the center of the waste
  site; and one at building 500, located right outside the southern fence.
\end{itemize}
This scenario is fully consistent with the \emph{actual} safeguards
access the IAEA had and, in particular, all detector deployment
locations reflect actual physical access. As a result, the detectors
at the IRT and 5\,MW$_\mathrm{e}$ have a standoff of 20\,m, the
detectors at the suspected waste site and reprocessing plant have a
distance of 25\,m, and the one at the building 500 has a distance of
80\,m.

Furthermore, we assume that in 1989 the DPRK discharged the complete
first core, which seems to be corroborated by the declaration of the
DPRK in 2008 that it possesses 30\,kg of plutonium~\autocite{Stock2012}.
After reprocessing of the spent fuel, the waste was stored somewhere
in the reprocessing plant~\autocite{HeinonenWaste}. Finally, we also
assume that the burn-up declared by the DPRK in 1992 is indeed
correct~\autocite{puzzle,HeinonenBurnup}. This completely specifies the
scenario.

The first relevant piece of data would be obtained by the IRT
detector during periods when the IRT is shut down, about
100 days out of each year. The neutrino signal stemming from the
operation of the 5\,MW$_\mathrm{e}$ is clearly detectable at this detector
location and provides a measurement of reactor power. In 1989, this signal
would have been recorded but would not have raised any special concern,
since the 5\,MW$_\mathrm{e}$ was not under safeguards at this time. At most,
this data would have helped to corroborate U.S. government analyses of
satellite imagery to ascertain the operational history of the
5\,MW$_\mathrm{e}$. However, soon after the DPRK had submitted its
initial declaration to the IAEA, in May 1992, this data would have
resulted in a discrepancy which, in combination with the results from
environmental sampling would have led to the conclusion that a large
amount of plutonium had been separated in 1989. This measurement,
according to table~\ref{table:TotalPuIndirect}, has a sensitivity which
corresponds to 2.55\,kg plutonium, or equivalently to a
$8.83/2.55=3.5\,\sigma$ detection, meaning the IAEA would have
known that a significant fraction of the first core had been discharged with
a confidence of 1 in 1\,900. Taking 4\,kg of plutonium as the quantity
needed for a nuclear bomb, this result translates into a 1 in 13
confidence that the DPRK has at least enough plutonium for one bomb.

In November of 1993, after a year of data collection, the detectors at
the suspected waste site would not have found anything nor would the
detector at building 500, the former result proving that only a small
amount of high-level waste could be present at the suspected waste
site and the latter being insignificant since the distance to the
waste is too large. The detector at the reprocessing plant would have
shown the presence of high-level radioactive
waste~\autocite{HeinonenWaste,Hecker2008}, corresponding to a plutonium
accuracy of 1.67\,kg. That is, with a confidence of 1 in 1,000,000,
the presence of reprocessing waste would have been
confirmed. Moreover, with a confidence now of 1 in 270, it would have
been known that enough plutonium for one weapon was processed. Six
months later, in May 1994, the 5\,MW$_\mathrm{e}$ detector would have
confirmed the burn-up declaration of the DPRK with an accuracy of
15\%. In combination, these results would have implied a 56\% chance
of there being enough plutonium for two or more bombs.

Overall, had the DPRK allowed the detectors to be installed and
operated, neutrino safeguards would, in the scenario considered, have
changed the state of information significantly. The existence of a
separate first core would have been established with very high
confidence and the fact that this core was reprocessed would have been
known to very high confidence. It would have been known, with very
high confidence, that at least enough plutonium for one bomb has been
separated. There would have been some indication that there would be
enough separated plutonium for two bombs. Again, if the DPRK had
allowed the use of neutrino detectors, all of this knowledge would
have been available by the end of 1993.

Assessing the impact this additional knowledge would have had on the
course of events and the policies driving those events is difficult
and will have to remain speculative. From an IAEA perspective, the
DPRK would have been found in substantial violation of the NPT and the
Director General would have to report his findings to the Board of
Governors and, as result, the matter would have been referred to the
UN Security Council. Furthermore, the history of the North Korean
plutonium production would have been known including a reasonably
accurate account of the amount of separated plutonium. The ability of
the IAEA to deliver such a detailed picture, despite extensive efforts
on the side of the DPRK to obfuscate, in a timely manner would have
been counted as a major success.

As far as the U.S. and the DPRK are considered, we note that the
primary goals and motivations would have remained invariant for both
sides. The U.S. still would have wanted to prevent any further
production of separated plutonium~\autocite{GallucciShutdown} and the DPRK
still would have wanted to obtain maximal material and political gains
for eventually accepting any U.S. demands. Both parties would have
remained keen to avoid war on the Korean peninsula, since for the
U.S. the number of casualties and financial burden would have appeared
difficult to justify and a war would clearly be against the interest
of its close ally South Korea. For the DPRK, or more specifically its
leadership, a war would constitute the ultimate catastrophe resulting
in regime change, at a minimum, and likely captivity or worse for the
very top echelons of government. These stakes remain for both sides to
this day and the mutual awareness thereof has provided some measure of
stability to the Korean peninsula.

In view of the information provided by neutrinos, both parties would
have lost the benefit of ambiguity, which would have forced the
U.S. to act decisively. For North Korea, it would have become much
more difficult to pretend that a mere accounting problem had
occurred. As a result, the tension would likely have risen much more
sharply resulting in the peak of the crisis at a much earlier time,
now the fall of 1993 instead of June 1994.

There are many arguments which can be levied against the scenario
described in the preceding paragraphs. On the technical side, neutrino
detectors which achieve the required level of background rejection did
not exist in the 90s and do not exist now, at least not with
demonstrated capabilities. In reply to this criticism we point to the
discussion in section~\ref{sec:backgrounds}, where also a brief
summary of the current R\&D efforts toward better detectors is
given. On the historical side, any attempt at counter-factual history
ultimately remains a piece of fiction and, while some story lines may
be more plausible than others, there is no way to really know what was
known by whom given the actual circumstances, or what differences
antineutrino monitoring, as described in this paper, would have made
in the interactions and outcomes.  On the policy side, safeguards
generally do not rely on secret technologies and black boxes. States
entering a safeguards agreement with the IAEA have the right to
require a complete specification and understanding of the employed
technologies to ascertain their right of protection of trade secrets
and information that is sensitive with respect to national
security. The DPRK is no exception to this rule~\autocite{GeshReid}
and its scientists are knowledgeable and competent experts, which
certainly would have the ability to understand all implications of
neutrino safeguards as outlined
here~\autocite{HeinonenInterview,GeshReid}. Given that they made every
effort to conceal the true history, they also would have tried to
thwart neutrino safeguards. Neutrino safeguards is not necessarily
more difficult to thwart than other means, it just requires different
counter measures. This back-reaction has been completely neglected, in
part because it would add another layer of speculation and because it
is a historical fact that the IAEA's trace analytic capabilities
caught the North Korean experts off guard.

%%%%%%%%%%%%%%%%%%%%%%%%%%%%%%%%%%%%%%%%%%%%%%%%%%%%%%%%%%%%%%%%%%%%%%

%%%%%%%%%%%%%%%%%%%%%%%%%%%%%%%%%%%%%%%%%%%%%%%%%%%%%%%%%%%%%%%%%%%%
\section{Summary \& Outlook}
%%%%%%%%%%%%%%%%%%%%%%%%%%%%%%%%%%%%%%%%%%%%%%%%%%%%%%%%%%%%%%%%%%%%
\label{sec:summary}

Neutrino reactor monitoring offers unique capabilities which seem to
make this method -- as proposed more than 30 years ago -- a useful
tool for safeguards. Also, neutrino detectors have been continually
refined since the days of Cowan and Reines and can be considered a
mature technology. Given the mechanisms of neutrino production and
detection, neutrino safeguards provide bulk measurements of reactor
core parameters like power or burn-up. This is to be contrasted with
the current safeguards approach which largely relies on item
accountancy and, in particular, neither power nor fuel burn-up are
actually measured by the IAEA~\autocite{doyle} or verified through
independent calculations for any reactor. The IAEA is apparently
satisfied that the existing arrangements are adequate for commercial
power reactors of the boiling and pressurized water types, especially
as long as a once-through fuel cycle without reprocessing is
considered. As a result, it has been difficult to show neutrino
safeguards would provide a decisive advantage in comparison to more
conventional techniques, especially since neutrino detectors are
larger and more expensive than most equipment currently used by the
IAEA. The fact that, in the literature, a rather diverse set of
results, in terms of the applicability to specific safeguards issues,
is
found~\autocite{Bernstein:2001cz,Nieto:2003wd,Huber:2004xh,Misner:2008,Bernstein:2009ab,Bulaevskaya:2010wy,Hayes:2011ci,huberINMM}
may have a further detrimental effect on the perception of neutrino
safeguards. These widely varying results can be attributed, to a large
degree, on differing assumptions about detector parameters and
different level of statistical treatment. The choice of detector
parameters often is inspired by the wish to be particularly realistic
or thrifty, which seems to be a classical case of what Donald Knuth
calls premature optimization~\autocite{Knuth}. Furthermore, many
analyses are rate-based which leads to serious deficiencies in
sensitivities since there is a pronounced degeneracy between reactor
power and fission fractions when spectral information is
ignored\endnote{cf.  compare the results
  in~\autocite{Bulaevskaya:2010wy} with the ones
  in~\autocite{huberINMM}}.

In this paper we have taken a different approach -- with the North
Korean nuclear crisis of 1994 we have identified a real-world scenario
in which traditional safeguard techniques ultimately were unable to
resolve the key questions and for which sufficient technical
information is publicly available to perform a detailed
analysis. Moreover, we base our detector parameters on the overall
acceptable size and weight of the entire detector system, which we
envisage to fit inside a standard 20\,ft intermodal shipping
container. We also assume that the detector will be able to operate at
the surface. Together with the standard packaging this will provide a
great deal of flexibility in the choice of deployment locations. The
price to pay is that the detector has to have excellent background
rejection, which seems to exclude single volume liquid scintillator
detectors and favors finely segmented solid detectors, see
section~\ref{sec:backgrounds}. Detectors with these capabilities
currently do not exist, but a basic science question related to the
possible existence of a new particle, a so-called sterile neutrino,
has triggered a large number of experimental efforts to perform
reactor neutrino experiments at a range of several meters from compact
reactor cores. These experiments face enormous challenges from
reactor-generated backgrounds and, therefore, have to solve the
background rejection issue. Many of the new designs do not rely on
large-area photo-detectors, which are typically hand-made in small
numbers and therefore are very expensive. New detector designs have
the potential to become more affordable in industrial production.
Therefore, there is ample reason to assume that within a few years
detectors with the required characteristics will be available at more
reasonable prices.

Spectral analysis of neutrino emissions from a reactor requires an
accurate understanding of the neutrino yields from fissile
isotopes. Recently, there have been a number of publications
addressing the issue of neutrino yields and many of the previously
accepted results from the late 1980s have been called into
question. In section~\ref{sec:fluxes}, we point out, that while the
absolute neutrino yields differ significantly between different
calculations, the differences in neutrino yields between the fissile
isotopes are predicted quite consistently. Ultimately, these questions
should be settled by a calibration measurement at a reactor with
well-known core composition.

In section~\ref{sec:pucontents}, we developed a quantitative framework
to determine the plutonium mass inventory of a number of different
reactor types from neutrino spectroscopy and based our study on
detailed reactor burn-up simulations. In figure~\ref{fig:scaling}, we
summarized these results as a function of the reactor thermal power for
an analysis which makes no assumptions about the history of reactor
power. In comparing various reactor types and their suitability for
neutrino safeguards, we identified the neutron flux density, which is
closely related to the power density, as the main parameter of
influence.

We showed that antineutrino safeguards could enable the IAEA to detect
unreported plutonium production or diversion of declared plutonium at
the one significant quantity amount, i.e. 8\,kg of
plutonium within 90 days at 90\% confidence level at light-water
moderated reactors producing less than 1\,GW$_\mathrm{th}$ power and at
heavy-water moderated reactors producing less than 0.1\,GW$_\mathrm{th}$
power. These results suggest that the heavy-water reactor at Arak in
Iran with an estimated thermal power of 0.04\,GW$_\mathrm{th}$ could
be an ideal target for neutrino safeguards, and a detection limit of
4.4\,kg plutonium within 90 days at 90\% confidence level seems
possible. Graphite moderated reactors, on the other hand, are more
difficult due to a relatively low power density.

We also developed an analysis method based on the fact that the
isotopic abundance of the various fissile isotopes as a function of
burn-up is governed by reactor physics and consequently these
quantities are correlated in a well-defined manner, as explained in
section~\ref{sec:reactors}. In this case, the problem can be rephrased in
terms of reactor power and burn-up and this improves the sensitivity
by roughly a factor of two. The reactor model required for this type
of analysis does not have to be extremely detailed or accurate, since
only the gross burn-up evolution is required.

For our purposes, the North Korean nuclear program consists of three
pieces: the IRT, an 8\,MW$_\mathrm{th}$ light-water research reactor
supplied by the USSR, which is fueled with HEU and has been under IAEA
safeguards since 1977; the 5\,MW$_\mathrm{e}$ reactor, a
20\,MW$_\mathrm{th}$ graphite moderated natural uranium fueled
reactor; and the Radiochemical Laboratory, a reprocessing facility
which can extract plutonium from the spent fuel using the PUREX
process. The central question was whether the fuel in the
5\,MW$_\mathrm{e}$ reactor was the original core load or whether there
was an earlier undeclared refueling during the shutdown in 1989. The
discharged fuel would have yielded about 8.8\,kg of plutonium,
sufficient for at least one nuclear bomb. Our technical analysis is to
a large degree based on the data presented in \citetitle{puzzle},
which we use as input for detailed reactor core simulations for both
the IRT and the 5\,MW$_\mathrm{e}$ reactors.

The North Korean declaration of the burn-up history of the
5\,MW$_\mathrm{e}$ is such that the core
configuration in terms of measurable quantities like burn-up and
reactor power is virtually identical for both the one-core and
two-core scenarios. Therefore, safeguards techniques, both
conventional and neutrino-based, have to resort to secondary
observables. In the case of neutrinos, secondary signatures focus on a
measurement of reactor power by a neutrino detector deployed to
implement the safeguards agreement for the IRT, which has been in force
since 1977. This signal is visible only when the much closer and, hence,
brighter neutrino source represented by the IRT is not in operation,
which occurs for about 100 days per year. This measurement would
provide evidence for the presence of a second core with a confidence
of 1 in 1\,900 (3.5\,$\sigma$), see section~\ref{sec:dprknu}.

Another secondary signature, which can be exploited with neutrinos is
the detection of the presence of reprocessing wastes, which contain
long-lived fission fragments, some of which emit detectable
neutrinos. Historically, three sites have been suspected to be the
potential disposal locations: Building 500, a suspected waste site,
and the Radiochemical Laboratory. The map in
figure~\ref{fig:NeutrinoMap} shows the relative locations. For the
latter two sites a neutrino safeguards detector would have been able
to detect the presence of the reprocessing wastes with a confidence of
better than 1 in 600, see section~\ref{sec:dprknu}. For the 1994
crisis, the application of neutrino safeguards could have resulted in
significantly reduced uncertainty about North Korean intentions.

In a more general context, we also studied the resulting sensitives
assuming that neutrino safeguards had been available from the start-up
of the 5\,MW$_\mathrm{e}$ reactor and showed that if a continuous
measurement of reactor power by neutrinos had been available, which
could then be compared to a measurement of the burn-up by neutrinos at
a later point, there would have been very little room for undeclared
plutonium production or refuelings; accuracies corresponding to
1-2\,kg of plutonium would have been achieved, see
table~\ref{table:TotalPuBudget}. Our work shows that even graphite
moderated reactors can be safeguarded successfully using neutrino
monitoring. Furthermore, we compared the abilities of neutrino
safeguards with conventional capabilities in
section~\ref{sec:conventional} and found that those conventional
techniques have to rely on a level of reactor physics modeling
comparable to the more advanced analysis techniques we have
presented. Provided the extensive effort and funds required for such
modeling can be expended, the overall accuracy of conventional
techniques should be in the 1-5\% range, whereas neutrino techniques
are in the 5-15\% range in terms of plutonium content. The crucial
advantage neutrino safeguards offer stems from the near real-time data
acquisition during reactor operation, whereas the conventional methods
require a reactor shutdown and a defueling of the reactor. Neutrino
safeguards also is entirely non-invasive; at 20\,m standoff the
detector can be deployed outside the reactor building. In the context
of break-out scenarios, deferring the ability to know how much
plutonium was produced or whether a diversion has taken place until a
later point in time, when there is no guarantee of safeguards access
at the required level, is problematic -- but this is what conventional
techniques have to rely on and have done in the case of the DPRK, with
the known result.

To summarize, using the North Korean nuclear crisis as a virtual
laboratory, we have found by detailed technical analysis that neutrino
safeguards for water moderated reactors with a thermal power less than
0.1-1\,GW$_\mathrm{th}$ can meet the IAEA detection goals in terms of
plutonium content and timeliness. This makes neutrinos a viable choice
for many research reactors, small, e.g. 40\,MW$_\mathrm{th}$,
plutonium production reactors, and for most of the planned commercial
small modular reactors. Small modular reactors would allow for the
inclusion of a neutrino safeguards system at the design stage. In the
specific North Korean case, we find that neutrinos provide an accuracy
which is marginally worse than conventional methods and,
qualitatively, the difference in accuracy seems to be irrelevant. At
the same time, neutrinos allow conclusions about the plutonium content
and potential diversion to be drawn in close to real-time, whereas
conventional methods provide the information only after the fact, once
the reactor is shut down and defueled. For all of these applications,
neutrino detectors have to work with minimal or no overburden and the
lower the residual background is, the more versatile the resulting
system will be. For very low background detectors, remote power
measurements and the detection of reprocessing wastes becomes an
attractive possibility, in particular for purposes of nuclear
archaeology.

%%%%%%%%%%%%%%%%%%%%%%%%%%%%%%%%%%%%%%%%%%%%%%%%%%%%%%%%%%%%%%%%%%%
\section*{Acknowledgements}

We thank R.~Gallucci, C.~Gesh, O.~Heinonen, H.~Menlove and B.~Reid for
their willingness to be interviewed for this project. We also thank
A.~Erickson, L.~Kalousis, J.~Link, C.~Mariani and in particular
T.~Shea for their expert opinions on many of the technical issues
involved regarding nuclear reactors, neutrino detectors, and
safeguards. We thank M.~Fallot for providing reactor neutrino fluxes
in machine readable format and we also acknowledge useful discussions
about solid, segmented neutrino detectors with A.~Vacharet and
A.~Weber. This work was supported by the U.S. Department of Energy
under contract \protect{DE-SC0003915} and by a Global Issues
Initiative grant by the Institute for Society, Culture, and Environment
at Virginia Tech.

\newpage

%%%%%%%%%%%%%%%%%%%%%%%%%%%%%%%%%%%%%%%%%%%%%%%%%%%%%%%%%%
\begin{appendix}

%%%%%%%%%%%%%%%%%%%%%%%%%%%%%%%%%%%%%%%%%%%%%%%%%%%%%%%%%
\section{The 5\,MW$_\mathrm{e}$ reactor}
\label{sec:5MWSCALE}

\begin{figure}[t]
\includegraphics[width=0.5\textwidth]{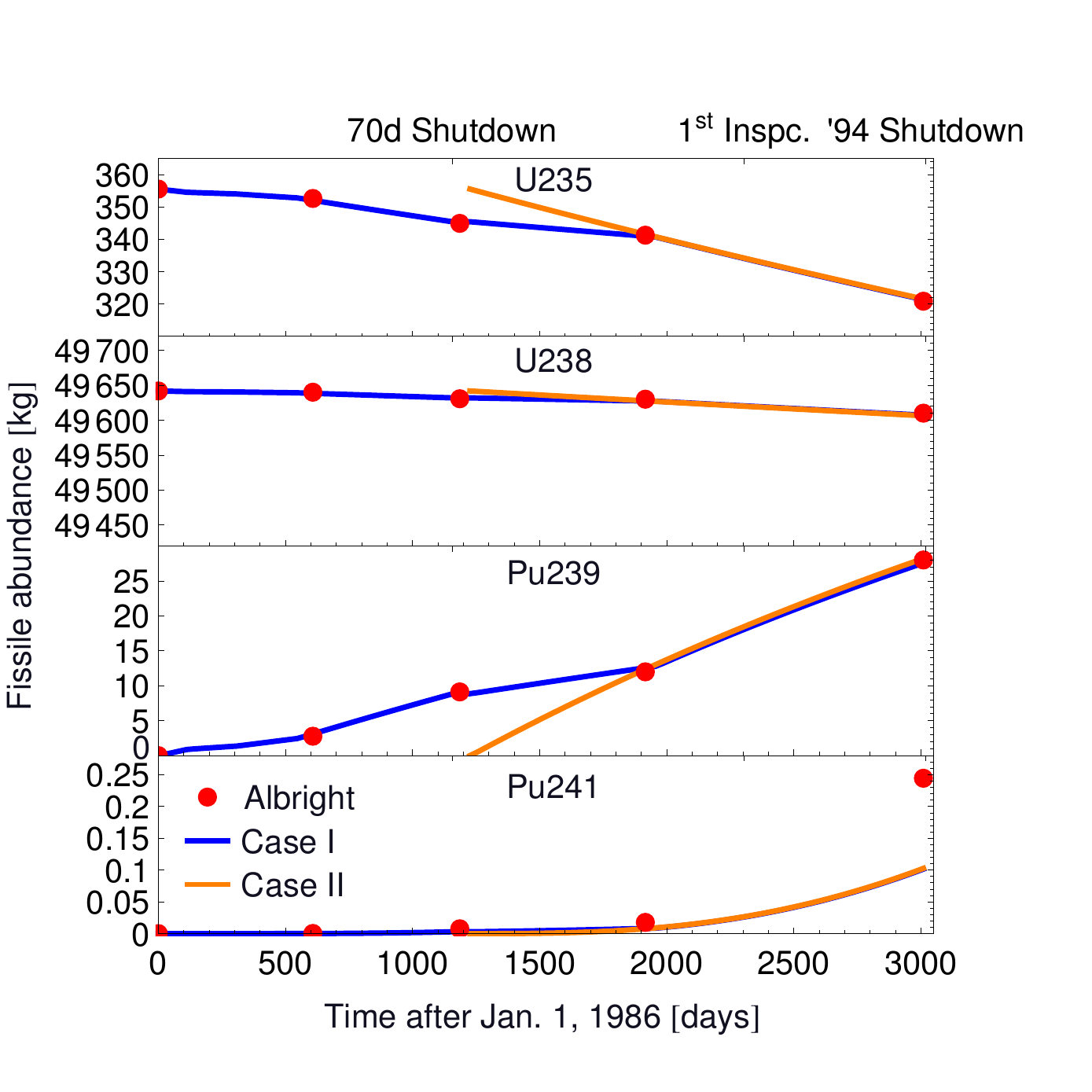}%
\includegraphics[width=0.5\textwidth]{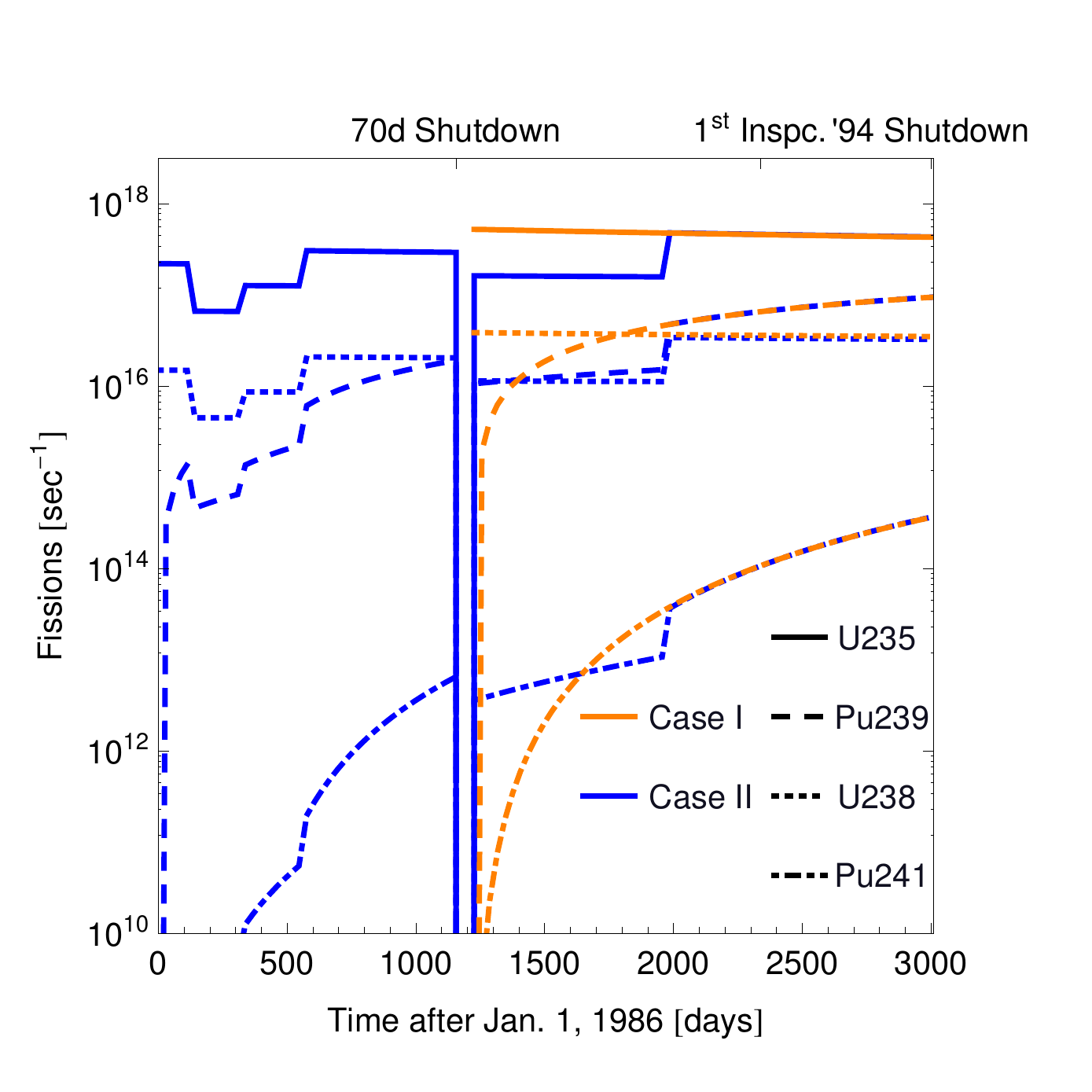}
\mycaption{\label{fig:5mwfissionrates} Fissile content of the
  5\,MW$_\mathrm{e}$ reactor (left hand panel) and the
  corresponding fission rates (right hand panel)}
\end{figure}

 The 5\,MW$_\mathrm{e}$ reactor is a graphite moderated and reflected
 Magnox reactor using natural uranium fuel, based on the British
 Calder Hall design, and runs with a nominal power capacity of
 20-25\,MW$_\mathrm{th}$. The average burn-up of the fuel elements is
 635\,MWd/t.  The 5\,MW$_\mathrm{e}$ reactor contains 812 vertical
 channels, each with up to 10 fuel elements per channel. The fuel
 elements are made of natural uranium in a magnesium-aluminum alloy
 (Magnox) and a full core consists of 50\,MTU. We have based most of our
 historical information regarding this reactor and its power history
 on figure  VI.2 in \citetitle{puzzle}. The 5\,MW$_\mathrm{e}$ reactor
 began operation in January 1986, experienced a 70 day shutdown in
 1989, and continued irradiation past the first IAEA inspections in
 1992 until the shutdown in April 1994. We are primarily concerned
 with two possible fueling histories: case I, or the no-diversion case,
 assumes that the same core was used in the 5\,MW$_\mathrm{e}$ reactor
 during the entire irradiation period from 1986 to 1994; case II, the
 core exchange case, is based on the assumption that, during the 70
 day shutdown of 1989, North Korea replaced the irradiated core with a
 fresh one and continued irradiation with a higher-than-declared power
 to reach the same burn-up as in case I by the time of the 1992
 inspection. The primary safeguards-relevant difference between these
 two cases is that in the second case an entire spent fuel load
 containing about 8.8\,kg of weapons-grade plutonium, is unaccounted
 for.

We simulate the 5\,MW$_\mathrm{e}$ reactor using the SCALE~6.1.1
software package~\autocite{scale} developed at Oak Ridge National
Laboratory. OrigenArp, a subset of SCALE, uses decay data from
ENDF/B-VII~\autocite{endf} and neutron information from the JEFF/A-3.0
~\autocite{jeff} databases to compute burn-up and isotopic
composition. OrigenArp is a deterministic approach which approximates
the structure of a reactor core into a zero-dimensional object by
using appropriately weighted cross section libraries. Some libraries,
including the Magnox type reactor, are predesigned and supplied with
the SCALE software.

OrigenArp begins by computing depletion equations for each individual
isotope in a given problem. The depletion or Bateman equation is given by
\begin{equation} \label{eq:depl}
\frac{dN_i}{dt} = \displaystyle\sum\limits_{j=1}^{m} l_{ij} \lambda_j N_j +
 \bar{\Phi} \displaystyle\sum\limits_{k=1}^{m}f_{ik}\sigma_{k}N_{k} -
 (\lambda_i + \bar{\Phi} \sigma_i)N_i \hspace{20mm} (i=1, ... ,m)
\end{equation}
This accounts for processes that produce nuclide $N_i$ in the first
two terms and for processes that destroy nuclide $N_i$ in the last,
negative term. The first term represents the decays of nuclide $j$ to
$i$ given by the decay constant, $\lambda_j$, the atom density of
nuclide $j$, $N_j$ and the branching fraction, $l_{ij}$, for decays
from nuclide $j$ to $i$. The second term indicates neutron captures
into nuclide $i$ given by the space and energy-averaged neutron flux,
$\bar{\Phi}$, the fraction of absorption on nuclide $k$ that produce
nuclide $i$, $f_{ik}$, and the spectrum-averaged neutron absorption
cross section of nuclide $k$, $\sigma_k$. The last term is the
collection of depletion modes consisting of the decay of nuclide $i$ via
decay constant, $\lambda_i$, and neutron absorption with a
spectrum-averaged neutron absorption cross section of nuclide $i$,
$\sigma_i$. The indices are summed over all branches including nuclide
$i$. SCALE solves this differential equation via a matrix exponential
method. Short-lived isotopes are removed to prevent loss of numerical
accuracy and calculated using Bateman chains.  With the input of
initial nuclide concentrations, the power history, and the reactor
configuration, OrigenArp can provide time-dependent fission rates,
radioactivity, and isotopic abundances during and after irradiation.

For our calculations, we have used the Magnox library provided with
SCALE. Our first check of SCALE is to investigate the production of
plutonium in comparison to earlier results~\autocite{AlbrightPu}. We
irradiate the Magnox fuel corresponding to 1\,MTU of natural uranium
for 1000 days at a constant power level to reach a given final
burn-up. For example, we can irradiate the core at 0.7\,MW for 1000
days to produce a final burn-up of 700\,MWd/t. We then extract the
total amount of plutonium produced and the percentage of $^{240}$Pu,
plutonium-241, and $^{242}$Pu. table~\ref{tab:magnoxburnup} summarizes
the results and compares them with table A.2 in
~\citetitle{AlbrightPu}\endnote{The numbers in~\citetitle{AlbrightPu}
  are originally taken from~\autocite{graphite}}.
\begin{table}[t]
\sf
\begin{tabular}{ccccc}

& SCALE & \citetitle{AlbrightPu} & SCALE & \citetitle{AlbrightPu} \\
Burn-up [MWd/t] & \%$^{240,241,242}$Pu & \%$^{240,241,242}$Pu & kg of Pu & kg of Pu \\ \hline
100 & 0.99 & 0.75 & 0.10 & 0.1 \\ 
200 & 1.9 & 1.5 & 0.20 & 0.19 \\ 
300 & 2.9 & 2.3 & 0.29 & 0.28 \\ 
400 & 3.8 & 3.1 & 0.38 & 0.36 \\ 
500 & 4.7 & 3.7 & 0.47 & 0.45 \\ 
600 & 5.5 & 4.4 & 0.56 & 0.535 \\
700 & 6.4 & 5.1 & 0.64 & 0.62 \\ 
800 & 7.2 & 5.7 & 0.72 & 0.7 \\ 
900 & 8.0 & 6.3 & 0.79 & 0.78 \\ 
1000 & 8.8 & 6.9 & 0.87 & 0.86 \\ 
1100 & 9.5 & $\sim$7.5 & 0.94 & $\sim$0.94 \\ 
1200 & 10 & $\sim$8.1 & 1.0 & $\sim$1.02 \\ 
\end{tabular}
\mycaption{\label{tab:magnoxburnup} Comparison of our SCALE-based results and
the numbers presented in \protect\citetitle{AlbrightPu} for the plutonium content of Magnox fuel as a function of burn-up.}%
\end{table}
We can see that our calculation and the results of~\citetitle{AlbrightPu}
consistently match in the total mass of plutonium produced for various
burn-ups.  SCALE predicts about $25\%$ more of the lesser plutonium
isotopes compared to~\citetitle{AlbrightPu}.

We also perform a comparison of the fissile abundances for the four
main fissile isotopes with the numbers quoted in table~VIII.5
in~\citetitle{puzzle}. We use the OrigenArp sequence with the Magnox
reactor library and the power history inferred from figure~VI.2
of~\citetitle{puzzle} with an initial fuel amount of 1\,MTU. The
x-axis of the plot shows the number of days that have passed after the
5\,MW$_\mathrm{e}$ reactor began irradiation on January 1, 1986.
Notable times are included, such as the 70 day shutdown in 1989
(beginning on $t=1156$\,d), the first IAEA inspection in 1992
($t=2337$\,d), and the shutdown on April 1, 1994 ($t=3012$\,d). In the
leftmost panel of figure~\ref{fig:5mwfissionrates} we have plotted the
results for case I and case II, as well as the data points found in
table~VIII.5 in~\citetitle{puzzle}, normalized to a 50\,MTU core and we
find excellent agreement. We note that immediately following the 70
day shutdown in 1989, the fissile abundances vary greatly between
cases.  The difference quickly disappears as the fresh fuel load is
burned at a higher power so that by the first IAEA inspection in 1992
the fissile abundance differences have vanished. The results in terms
of mass inventory and fission rates are shown in
figure~\ref{fig:5mwfissionrates}.

%%%%%%%%%%%%%%%%%%%%%%%%%%%%%%%%%%%%%%%%%%%%%%%%%%%%%%%%%
\section{CANDU and LEU reactors}
\label{sec:CANDULEU}

The ``$\mathrm{H}_2$O, LEU'' reactor and the ``$\mathrm{D}_2$O, NU''
CANDU reactor are both calculated in the same fashion as the
5\,MW$_\mathrm{e}$ reactor. The LEU reactor calculation is done for a
typical pressurized light water reactor. Specifically, we have taken a
power history from one such reactor, namely Ling Ao I, located in the
Daya Bay complex in China. Ling Ao I is a Framatome M310 reactor,
which uses a 17x17 AFA 3G fuel assembly. SCALE does not have this
specific library, but does contain the very similar Westinghouse 17x17
array, which we have used. Details of the Ling Ao I reactor history
and fuel composition are taken from the yearly power histories
published via IAEA Operation Experience in Member States documents
~\autocite{opex}. To summarize, Ling Ao I has a total fuel load of
72.4\,MTU enriched to 3.7\%. The yearly power histories are converted
into OrigenArp input files. The Ling Ao I reactor runs on a 12 month
refueling cycle, meaning that it must shut down about every 12 months
to refuel. Typically, one third of the fuel is replaced with a fresh
third, and the fuel rods are shuffled within the core to reach a
flatter burn-up distribution. To simulate this, we produce three SCALE
computations: one third of a core that has been irradiated once,
another third that has been irradiated twice, and the last third that
has been irradiated three times. These three output files are then
summed resulting in the final full core. Each third has been
irradiated for approximately 335 days per cycle at an average power of
about 965\,MW$_\mathrm{th}$, resulting in a total power of
2.9\,GW$_\mathrm{th}$. We acquire the fission rates and the fissile
abundances for the four main fissiles from this final full core sum.
The ``$\mathrm{D}_2$O, NU'' is also calculated via SCALE. We use the
CANDU 37-element cross section library in SCALE. The 37-element, as
opposed to the 28-element, design was chosen as it is a more common
design for newer CANDU reactors. This simulation was performed with a
three year irradiation time at an average power of
40\,MW$_\mathrm{th}$ with a 8.6\,MTU natural uranium fuel load. The
CANDU reactor is run continuously with no refueling periods. From the
SCALE output we obtain the fission rates and the fissile abundances
for the four main fissiles. The specifications for this calculation
are intended to mirror the specifications of the Arak reactor in Iran~\autocite{Arak}.

%%%%%%%%%%%%%%%%%%%%%%%%%%%%%%%%%%%%%%%%%%%%%%%%%%%%%%%%%
\section{The IRT reactor}
\label{sec:IRT}

The IRT is a light-water pool-style research reactor and was supplied
to the DPRK by the Soviet Union in the 1960s. First criticality
occurred in the IRT reactor on August 15, 1965. The IRT contains 56
core grid compartments during the time of interest here, i.e.
after 1986~\autocite{puzzle}.  The exact configuration of the IRT is
unknown, but we can base possible configurations on similar IRT
reactors such as the IRT-Sofia~\autocite{iaeabook} located in
Bulgaria. These compartments can contain a driver or target
element. Drivers are the primary fission source in this research
reactor and are made of highly enriched uranium. The targets, which
are primarily composed of fertile isotopes, such as natural uranium,
will experience a small number of fissions. In 1974, the IRT was
upgraded in power from 2\,MW$_\mathrm{th}$ to 4\,MW$_\mathrm{th}$ and
later in 1986 it was upgraded from 4\,MW$_\mathrm{th}$ to
8\,MW$_\mathrm{th}$. The driver element enrichment also increased
during this time from 10\% in 1967 to 80\% by 1986~\autocite{puzzle}.
Exact inventories are unavailable, but estimates indicate the DPRK may
have had access to at least 92 of these 80\% enriched driver
elements~\autocite{puzzle}. From 1986 on-wards, 30 driver elements were
typically loaded and the IRT could run for about 250 days out of the
year.

SCALE computations for the IRT are more difficult as there is no
available cross section library provided with SCALE. Therefore, we use
the Triton and NEWT modules to produce custom cross section libraries
for the IRT calculations. NEWT generates the neutron transport
calculation for a user-defined core configuration, which can then be
used by Triton over a sample burn-up history to produce decay and
cross section libraries. The input information we provide consists of
detailed isotopic compositions of the driver and target elements as
well as physical parameters of these elements and a core
configuration. Information concerning the driver and target elements
is taken from table~VIII.6 in~\citetitle{puzzle}. To summarize, we are
using 80\% enriched U-Al alloy drivers with an aluminum cladding in a
light-water moderator surrounded by a reflector. The target elements
are natural uranium metal in an aluminum cladding. We utilize SCALE
and the IRT power history provided in table~VIII.7 in
\citetitle{puzzle} along with the initial loading of 6\,kg for the
drivers and 633\,kg for the targets.

\begin{figure}[t]
\includegraphics[width=0.5\textwidth]{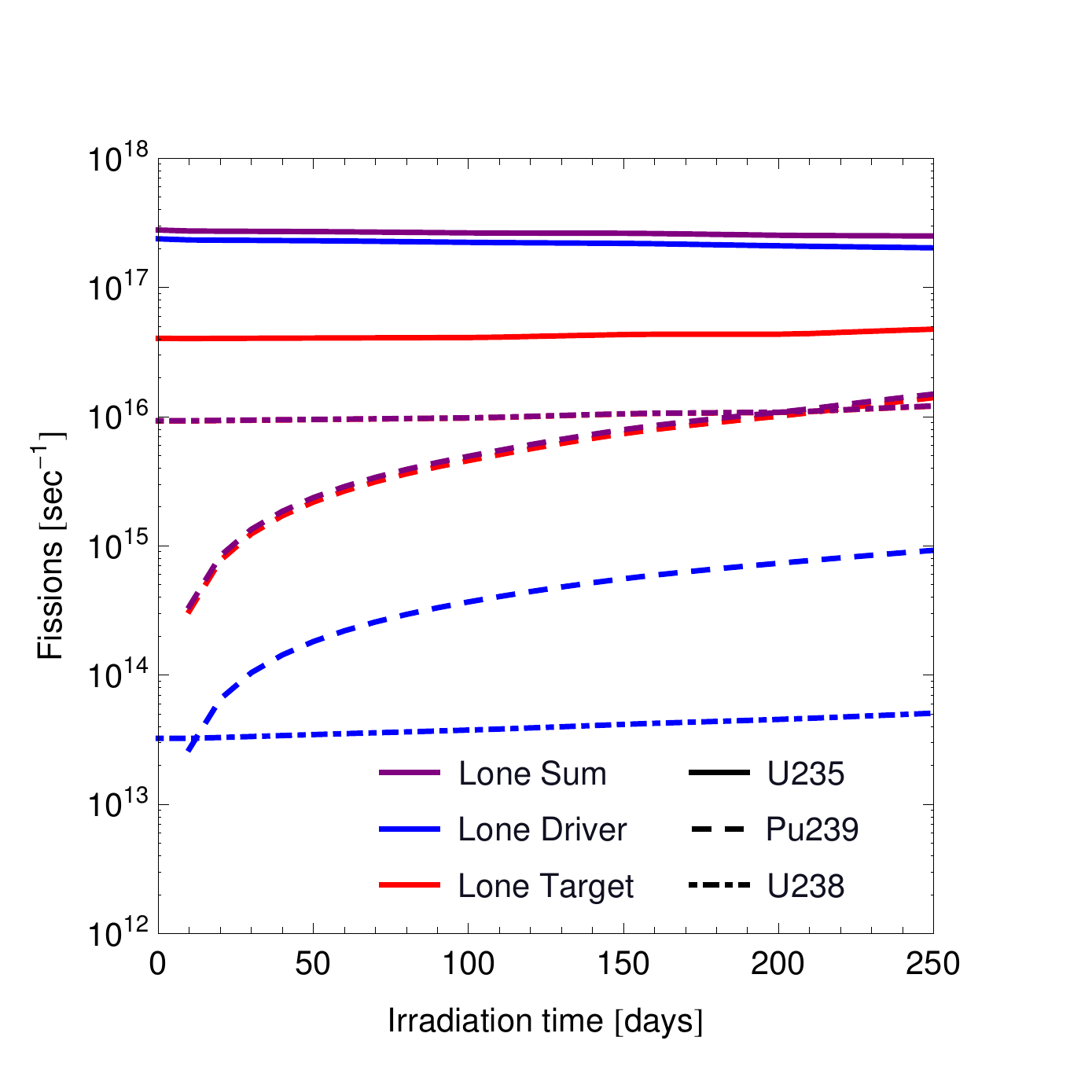}%
\includegraphics[width=0.5\textwidth]{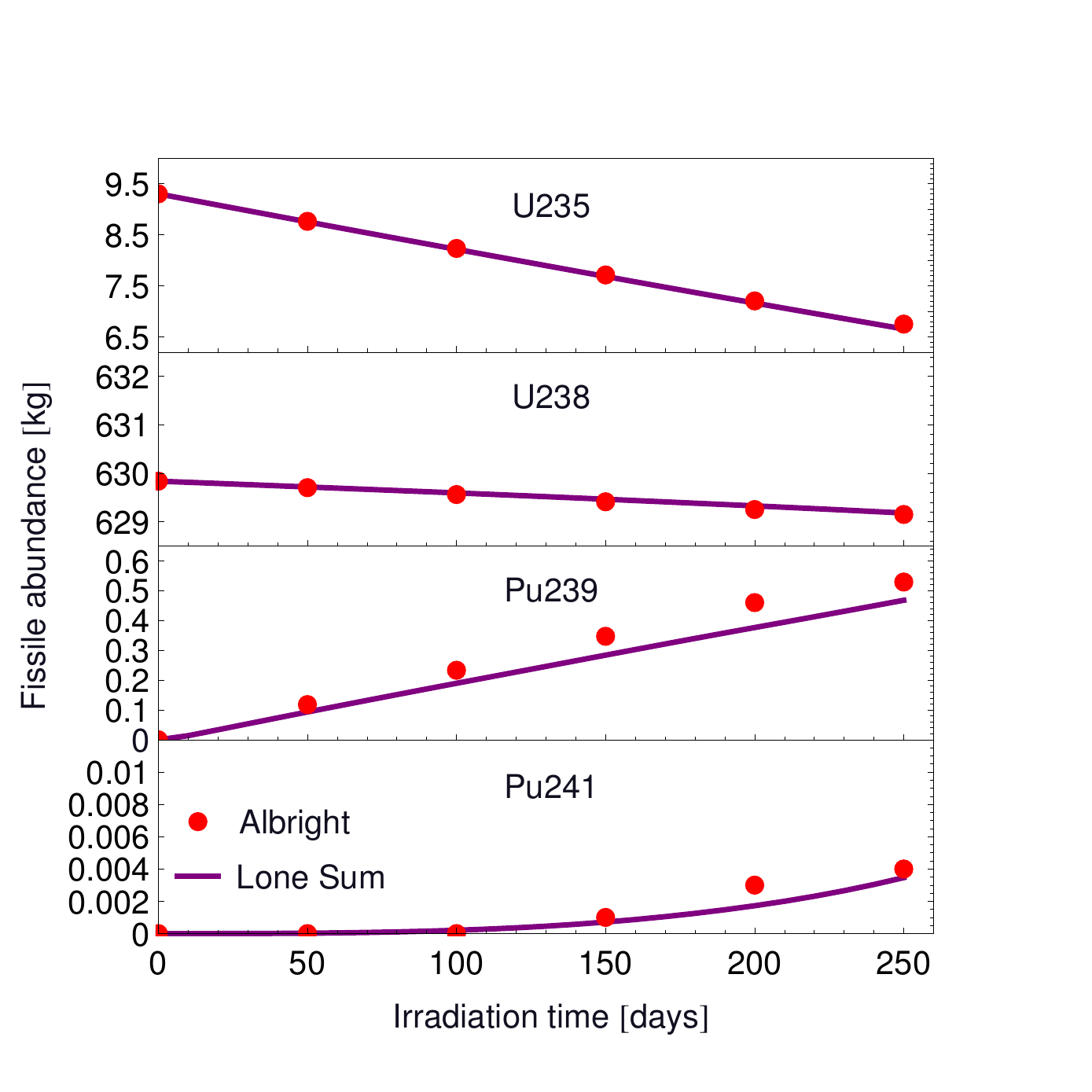}
\mycaption{\label{fig:IRTfissrates} IRT fission rates of a lone driver and
  target element separately and their sum (left panel). The abundances
  of fissile isotopes in the IRT as calculated by the lone sum
  method (right hand panel).}
\end{figure}

\begin{figure}[h]
\includegraphics[width=0.40\textwidth]{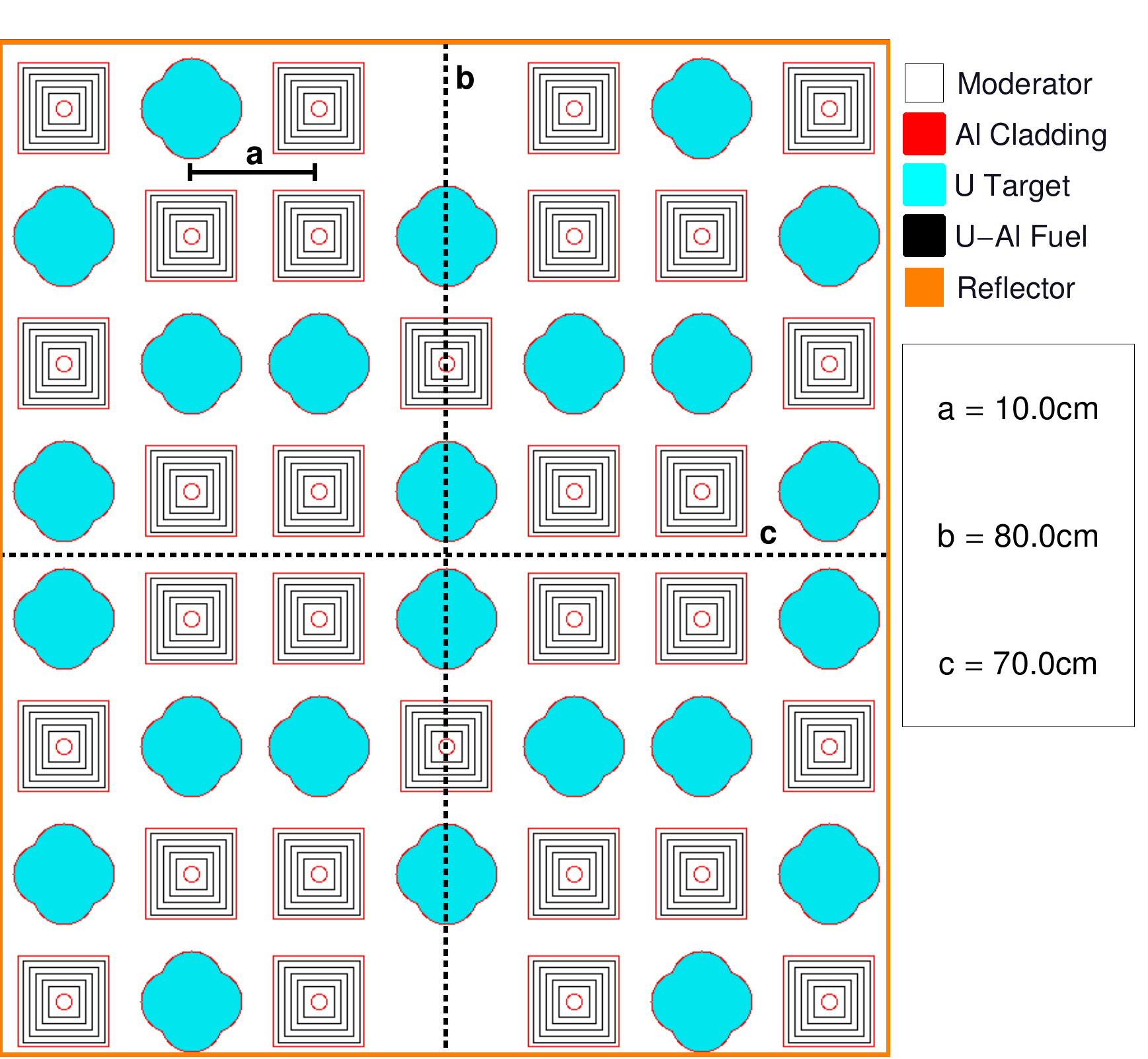}%
  \includegraphics[width=0.4\textwidth]{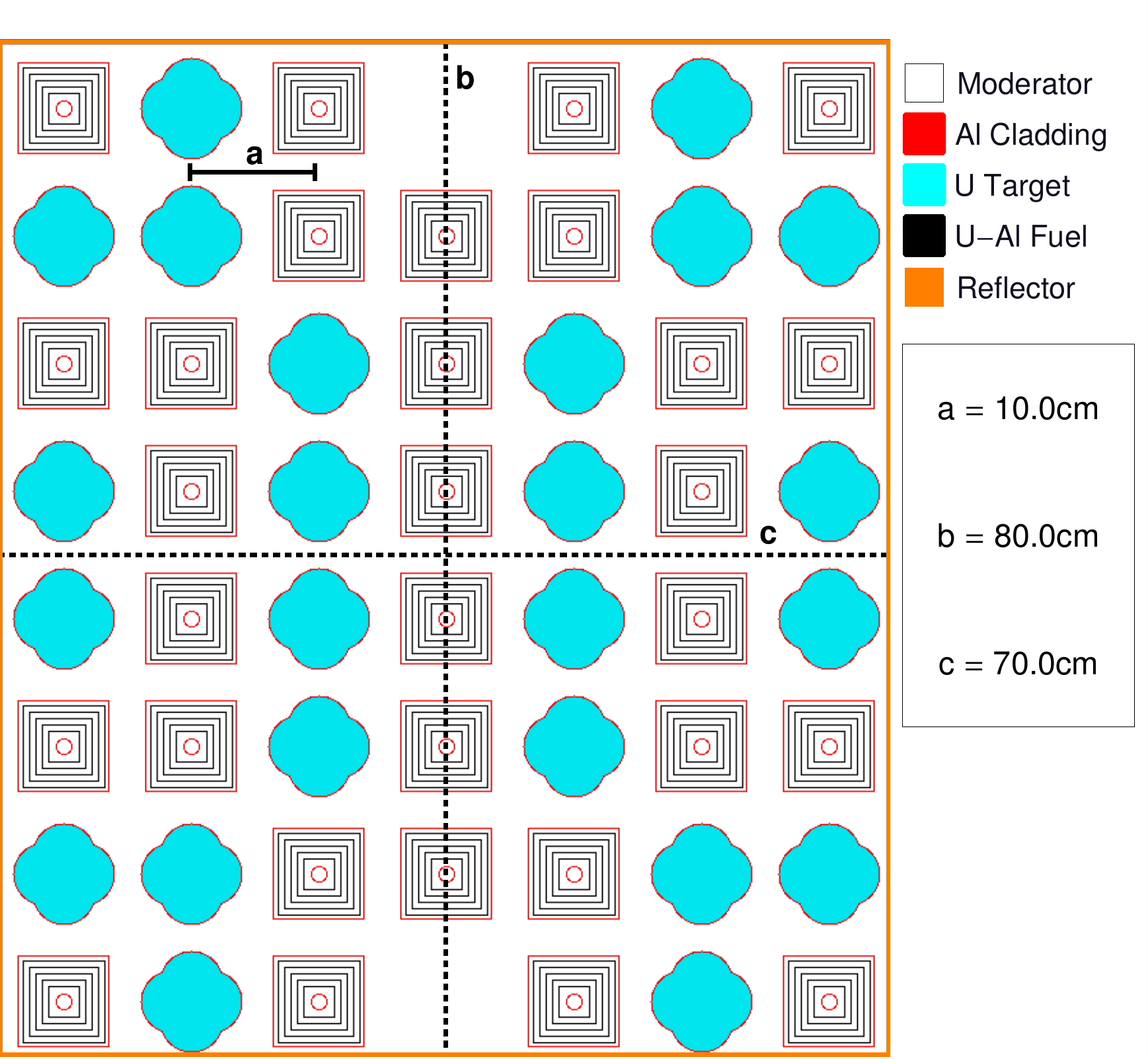}%
  \mycaption{\label{fig:coreconfigurations}  The left hand panel shows the IRT full core v1 material diagram, whereas the right hand panel shows the IRT full core v2 material diagram.}%
\end{figure}

The actual core configuration is not known and detailed designs are
unavailable so we will first determine the impact of the unknown core
configuration on the fission rates, as these rates determine the
neutrino spectrum. We tested several methods of computing the fission
rates of the IRT research reactor. The first group of methods
considered a calculation of a full IRT core with both drivers and
targets present. Two core configurations were tested and are
illustrated in figure~\ref{fig:coreconfigurations}.

\begin{figure}[h]
  \includegraphics[width=0.40\textwidth]{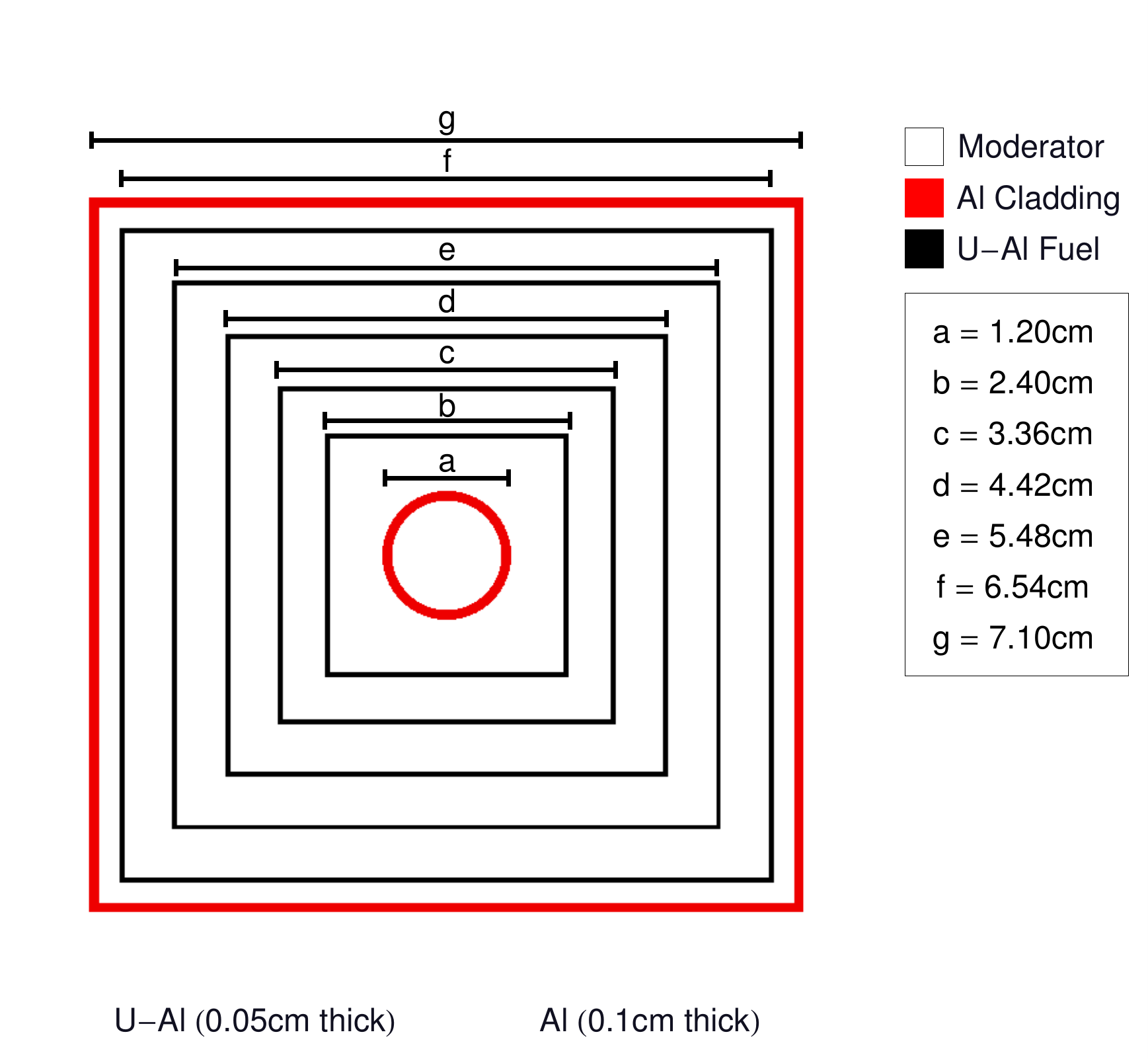}%
 \includegraphics[width=0.4\textwidth]{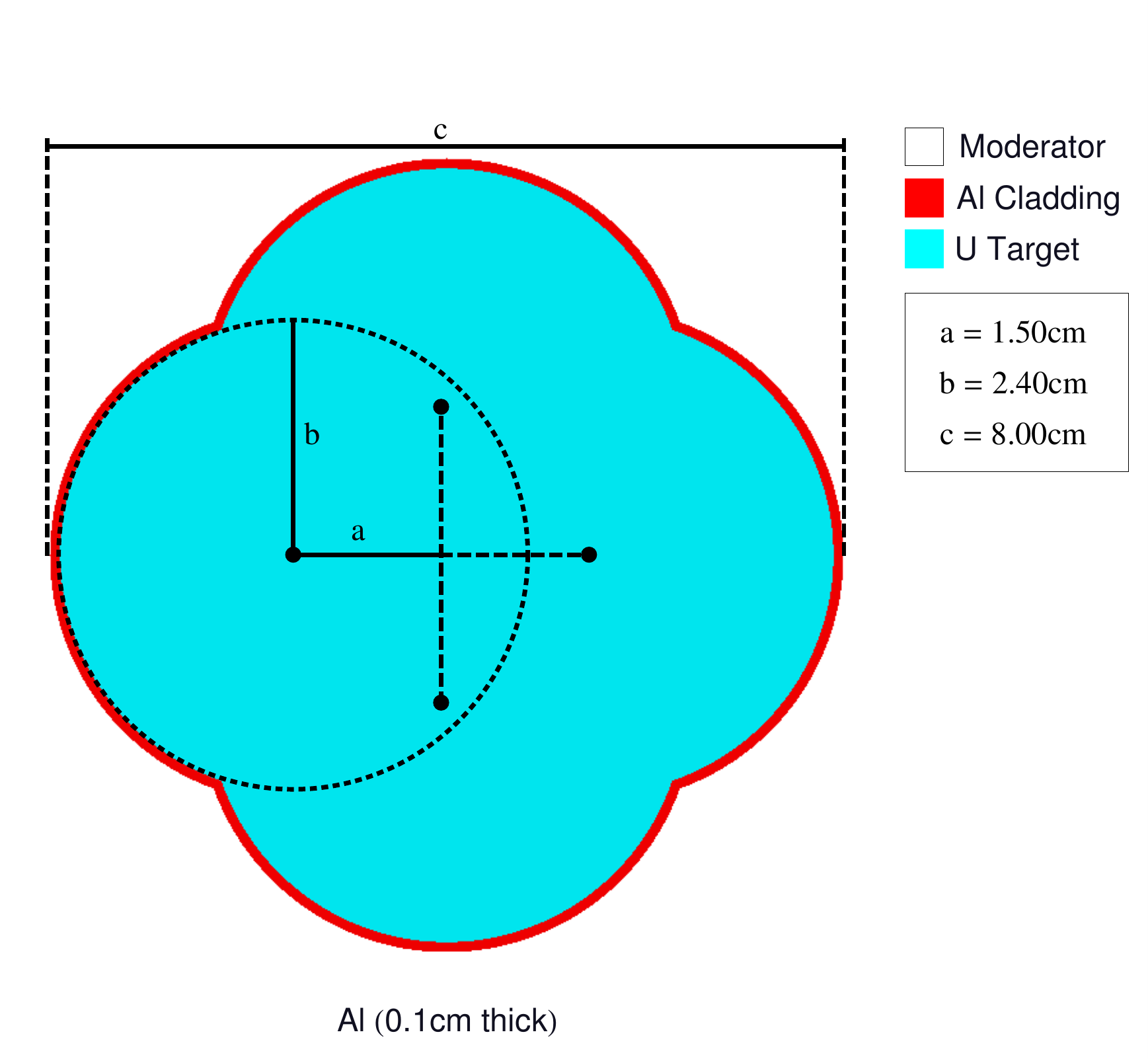}
  \mycaption{\label{fig:targetdriverelement}  The left hand panel shows the IRT driver element material diagram, whereas the right hand panel shows IRT target element material diagram.}%
\end{figure}

The second group of methods consisted of calculating the drivers and
targets separately and then summing the fission rates and fission
yields for a full core. For this group we considered three
sub-cases. We calculated one lone driver and one lone target
separately, multiple drivers and multiple targets separately in a
first core configuration and multiple drivers and multiple
targets separately in a second, different, core configuration. The
specific dimensions of each element and the two different core
configurations can be found in figure~\ref{fig:targetdriverelement} and
figure~\ref{fig:coreconfigurations}.  For the multiple drivers and
targets separately we simply removed either the target or driver
elements in figure~\ref{fig:coreconfigurations}.

The fission rates across these several methods were found to be nearly
identical.  The full core of both targets and drivers produced
slightly higher plutonium-239 and uranium-238 fission rates. This is a
result of the additional fission neutrons produced from the interplay
between drivers and targets, which combines with the total neutron
flux. This increase is a very small effect, especially when compared
to the uranium-235 fission rate, which dominates the IRT. From this, we
conclude that a specific core configuration has only a minor
impact on the fission rates and we use the lone element and
lone driver calculations. The primary reasoning behind this is to
avoid any unverifiable core configuration bias. The fission rates for
the lone element method are given in figure~\ref{fig:IRTfissrates}.
Blue curves indicate fission rates for the driver, red curves indicate
fission rates for the targets and purple curves are the sum of
these. We have only illustrated the three main fissiles that
contribute to the overall fission rates. The drivers comprise the
majority of uranium-235 fissions and, thus, the majority of fissions
over all fissiles.  The addition of targets increases the plutonium-239
and uranium-238 fission rates, but these are still an order of magnitude
lower than the uranium-235 fission rates. Again, we see the burn-up
effect by the decreasing fission rate of uranium-235 along with an
increase in the plutonium-239 fission rate.

We also wish to track the abundances of the four main fissile isotopes
as a function of time over the given 250 day irradiation cycle of the
IRT. The abundances have been calculated using SCALE in the same
fashion as described before. We attempt three different methods, a
lone driver and lone target core individually calculated and then
summed, a full core with drivers and a full core with targets individually
calculated and then summed, and a full core containing both drivers
and targets. Again, the differences in these three methods were
extremely minor resulting in the similar conclusion that the core
configuration is negligible when considering the fissile abundances
for the IRT. In the right panel of figure~\ref{fig:IRTfissrates} we
compare the fissile abundances as produced via the lone target and
element separately calculated and then summed to a full core with
table~VIII.7 in~\citetitle{puzzle}. We find that there is an excellent
match between the two results. The largest deviation occurs in
plutonium-241 as there is little of this isotope being produced.

To summarize, we conclude that all methods we tested for computing the
fission rates, isotopic abundances, and neutrino fluxes for the IRT
yield very similar results.  The main effect in the IRT is the addition of
breeding targets and we reproduce previous results in the literature.


\end{appendix}

%%%%%%%%%%%%%%%%%%%%%%%%%%%%%%%%%%%%%%%%%%%%%%%%%%%%%%%%%%%%%%%%%%%
%\bibliographystyle{plain} 
\newpage
%\theendnotes
%\printbibliography
%%%%%%%%%%%%%%%%%%%%%%%%%%%%%%%%%%%%%%%%%%%%%%%%%%%%%%%%%%%%%%%%%%%

%\listoffigures

\end{document}